\documentclass[superscriptaddress,groupedaddress,nofootnoteinbib,11pt]{article}
\pdfoutput=1 

\usepackage{jheppub} 

\usepackage{amsmath, amsfonts, amsthm, amssymb, graphicx, color, hyperref,braket}

\usepackage{float, subfig}
\usepackage{pstool}
\usepackage[utf8]{inputenc}
\usepackage[normalem]{ulem}
\usepackage[english]{babel}
\usepackage{blkarray}
\usepackage{dsserif}
\usepackage{mathtools}
\usepackage{makecell}
\usepackage{multicol}
\usepackage{cleveref}
\usepackage{url}
\usepackage{dsfont}
\usepackage{hyperref}
\usepackage{multirow}
\usepackage{longtable}
\usepackage{diagbox}

\allowdisplaybreaks[3]

\def \bal#1\eal  {\begin{align} #1 \end{align}}
\def\({\left(}
\def\){\right)}
\def\[{\left[}
\def\]{\right]}
\def\<{\langle}
\def\>{\rangle}

\newcommand{\eref}[1]{Eq.~\eqref{#1}}
\newcommand{\f}[2]{\frac{#1}{#2}}
\newcommand{\bim} {\begin{itemize}[noitemsep]}

\newcommand{\eim}{\end{itemize}}
\newcommand{\be} {\begin{equation}}
\newcommand{\ee} {\end{equation}}
\newcommand{\bc}{\begin{center}}
\newcommand{\ec}{\end{center}}

\newcommand{\nn} {\nonumber\\}

\newcommand{\blue}[1]{{\color{blue}{#1}}}

\newcommand{\ie}{{\it i.e.,}~}

\newcommand{\dd} {\delta}

\newcommand{\mc} {\mathcal}

\newcommand{\bfa} {{\bf a}}   \newcommand{\bfb} {{\bf b}}   
      
      \newcommand{\bfi} {{\bf i}}
\newcommand{\bfj} {{\bf j}}      
\newcommand{\bfm} {{\bf m}}   \newcommand{\bfn} {{\bf n}}

\newcommand{\ai}{{\alpha}}
\newcommand{\bi}{{\beta}}
\newcommand{\gi}{{\gamma}}

\newcommand{\ti}{{\tau}}

\newcommand{\Li}{\Lambda}
\usepackage[dvipsnames]{xcolor}
\definecolor{c1}{HTML}{D9A1A7}
\definecolor{c2}{HTML}{A1B4A5}
\definecolor{c3}{HTML}{FCD38F}
\definecolor{c4}{HTML}{309F96}

\title{Full positivity bounds for anomalous quartic gauge couplings in SMEFT}

\author[a,b]{Fu-Ming Chang}
\author[a,b]{, Zhuo-Yan Chen}
\author[a,b]{and Shuang-Yong Zhou}

\affiliation[a]{Interdisciplinary Center for Theoretical Study, University of Science and Technology of China, Hefei, Anhui 230026, China}
\affiliation[b]{Peng Huanwu Center for Fundamental Theory, Hefei, Anhui 230026, China}

\emailAdd{changfum@mail.ustc.edu.cn}
\emailAdd{kd511025@mail.ustc.edu.cn}
\emailAdd{zhoushy@ustc.edu.cn}

\preprint{{\footnotesize USTC-ICTS/PCFT-26-11}}

\date{\today}

\abstract{Electroweak boson scattering at the LHC provides a crucial avenue for probing physics beyond the Standard Model, particularly regarding deviations in quartic gauge couplings. We derive the complete set of positivity bounds for the $22$ dimension-$8$ anomalous quartic gauge coupling (aQGC) coefficients within the Standard Model Effective Field Theory (SMEFT). Moving beyond previous studies limited to transverse vector bosons, our analysis incorporates all electroweak boson modes, explicitly constructing the extremal rays (ERs) of the positivity cone through a group theoretic framework. We utilize two independent methods--direct construction and Casimir operator analysis--to determine these rays, addressing complexities such as parity-violating operators and continuous parameter degeneracies. Our results indicate that the positivity bounds impose severe constraints, restricting the physically viable parameter space to approximately $0.05255\%$ of the naive total space. Furthermore, we derive linear analytical bounds for various operator combinations and provide an easy-to-use Python package, {\tt SMEFTaQGC}, which implements algorithms to numerically verify positivity and compute the optimized positivity bounds for general aQGC configurations.}

\begin{document}
\maketitle
\flushbottom

\section{Introduction}

The Standard Model Effective Field Theory (SMEFT) provides a systematic, bottom-up extension of the successful Standard Model (SM) by incorporating higher-dimensional operators that parameterize possible effects of new physics~\cite{Weinberg:1978kz,Buchmuller:1985jz,Leung:1984ni}. In recent years, in the absence of direct evidence for new particles at high-energy colliders, the SMEFT has emerged as a unified framework for interpreting collider measurements, providing a model-independent bridge between experimental data and possible UV theories \cite{Brivio:2017vri,deBlas:2022ofj,Contino:2013kra}.

However, owing to the rich particle content of the Standard Model, the corresponding SMEFT contains a large number of effective operators and independent Wilson coefficients. Naively, this leads to an extremely high-dimensional parameter space, making a comprehensive analysis in the SMEFT and the extraction of UV information both computationally demanding and conceptually challenging. Fortunately, theoretical consistency conditions can dramatically restrict this vast parameter space to physically viable regions~\cite{Zhang:2018shp,Low:2009di, Bellazzini:2018paj, Gomez-Ambrosio:2018pnl, Bi:2019phv,Remmen:2019cyz, Zhang:2020jyn, Yamashita:2020gtt, Trott:2020ebl, Remmen:2020vts, Remmen:2020uze, Gu:2020thj, Fuks:2020ujk, Gu:2020ldn, Bonnefoy:2020yee, Li:2021lpe, Davighi:2021osh, Chala:2021wpj, Zhang:2021eeo, Chala:2021pll, Boughezal:2021tih, Ghosh:2022qqq, Remmen:2022orj, Li:2022tcz, Li:2022rag, Li:2022aby, Altmannshofer:2023bfk, Yang:2023ncf,Davighi:2023acq,Chala:2023xjy,Chen:2023bhu,Hong:2024fbl,Remmen:2024hry,Chakraborty:2024ciu,Liu:2025deo}. These are positivity bounds for Lorentz-invariant EFTs, which arise from imposing fundamental principles of quantum field theory/the S-matrix such as causality and unitarity ~\cite{Adams:2006sv,Tolley:2020gtv,Caron-Huot:2020cmc,deRham:2017avq,deRham:2017zjm,Chiang:2022ltp,Arkani-Hamed:2020blm,Bellazzini:2020cot,Sinha:2020win,Alberte:2020jsk,Guerrieri:2021tak,Alberte:2021dnj,Caron-Huot:2021rmr,Du:2021byy,Pham:1985cr,Pennington:1994kc,Ananthanarayan:1994hf,Comellas:1995hq,Manohar:2008tc,Chiang:2021ziz,Sanz-Cillero:2013ipa,Bellazzini:2016xrt,deRham:2018qqo,deRham:2017imi,Bonifacio:2016wcb,Du:2016tgp,Bellazzini:2017fep,Hinterbichler:2017qyt,Bellazzini:2017bkb,Bonifacio:2018vzv,Bellazzini:2019xts,Melville:2019wyy,Melville:2019tdc,deRham:2019ctd,Alberte:2019xfh,Alberte:2019zhd,Ye:2019oxx,Huang:2020nqy,Wan:2024eto,Eliasmiro:2022xaa,deRham:2025vaq,deRham:2025htd,Cheung:2025nhw,Bellazzini:2015cra,Cheung:2016yqr,Tokuda:2020mlf,Caron-Huot:2022ugt,Chiang:2022jep,Henriksson:2022oeu,Hong:2023zgm,Xu:2024iao,Bellazzini:2023nqj,Huang:2025icl,Wang:2020jxr,Fernandez:2022kzi,Albert:2023jtd,Ma:2023vgc,Li:2023qzs,Albert:2023seb}(see~\cite{Derham:2022hpx} for a review). 

The key ingredient is causality, which implies the analyticity of scattering amplitudes and thereby allows one to derive nonperturbative dispersion relations. More precisely, positivity bounds are UV unitarity constraints passed down to the IR via the dispersion relations, without reference to any concrete UV completion. As a result, they are robust, model-independent and easy-to-use theoretical bounds on the Wilson coefficients, and their application within the SMEFT framework aligns well with the underlying spirit of effective field theory. Any point in the naive parameter space that violates positivity bounds cannot originate from a standard Wilsonian UV completion.

Electroweak boson scattering plays a central role in probing the high-energy behavior of the symmetry-breaking sector of the Standard Model and its possible extensions \cite{He:1994br,ATLAS:2020nlt,Covarelli:2021gyz}. Deviations from the SM predictions in vector boson scattering, as well as in vector boson-Higgs scattering, can be systematically parameterized by anomalous quartic gauge couplings (aQGCs), which involve both dim-6 and dim-8 operators~\cite{Eboli:2016kko,Durieux:2024zrg}. However, in the SMEFT, dim-6 operators cannot contribute to quartic gauge couplings without simultaneously affecting triple gauge couplings~\cite{Green:2016trm}, which can be tightly constrained in other scattering channels~\cite{Chang:2013aya}. As a result, genuine aQGCs arguably first arise at dim-8, where they can induce scattering amplitude terms that grow as $s^2$ (with $s$ denoting the center-of-mass energy squared). These contributions can be systematically constrained by the leading order positivity bounds.

Concretely, by applying the optical theorem to the dispersion relations, physically viable $s^2$ coefficients $M_{ijkl}$ of the scattering amplitudes are found to lie within a convex cone ${\cal C}$ \cite{Zhang:2020jyn,Bellazzini:2014waa,Bonnefoy:2025uzf,Zhang:2021eeo},
\begin{equation}
    M_{ijkl}\in \mathcal{C}\equiv \mathrm{cone}\left( \{  A_{ij\to X}A^*_{kl\to X}+ A_{i\bar{l}\to X}A^*_{k\bar{j}\to X} \} \right) ,
\end{equation}
where $i,j,k,l$ denote the scattering particles, $\bar l,\bar j$ denote the corresponding anti-particles, and $X$ denotes a generic intermediate state arising from the optical theorem. 
The physically-allowed space of the Wilson coefficients can be thus characterized by the extreme rays (ERs) of this {\it amplitude cone}, which can be constructed straightforwardly using the projectors of the symmetries of the theory. More precisely, the positivity bounds correspond to the facets (or co-dimension-1 boundaries) of the amplitude cone, providing the cone's inequality representation. These facets can be computed from the cone’s extremal rays using a vertex-enumeration algorithm, and the resulting positivity bounds obtained within this convex-cone framework provide the strongest constraints that follow directly from the fundamental principles of the S-matrix, while remaining completely agnostic about the details of the UV completion. 

This extremal positivity approach has been applied to extract optimal positivity bounds from scatterings involving only transversal electroweak vector bosons~\cite{Yamashita:2020gtt} (see \cite{Zhang:2018shp, Bi:2019phv, Remmen:2019cyz} for earlier (non-optimal) positivity bounds on vector boson scattering and \cite{Remmen:2024hry} for extremal positivity bounds on Higgs scattering in Higgs EFT). 
However, scattering processes involving transversal electroweak vector bosons constitute only a limited subspace of the broader LHC electroweak program. A comprehensive exploration of electroweak dynamics, particularly in the context of new physics searches, requires going beyond the transverse sector to include the Higgs boson related contributions and their interference effects, which are often more sensitive to deviations from the SM.

In this paper, we generalize this approach to compute the optimal positivity bounds on the full set of aQGC coefficients, which can be extracted from scattering amplitudes involving all electroweak bosons. Specifically, we explicitly construct the primal positivity cone in a group-theoretical manner by carefully analyzing the convex structure of all 2-to-2 scatterings among the following low-energy modes,
\begin{equation}
B_x,B_y,W^1_x,W^1_y,W^2_x,W^2_y,W^3_x,W^3_y,H^1,H^2,H_1^{\dagger},H_2^{\dagger},    
\end{equation}
where $x$ and $y$ label the polarization modes of the vector bosons, and the remaining indices denote the relevant gauge-group indices. Owing to the large number of modes involved, extracting the ERs in the present case is considerably more involved than in previous studies. We first revisit and clarify the group-theoretical framework for constructing ERs from the Clebsch–Gordan (CG) coefficients of the symmetry groups, including aspects of discrete symmetries, as our analysis now involves P-violating but CP-even coefficients. We then derive these ERs using two independent methods: direct construction via mapping to the standard CG coefficients and the computation of eigenstates of the relevant Casimir operators. Along the way, we clarify a previously imprecisely stated point concerning the predictive power of ERs for the spin–parity properties of UV states.

\begin{figure}[H]
    \centering
    \includegraphics[width=0.45\linewidth]{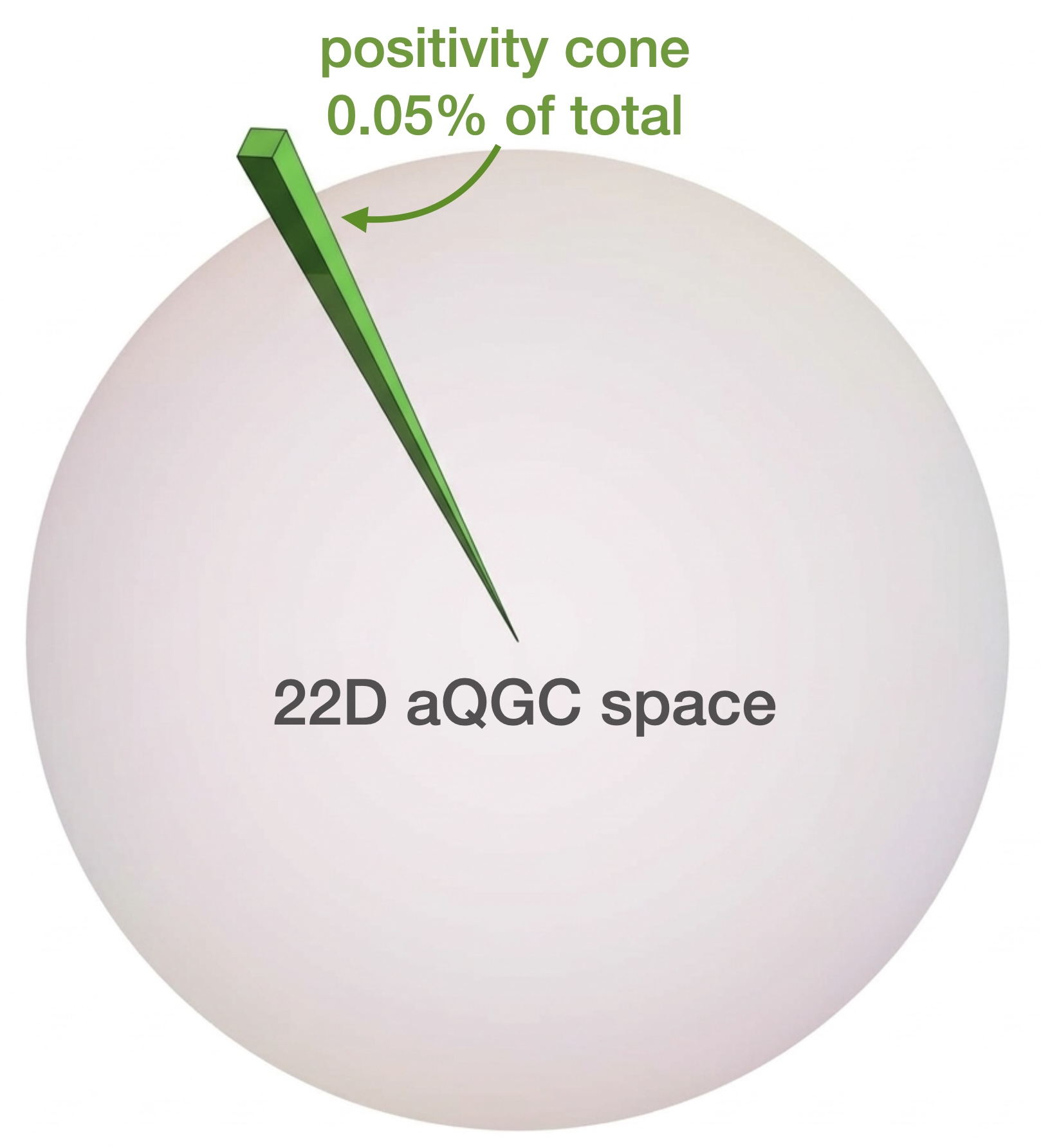}
    \caption{Cartoon representation of the positivity (convex) cone within the 22D SMEFT aQGC parameter space. Imposing the fundamental principles of the S-matrix, we find that the theoretically consistent aQGC region, {\it i.e.}, the positivity cone, occupies only $0.05255\%$ of the total naive parameter space.}
    \label{fig:solidfig}
\end{figure}

In this construction, in addition to isolated ERs, we also encounter parameterized, continuous ERs, reflecting degeneracies among different low-energy configurations. The increased computational complexity in the full aQGC case arises from the presence of two independent continuous parameters in some ERs, which makes an analytical treatment challenging. Nevertheless, through suitable discretization, we are able to obtain the optimal bounds numerically. We devise several algorithms to probe the properties of the amplitude cone, including testing the positivity of a given set of coefficients and extracting the optimal bound along a specified direction. These algorithms have been efficiently implemented using linear programming and are provided in the accompanying Python package {\tt SMEFTaQGC}. By working in the dual cone, we are also able to derive some optimal linear analytic bounds, which are particularly easy to use. In general, however, the optimal bounds are nonlinear, while additional linear bounds can be readily obtained numerically. These bounds are found to significantly reduce the allowed physical region in the naive aQGC parameter space (see Figure.~\ref{fig:solidfig}), providing robust, model-independent guidance for future theoretical and experimental searches for potential aQGCs.

The structure of this paper is as follows. In \Cref{sec:aQGC}, we list all the relevant dim-$8$ CP-even effective aQGCs operators, noting that two of them, $O_{M,8}$ and $O_{M,9}$ are parity violating. In \Cref{sec:congeoEFTamp}, we introduce the theoretical foundations of our work. In \Cref{sec:disper_rel}, we show that dispersion relations imply a cone structure for the $s^2$ coefficients in $2\to2$ forward (IR-subtracted) EFT amplitudes, and the corresponding ERs can be constructed using the projectors of the symmetry group, as discussed in \Cref{sec:Perssymmetry}. \Cref{sec:CP} and \Cref{sec:exchangeCG} clarify the CP properties of the asymptotic states and the spin-parity of the UV states. In \Cref{sec:aQGCER}, we describe the two methods we use to determine the ERs of the aQGC positivity cone, and present the final form of ERs in~\Cref{sec:ERs}. Finally, we convert the ER representation of the amplitude cone to its inequality representation to obtain the optimal positivity bounds for aQGC coefficients in~\Cref{sec:bounds}, and summarize our findings in~\Cref{sec:conclusion}.

\section{Effective aQGC operators}
\label{sec:aQGC}

We begin by listing the effective operators most relevant to our analysis. Our focus will be on scatterings among electroweak vector bosons, as well as scatterings between vector bosons and the Higgs boson, both of which are directly connected to the anomalous quartic gauge couplings (aQGCs). In the SMEFT, aQGCs can arise from both dim-6 and dim-8 operators,
\begin{equation}
\label{eq:smeftLagGen}
  \mathcal{L}_{\rm SMEFT}=\mathcal{L}_{\rm SM} + \sum_{i} \frac{c_{i}^{(6)}O_{i}^{(6)}}{\Lambda^2} + \sum_{i} \frac{c_{i}^{(8)}O_{i}^{(8)}}{\Lambda^4} + \cdots,
\end{equation}
where the $SU(2)_L\otimes U(1)_Y$ gauge symmetry is linearly realized in the massless limit. However, the dim-6 QGCs are fully correlated with the three dim-6 TGC operators, which are loop-induced and thus suppressed. Genuine aQGC effects are therefore expected to first appear at dim-8. If sizable, dim-6 effects would more likely manifest in other observables first. For this reason, we neglect the contributions from dim-6 operators and focus on those of dim-8\,\footnote{An explicit class of (partial) UV scenarios where dimension-6 operators are not generated to leading order is provided by a massive spin-2 resonance, with a Kaluza–Klein graviton in extra-dimensional models furnishing a concrete realization \cite{Han:1998sg}. Integrating out such a resonance at tree level generates dim-8 operators without inducing the relevant dim-6 operators. For fixed-helicity amplitudes, dim-6 contributions can vanish in the forward limit, and the leading BSM effects arise from dim-8 operators; see, e.g.,~\cite{Gu:2020ldn,Liu:2025deo}. }. This choice is further motivated by the fact that linear dimension-6 terms do not contribute to positivity bounds, while quadratic dimension-6 terms contribute negatively (see Appendix \ref{appd:dim6part}), so neglecting the dimension-6 effects thus yields bounds that are valid but conservative \cite{Zhang:2018shp, Li:2021lpe}.

The dim-8 aQGC operators are usually categorized into three classes: 1) S-type operators, which are quartic in the Higgs field, 2) M-type operators, 
which are bi-quadratic in the Higgs and gauge field strengths, mixing the $\Phi$ scalar and the $W$ and $B$ tensors, and 3) T-type operators, which are quartic in the gauge field strengths. 
The independent set of CP-even dim-8 aQGC operators has been constructed over the years \cite{Remmen:2019cyz, Durieux:2024zrg,Almeida:2020ylr} (see also \cite{Li:2020gnx,Murphy:2020rsh} for the complete set of all dim-8 SMEFT operators). We adopt the basis newly proposed in~\cite{Durieux:2024zrg}, which extends the basis of~\cite{Almeida:2020ylr} by incorporating two parity-violating but CP-even operators ({\it i.e.}, $O_{M,8}$ and $O_{M,9}$) that were not identified in~\cite{Almeida:2020ylr}. In this basis, the 22 independent, CP-even dim-8 aQGC operators are given by
\begin{align}
  O_{S,0} & =\left[  \left(  D_{\mu}\Phi \right)  ^{\dagger}D_{\nu}\Phi \right] \times \left[  \left(  D^{\mu}\Phi \right)  ^{\dagger}D^{\nu}\Phi \right], &\quad O_{M,9} & =-\frac{i}{2} D^{\mu}\Phi^{\dagger}\hat{W}_{\mu \rho}\tilde{W}^{\nu \rho}D_{\nu}\Phi+\mathrm{h.c.},\nn
  O_{S,1} & =\left[  \left(  D_{\mu}\Phi \right)  ^{\dagger}D^{\mu}\Phi \right] \times \left[  \left(  D_{\nu}\Phi \right)  ^{\dagger}D^{\nu}\Phi \right], &\quad  O_{T,0} & =\mathrm{Tr}\left[  \hat{W}_{\mu \nu}\hat{W}^{\mu \nu}\right]\mathrm{Tr}\left[  \hat{W}_{\alpha \beta}\hat{W}^{\alpha \beta}\right],\nn
  O_{S,2} & =\left[  \left(  D_{\mu}\Phi \right)  ^{\dagger}D_{\nu}\Phi \right] \times \left[  \left(  D^{\nu}\Phi \right)  ^{\dagger}D^{\mu}\Phi \right],&\quad O_{T,1} & =\mathrm{Tr}\left[  \hat{W}_{\alpha \nu}\hat{W}^{\mu \beta}\right] \mathrm{Tr}\left[  \hat{W}_{\mu \beta}\hat{W}^{\alpha \nu}\right],\nn
  O_{M,0} & =\mathrm{Tr}\left[  \hat{W}_{\mu \nu}\hat{W}^{\mu \nu}\right] \times \left[  \left(  D_{\beta}\Phi \right)  ^{\dagger}\left(  D^{\beta}
  \Phi \right)  \right]  ,&\quad O_{T,2} & =\mathrm{Tr}\left[  \hat{W}_{\alpha \mu}\hat{W}^{\mu \beta}\right]  \mathrm{Tr}\left[  \hat{W}_{\beta \nu}\hat{W}^{\nu \alpha}\right],\nn
  O_{M,1} & =\mathrm{Tr}\left[  \hat{W}_{\mu \nu}\hat{W}^{\nu \beta}\right]  \times \left[  \left(  D_{\beta}\Phi \right)  ^{\dagger}D^{\mu}\Phi \right] &\quad    O_{T,3} & =\mathrm{Tr}\left[ \hat{W}_{\mu\nu}\hat{W}_{\alpha\beta} \right]\mathrm{Tr}\left[ \hat{W}^{\alpha\nu}\hat{W}^{\mu\beta} \right],\nn
  O_{M,2} & =\hat{B}_{\mu \nu}\hat{B}^{\mu \nu}\times \left[  \left(
  D_{\beta}\Phi \right)  ^{\dagger}D^{\beta}\Phi \right]  ,&\quad  O_{T,4} & =\mathrm{Tr}\left[ \hat{W}_{\mu\nu}\hat{W}_{\alpha\beta} \right] \hat{B}^{\alpha \nu}\hat{B}^{\mu \beta},\nn
  O_{M,3} & =\hat{B}_{\mu \nu}\hat{B}^{\nu \beta}\times \left[  \left(  D_{\beta}\Phi \right)  ^{\dagger}D^{\mu}\Phi \right]   &\quad  O_{T,5} & =\mathrm{Tr}\left[  \hat{W}_{\mu \nu}\hat{W}^{\mu \nu}\right]\hat{B}_{\alpha \beta}\hat{B}^{\alpha \beta},\nn
  O_{M,4} & =\frac{1}{2}\left[  \left(  D_{\mu}\Phi \right)  ^{\dagger}\hat{W}_{\beta \nu}\left(  D^{\mu}\Phi \right)  \right]  \times \hat{B}^{\beta \nu},&\quad  O_{T,6} & =\mathrm{Tr}\left[\hat{W}_{\alpha \nu}\hat{W}^{\mu \beta}\right]  \hat{B}_{\mu \beta}\hat{B}^{\alpha \nu}, \nn
  O_{M,5} & =\frac{1}{4}\left[  \left(  D_{\mu}\Phi \right)  ^{\dagger}\hat{W}_{\beta \nu
  }\left(  D^{\nu}\Phi \right)  \right]  \times \hat{B}^{\beta \mu}+\mathrm{h.c.}, &\quad  O_{T,7} & =\mathrm{Tr}\left[  \hat{W}_{\alpha \mu}\hat{W}^{\mu \beta}\right]  \hat{B}_{\beta \nu}\hat{B}^{\nu \alpha},\nn
  O_{M,7} & =\frac{1}{4}\left[  \left(  D_{\mu}\Phi \right)  ^{\dagger}\hat{W}_{\beta \nu
  }\hat{W}^{\beta \mu}\left(  D^{\nu}\Phi \right)  \right]  ,&\quad  O_{T,8} &  =\hat{B}_{\mu \nu}\hat{B}^{\mu \nu}\hat{B}_{\alpha \beta}\hat{B}^{\alpha \beta},\nn
  O_{M,8} & =\frac{i}{4} \left[ D_{\mu}\Phi^{\dagger}\tilde{W}_{\nu \rho}D^{\nu}\Phi \right]  \hat{B}^{\mu \rho}+\mathrm{h.c.}, &\quad   O_{T,9} & =\hat{B}_{\alpha \mu}\hat{B}^{\mu \beta}\hat{B}_{\beta \nu}\hat{B}^{\nu \alpha}. 
\end{align}
where $\epsilon_{\mu\nu\rho\sigma}$ is the Levi-Civita tensor with $\epsilon_{0123}=1$, $D_\mu=\partial_\mu-ig W^I_\mu\sigma^I-\frac{i}{2}g'B_\mu$
\begin{align}
  \hat{W}^{\mu\nu}&\equiv ig\frac{\sigma^I}{2}W^{I,\mu\nu}, \quad \hat{B}^{\mu\nu}\equiv ig'\frac{1}{2}B^{\mu\nu},\quad \tilde{W}^{\mu\nu}\equiv ig\frac{\sigma^I}{2}\left(\frac{1}{2}\epsilon_{\mu\nu\rho\sigma}W^{I,\rho\sigma}\right),\nonumber\\
  W^{I}_{\mu\nu}&=\partial_\mu W^I_{\nu}-\partial_\nu W^I_{\mu}+g\epsilon_{IJK}W^J_{\mu} W^K_{\nu},\quad B_{\mu\nu}=\partial_\mu B_\nu-\partial_\nu B_\mu,\label{eq:WBfield}
\end{align}
and $g$ and $g'$ are $SU(2)_L$ and $U(1)_Y$ gauge couplings, respectively.
Accordingly, the effective Lagrangian for the aQGC interactions will be written as
\begin{equation}
  \mathcal{L}_{\rm aQGC}=\sum_{i} \frac{f_{S,i}}{\Lambda^4}O_{S,i}+\sum_{i} \frac{f_{M,i}}{\Lambda^4}O_{M,i}+\sum_{i} \frac{f_{T,i}}{\Lambda^4}O_{T,i},
\end{equation}
we will find it convenient to absorb the electromagnetic coupling constant $e$, as well as $\cos\theta_w$ and $\sin\theta_w$ of the Weinberg angle $\theta_w$, into the Wilson coefficients as follows
\begin{equation}
  F_{S,\mathcal{I}_1}\equiv f_{S,\mathcal{I}_1}, \quad F_{M,\mathcal{I}_2}\equiv \frac{e^2}{s_w^nc_w^{2-n}} f_{M,\mathcal{I}_2}, \quad F_{T,\mathcal{I}_3}\equiv \frac{e^4}{s_w^n c_w^{4-n}} f_{T,\mathcal{I}_3}.
\end{equation}
Here, $n=0$ for $\mathcal{I}_2=\{2,3\}$ and $\mathcal{I}_3=\{8,9\}$; $n=1$ for $\mathcal{I}_2=\{4,5,8\}$; $n=2$ for $\mathcal{I}_2=\{0,1,7,9\}$ and $\mathcal{I}_3=\{4,5,6,7\}$; and $n=4$ for $\mathcal{I}_3=\{0,1,2,3\}$.

All these dim-8 aQGC coefficients linearly enter the tree-level 2-to-2 VBS (+ Higgs) amplitudes, appearing as coefficients of the $s^2$ terms. The next step is to derive the optimal positivity bounds on these $s^2$ amplitude coefficients, based on the fundamental principles of quantum field theory, particularly dispersion relations, and to use these bounds to carve out the theoretically consistent region of the aQGC parameter space. As we will see, this consistent region forms a  convex cone, and the cone only accounts for a tiny fraction of the naive total space. As these aQGC operators involves many symmetries, the problem of finding the convex cone can be largely achieved by enumerating the extremal rays of the cone with a group-theoretical method.

\section{Convex geometry of EFT amplitudes}\label{sec:congeoEFTamp}

The fundamental principles of the S-matrix, such as unitarity and analyticity, impose strong constraints on the SMEFT parameter space in the form of positivity bounds \cite{Zhang:2018shp}. As emphasized in \cite{Zhang:2020jyn}, the optimal positivity bounds on the $s^2$ coefficients take a natural convex-geometric form, especially in settings with substantial symmetry where the positivity cone can be determined by the group projectors of the scatterings involved. In this section, we shall briefly review the theoretical framework and the procedure for deriving the convex positivity bounds.

\subsection{From dispersion relations to the amplitude cone}\label{sec:disper_rel}

The first step is to derive the dispersion relations (or sum rules) between the $s^2$ coefficients ($s,t,u$ being the standard Mandelstam variables) and potential UV spectra. The dispersion relations encode causality or analyticity of the S-matrix, and positivity bounds on the EFT coefficients are exactly the positivity part of the unspecified UV theory's unitarity conditions, transmitted to the IR by the dispersion relations.

\begin{figure}
    \centering
    \includegraphics[width=0.6\linewidth]{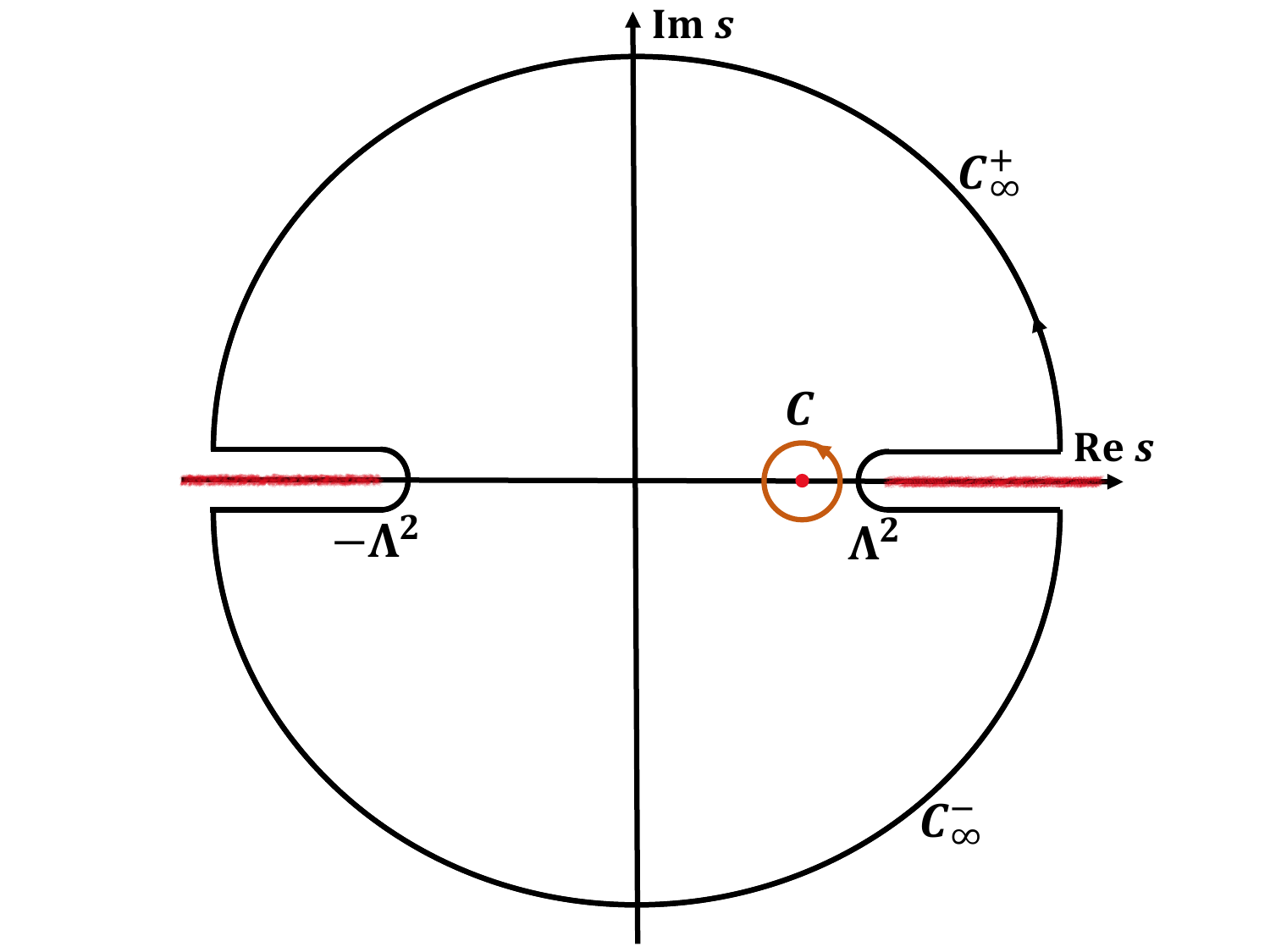}
    \caption{Contours to derive dispersion relations for the pole-subtracted, forward amplitude $\bar A_{ijkl}(s)$. The red branch cuts encode possible UV states.}
    \label{fig:contour}
\end{figure}

We shall consider all possible $2$-to-$2$ scattering amplitudes $A_{ij kl}$ of (electroweak) vector bosons and the Higgs, with initial-state particles labeled as $i$ and $j$ and final-state particles labeled as $k$ and $l$. We shall take all SM particles to be massless, as appropriate in the SMEFT framework where the electroweak symmetry is linearly realized.  For the high-energy scatterings of interest, these masses are negligible compared to the characteristic scales. In the massless limit, SMEFT amplitudes exhibit singularities near $s,t=0$, which obstruct the construction of the dispersion relations required for deriving positivity bounds. For the leading tree-level amplitudes, there are only simple poles at $s,t=0$, which we can easily subtract. 
Since we are only interested in the positivity of the $s^2$ coefficients here, let us define the pole-subtracted, forward-scattering amplitudes:
\bal
\label{eq:AmodDef}
\bar {A}_{ijkl}(s) &\equiv \bigg[ A_{ij kl}(s,t) - (\text{IR poles}) \bigg]_{t\to 0} 
\eal
More explicitly, the low-energy $s$- and $u$-channel poles can be safely subtracted after deforming the contour, since their residue contributions can be explicitly isolated. For a $t$-channel pole arising from the exchange of a particle with spin $J<2$, the pole contribution grows at most linearly with $s$ and can therefore be removed without conflicting with the twice-subtracted dispersion relation. This argument does not apply to massless states with $J\geq 2$ \cite{Alberte:2020jsk}. Since the relevant SM exchanges in our analysis have $J=0$ or $J=1$, the IR-pole subtraction in Eq.(3.1) is well defined.

Thank to Martin's analyticity \cite{Martin:1965jj}, the pole-subtracted amplitude $\bar A_{ijkl}(s)$ is analytical away from the real $s$ axis, and below the EFT cutoff $|s|<\Lambda^2$.
Then, by the Cauchy integral formula, the IR-subtracted amplitude can be written as
\begin{align}
    \f{\mathrm{d}^2}{2!\mathrm{d}s^2}\bar{A}_{ijkl}(s) &=\oint_C \f{\mathrm{d}\mu}{2\pi i}\frac{\bar{A}_{ijkl}(\mu)}{(\mu-s)^3}
    \\
    &=\int_{C^{\pm}_\infty}\f{\mathrm{d}\mu}{2\pi i}\frac{\bar{A}_{ijkl}(\mu)}{(\mu-s)^3} +\left(\int_{\Li^2}^{\infty}+\int^{-\Li^2}_{-\infty}\right)\f{\mathrm{d}\mu }{2\pi i}\frac{\mathrm{Disc}\bar{A}_{ijkl}(\mu )}{(\mu -s)^3} 
    \\
    &=\left(\int_{\Li^2}^{\infty}+\int^{-\Li^2}_{-\infty}\right)\f{\mathrm{d}\mu }{2\pi i}\frac{\mathrm{Disc}A_{ijkl}(\mu )}{(\mu -s)^3} ,
\end{align} 
where in the second line above the contour $C$ is deformed into two semicircular contours at infinity together with the two UV branch cuts (see Figure.~\ref{fig:contour}), and in the third line we have used the Froissart bound $|A_{ijkl}(s\to \infty)|< const\cdot |s \ln^2 s|$ to drop the semicircular contours and also utilized the fact that $\mathrm{Disc}\bar{A}_{ijkl}(\mu )=\mathrm{Disc}A_{ijkl}(\mu,t\to0 )$ for $\mu >\Li^2$ or $\mu <-\Li^2$ in our SMEFT setup. Invoking $s$-$u$ crossing symmetry of the amplitude, we get $\mathrm{Disc}A_{ijkl}(s) =-\mathrm{Disc}A_{i\bar{l}\to k\bar{j}}(-s)$, where the barred indices denote the corresponding anti-particles. (It is worth emphasizing that the indices $i,j,k,l$ run over all particle states, including their polarizations and other quantum numbers such as gauge indices.) Thus, the amplitude can then be expressed as
\begin{align}
    \f{\mathrm{d}^2}{2!\mathrm{d}s^2}\bar{A}_{ijkl}(s) &=\int_{\Li^2}^{\infty}\f{\mathrm{d}\mu }{2\pi i}\frac{\mathrm{Disc}A_{ijkl}(\mu )}{(\mu -s)^3}+\int^{\infty}_{\Li^2}\f{\mathrm{d}\mu }{2\pi i}\frac{\mathrm{Disc}A_{i\bar{l} k\bar{j}}(\mu )}{(\mu +s)^3}.
\end{align}
For notational simplicity, we define $M_{ijkl}$ to be the $s^2$ Taylor coefficients and get the sum rules
\begin{align}
M_{ijkl} &=\f{\mathrm{d}^2}{2!\mathrm{d}s^2}\bar{A}_{ijkl}(s)\Big|_{s=0} =\int_{\Lambda^{2}}^{\infty}\frac{\mathrm{d}\mu}{2\pi i}\frac{\mathrm{Disc}A_{ij kl}\left(\mu\right)  }{ \mu^{3}}+\left( j\to \bar l, l\to \bar j\right) \label{eq:disrel}
\end{align}

To get positivity bounds, we shall make use of the UV unitary conditions. Recall that the unitarity of the S-matrix can be formulated as the generalized optical theorem: $A_{ijkl}(s)-A_{kl ij}^*(s) =i\sum_{X} A_{ij\to X}(s) A_{kl\to X}^{*}(s)$, where the sum over the intermediate state $X$ is schematic and includes the integration of the phase space, and $A_{ij\to X}$ depends on $s$ as well as the momenta of $X$. The left-hand side can be written as a discontinuity thanks to the hermitian analyticity $A_{kl ij}(s+i\varepsilon)^*=A_{ijkl}(s-i\varepsilon)$, so we have 
\be
{\rm Disc}A_{ijkl}(s) =i\sum_{X} A_{ij\to X}(s) A_{kl\to X}^{*}(s)
\ee
A consequence of this unitarity condition is that, when viewing $ij$ as one index and $kl$ as another, $A_{ij\to X} A_{kl\to X}^{*}$ (and thus $\sum_X A_{ij\to X} A_{kl\to X}^{*}$) is a semi-positive definite matrix, which is the only part of unitarity that will be used in this paper. Thus, the final dispersion relations/sum rules we will use to derive the positivity bounds are 
\begin{equation}
M_{ijkl}  =\int_{\Lambda  ^{2}}^{\infty}\frac{\mathrm{d}\mu}{2\pi}\frac{\sum_{X}A_{ij\to X}(\mu)A_{kl\to X}^{*}(\mu)}{\mu^{3}}+\left( j\to \bar l, l\to \bar j\right)   .\label{eq:disp_opti}
\end{equation}
These sum rules represent a profound, un-decoupled IR-UV connection: $M_{ijkl}$'s are the $s^2$ Taylor coefficients of the low energy EFT amplitudes $\bar A_{ijkl}(s)$, which are directly linked to the high energy amplitudes on the right hand side where the integration runs over energies above $\Li^2$.

The philosophy of positivity bounds is to remain agnostic about the specific UV completion. Without specifying the UV theory, $A_{ij\to X}(\mu)(\equiv m_{ij}(\mu,X))$ is just an $n\times n$ matrix that depends on $\mu$ and $X$, where $n$ denotes the total number of particle states that $i$ or $j$ enumerates. To facilitate a smooth transition to the convex geometry treatment of the sum rules, it is instructive to write \eref{eq:disp_opti} in a more abstract form
\begin{equation}
  M_{ijkl}=\sum_{\mu,X} \big[ m_{ij}(\mu,X)m^{*}_{kl}(\mu,X)+m_{i\bar{l}}(\mu,X)m^{*}_{k\bar{j}}(\mu,X) \big] 
  \label{eq:diprelmat}
\end{equation}
where the sum over $\mu$ and $X$ is highly schematic and includes some unimportant positive factors. That is, $M_{ijkl}$ is a positively weighted sum of $m_{ij}m^*_{kl}+m_{i\bar{l}}m^*_{k\bar{j}}$.

Now, it is natural to interpret $M_{ijkl}$ as an element of a convex cone $\mathcal{C}$, 
\begin{equation}
    M_{ijkl}\in \mathcal{C}\equiv \mathrm{cone}\left(\{  m_{ij}m^*_{kl}+ m_{i\bar{l}}m^*_{k\bar{j}} \} \right),
\end{equation}
where cone$(S)$ denotes the semi-positive (or conical) combination of the elements in set $S$. (A brief introduction to the fundamentals of convex geometry is provided in Appendix~\ref{appd:convgeo}.) 
Recognizing that $m_{ij}m^*_{kl}$ is a semi-positive matrix\,\footnote{Of course, the unitarity condition or the generalized optical theorem extends beyond the mere positivity of $A_{ij\to X} A_{kl\to X}^{*}$. The full unitarity condition becomes particularly important when delineating the allowed parameter space of strongly coupled theories. In particular, the non-positivity components imply that the positivity cone must be capped from above~\cite{Chen:2023bhu,Hong:2024fbl}, as one might expect.}, the task of determining the allowed space for the $s^2$ coefficients $M_{ijkl}$, {\it i.e.}, the positivity bounds, amounts to carving out the amplitude convex cone $\mc{C}$ \cite{Zhang:2020jyn}. This connection enables us to utilize tools from convex geometry, together with group-theoretical constructions, to obtain the optimal positivity bounds, particularly in situations with sufficient symmetries and not too many degrees of freedom. 

On the other hand, by dealing with the dual cone, this geometric perspective also implies that a generic positivity-bounds problem can be efficiently solved using convex optimization techniques~\cite{Li:2021lpe}. The dual cone of the $\mc{C}$ cone is defined as $\mathcal{C}^*=\{T|M\cdot T\geq0,\forall M\in \mathcal{C}\}$, where $M\cdot T=\sum_{i,j,k,l}M_{ijkl}T_{ijkl}$. In the dual cone approach, the optimal positivity bounds can be obtained by enumerating all extremal rays (ERs) of $\mc{C}^*$: $T^{(\rm ER)}_I \cdot M\geq 0$, with $I$ indexing the extremal rays. (An ER is an element of a cone that can not be positively split into two other elements, and a convex cone can be defined by its ERs.) This is simply because any element of $\mc{C}^*$ can be obtained by a semi-positive sum of its ERs.

However, for our current case of constraining the aQGC coefficients, we will directly construct the amplitude cone $\mc{C}$ itself to extract the positivity bounds. To understand the structure of the $\mc{C}$ cone, note that we can decompose $2m_{ij}m^*_{kl} = (m_{ij}m^*_{kl} + m_{i\bar{l}}m^*_{k\bar{j}})+  (m_{ij}m^*_{kl}  -m_{i\bar{l}}m^*_{k\bar{j}})$. This observation shows that the amplitude cone $\mathcal{C}$ can be regarded as the projection of $\mathcal{C}_s$ onto $\mathcal{S}$
\be
\mathcal{C} =\mathrm{Proj}_{\mc{S}}(\mathcal{C}_{\rm s}), 
\ee
where
\be
\mathcal{C}_{\rm s}\equiv \mathrm{cone}\left(\{  m_{ij}m^*_{kl}\}\right)
\ee
is the cone generated by all rank-one matrices of the form $m_{ij}m^*_{kl}$ and  
\begin{equation}
\mc{S} =  \{ S_{ijkl}\mid  S_{ijkl}  =S_{i\bar{l}k\bar{j}} \} 
\end{equation}
is the linear subspace enforcing the required crossing symmetry. So the problem of finding the best positivity bounds can be separated in two steps: first determine the $\mc{C}_{\rm s}$ and then perform a crossing-symmetric projection. A standard result in convex geometry states that the ERs of the cone of positive semi-definite matrices $M_{mn}$ are precisely the rank-1 matrices $u_m u_n^*$. Thus, in the absence of any relations between the particle labels $i$ and $j$ in $m_{ij}$, all matrices of the form $m_{ij}m^*_{kl}$ would generate ERs of the $\mc{C}_{\rm s}$ cone. However, in our current setting, $m_{ij}m^*_{kl}$ need not be an ER, as will be demonstrated in the next subsection.

For scatterings between self-conjugate states $\bar i=i$, it is convenient to separate the UV amplitude $A_{ij\to X}$ into the real and imaginary part: $m_{ij}=m^{R}_{ij}+im^{I}_{ij}$, which leads to
\be
m_{ij}m^*_{kl}+m_{i\bar{l}}m^*_{k\bar{j}} = \sum_{K=R,I}m^K_{ij}m^K_{kl} + (j\leftrightarrow l) + i(\cdots)
\ee
Since we focus on the CP-even sector of the SMEFT, the imaginary part above must vanish. In this case, re-interpreting the sum over $K$ as part of the sum over $X$, we have $M_{ijkl}=\sum ( m_{ij}m_{kl}+m_{il}m_{kj})$ and the amplitude cone reduces to
\begin{equation}
    M_{ijkl}\in \mathcal{C}\equiv \mathrm{cone}\Big(\{  m_{ij}m_{kl}+ m_{il}m_{kj} \}\Big) .
\end{equation}
For scalars and vector bosons, this form is particularly convenient, as it allows the amplitude $M_{ijkl}$ to be expressed entirely in terms of real quantities.

At tree-level, an $s^2$ coefficient $M_{ijkl}$ depends linearly on the dim-8 coefficients: $M_{ijkl}=\sum_{\alpha=1} F_\alpha M^\alpha_{ijkl}$, where $F_\alpha$ are the Wilson coefficients and the set $\{M^\alpha_{ijkl}\}$ provides a basis in which each element corresponds to the amplitude associated with a specific dim-$8$ operator. For dim-8 aQGC coefficients, $\ai$ runs from $1$ to $22$, as enumerated in Section \ref{sec:aQGC}. It is straightforward to extract the bounds on $\vec{F}=(F_1,F_2,\dots)$ from the bounds on $M_{ijkl}$.

\subsection{Potential extremal rays from symmetries}\label{sec:Perssymmetry}

To understand why $m_{ij}m^*_{kl}$ is generally not an ER of the $\mc{C}_{\rm s}$ cone, recall that $m_{ij}$ is the amplitude from $| ij \rangle$ to $|X \rangle$:  
\be
m_{ij}=A_{ij\to X}=\langle X| \bar T | ij \rangle ,
\ee
where the transfer-matrix operator $\bar T$ is defined to absorb unimportant factors. Since the scattering particles are charged under the internal and spacetime symmetries of the SMEFT, the amplitudes $m_{ij}$ are covariant under the transformations of the internal symmetry group $SU(2)_L\otimes U(1)_Y$ and the little group $SO(2)$ scalings\,\footnote{The translation part of the $ISO(2)$ group for massless particles acts trivially on physical one-particle states and thus is neglected.} of the external momenta, along with the discrete CP symmetry. For forward scattering, all momenta are collinear, so the little groups can be identified up to flipping the directions of half of the external momenta. (Although collinear, there can still be a nonzero impact parameter/angular momentum; see \Cref{sec:exchangeCG}.) The tensor product states $| ij \rangle$ are generally not in the irreps of the total symmetry group, which now includes the identified little group scaling. That is, denoting the irrep states as $\left\vert \bfm,\alpha\right\rangle$, where $\alpha$ labels the state in the irrep $\mathbf{m}$, the tensor product state $|i j\rangle$ is generally reducible and can be decomposed into these irreps with the Clebsch–Gordan (CG) coefficients $C_{i,j}^{\mathbf{m},\alpha}$:
\be
|i j\rangle = \sum_{\bfm, \alpha} \left|\bfm, \alpha\right\rangle \langle\bfm, \alpha|i j\rangle= \sum_{\bfm, \alpha} C_{i, j}^{\bfm, \alpha}\left|\bfm, \alpha\right\rangle 
\label{eq:CGdecomp}
\ee
Note here that we have omitted the information of the particle momenta, as the CG coefficients do not depend on it. (Although the $|X\rangle$ states may be charged under a larger symmetry group, we only restrict to the symmetry group of the SMEFT, and organize the states according to the irreps of this symmetry group.) Thus, for general $X$, $m_{ij}=\langle X| \bar T | ij \rangle$ is a sum of transition amplitudes between irrep states, which, as we will see shortly, in turn means that $m_{ij}m^*_{kl}$ is generally not an ER. It is thus convenient to choose $|X\rangle=\left\vert \bfm,\alpha\right\rangle|w\rangle=\left\vert \bfm,\alpha\right\rangle_w$, where $w$ indexes the multiplicity/degeneracy space.

Recall that the positivity cone $\mc{C}$ is essentially the $\mc{C}_{\rm s}$ cone, up to a crossing projection. We can first construct $\mc{C}_{\rm s}$'s ERs, for which we simply need to decompose $| ij \rangle$ into its irreps. A fundamental result in the representation theory of compact or finite groups is Schur’s lemma, often known as the Wigner–Eckart theorem in physics, which for the transfer-matrix operator implies 
\be
\left\langle \bfm,\alpha|\bar T|\bfn,\beta \right\rangle=\delta_{\bfm,\bfn}\delta_{\alpha,\beta}\left\langle \bfm\right\Vert \bar T\left\Vert\bfm\right\rangle ,
\ee
where $\left\langle \bfm\right\Vert \bar T\left\Vert \bfm\right\rangle$ is the reduced transfer matrix. By the tensor-product decomposition, the amplitude $m_{ij}$ can be expressed as 
\begin{equation}
\label{eq:mijCijm}
 m_{ij}={}_w\!\left\langle \bfm,\alpha\right\vert \bar T\left\vert ij\right\rangle=\sum_{\bfn,\bi}{}_w\!\left\langle \bfm,\alpha|\bar T|\bfn,\beta \right\rangle \left\langle \bfn,\beta | ij \right\rangle  ={}_w\!\left\langle \bfm\right\Vert \bar T\left\Vert\bfm\right\rangle C_{i,j}^{\bfm,\alpha}.
\end{equation}
That is, $m_{ij}(\mu, X)$ is proportional to the CG coefficient, differing only by a scaling factor that will not be important for our purposes. To see this, substituting this relation into Eq.~\eqref{eq:diprelmat}, we get
\begin{align}
M_{ijkl} &  =\sum_{\mu,w,\bfm,\alpha}\left\vert {}_w\!\left\langle \bfm\right\Vert \bar T\left\Vert
\bfm\right\rangle \right\vert ^{2}\left(  C_{i,j}^{\bfm,\alpha}\left(  C_{k,l}^{\bfm,\alpha}\right)  ^{*}+C_{i,\bar{l}}^{\bfm,\alpha}\left(  C_{k,\bar{j}}^{\bfm,\alpha}\right)  ^{*}\right)\nn
&  =\sum_{\mu,w,
\bfm}\left\vert {}_w\!\left\langle \bfm \right\Vert \bar T\left\Vert \bfm\right\rangle \right\vert ^{2}
\bar P_{\bfm}^{ijkl} \label{eq:procone}
\end{align}
where we have defined $\bar P_\bfm^{ijkl}\equiv (P_\bfm^{ijkl}+P_\bfm^{i\bar{l}k\bar{j}})/2!$ with
\begin{equation}
P_{\bfm}^{ijkl}\equiv\sum_{\alpha}C_{i,j}^{\bfm,\alpha}\left(C_{k,l}^{\bfm,\alpha}\right)  ^{*} 
\end{equation}
being the projector that projects into the $\bfm$ irrep. Since $| {}_w\!\left\langle \bfm \right\Vert \bar T\left\Vert \bfm\right\rangle |^{2}$ is non-negative, this means that the amplitude cone can be expressed as 
\be
M_{ijkl}\in \mc{C}=\text{cone}(\{\bar P_\bfm^{ijkl}\})
\ee
and its ERs are in the form of $\bar P_\bfm^{ijkl}$. Note that, while $P_\bfm^{ijkl}$ are necessarily the ERs of the $\mc{C}_s$ cone, $\bar P_\bfm^{ijkl}$ are potential ERs of the $\mc{C}$ cone, due to the crossing projection imposed by the set $\mc{S}$. Thus, while the set $\{\bar P_\bfm^{ijkl}\}$ contains redundancies, its conical hull nevertheless generates the optimal positivity bounds.

Notice that the $i,j$ indices and thus $C_{i,j}^{\bfm,\alpha}$ contain all the symmetries in the problem. If the total symmetry group, as in our case, is a tensor product of multiple groups $\bigotimes_g G_g$:
$|i,j\rangle = \bigotimes_g |i_g,j_g \rangle$,
then we have $C_{i,j}^{\bfm,\alpha} = (\bigotimes_g\langle\bfm_g, \alpha_g|) (\bigotimes_g |i_g,j_g \rangle) = \prod_g C_{i_g,j_g}^{\bfm_g,\alpha_g}$.
Alternatively, for the finite groups as well as the abelian groups $U(1)$ and $SO(2)$, the decomposition into the irreps can be easily performed without explicitly invoking the CG coefficients.

\subsection{CP symmetry}
\label{sec:CP}

Before we proceed to construct the group projectors for the amplitude cone in the next section, let us first spell out the CP properties of the scattering states. (Note that when constructing the dim-8 operators in \Cref{sec:aQGC}, we only imposed the combined conservation of charge conjugation and parity, and thus some operators are parity-violating.)

For our cone construction later, we will use the real two-dimensional irrep of the $SO(2)$ little group for massless gauge bosons, {\it i.e.}, the linear polarization states, which can be expressed in terms of the helicity eigenstates: $|+\rangle, ~|-\rangle$. That is, the irrep  ${\bf 2}$ is given by
\begin{align}
    \left| \mathbf{2},1 \right\rangle &= \left|+\right\rangle + \left|-\right\rangle, \nn
    \left| \mathbf{2},2 \right\rangle &= -i\left(\left|+\right\rangle - \left|-\right\rangle\right) .\label{eq:Pxy}
\end{align}
We see that the first component $\left| \mathbf{2},1 \right\rangle$ has even parity and $\left| \mathbf{2},2 \right\rangle$ has odd parity, as under parity $|+\rangle\leftrightarrow |-\rangle$.
For a two-particle vector state with helicities $h_1$ and $h_2$, its tensor product decomposition can be obtained by a simple matrix diagonalization $\mathbf{2}\otimes \mathbf{2}=\mathbf{1}_S\oplus\mathbf{1}_A\oplus\mathbf{2}$, where the irreps are given by (see, {\it e.g.}, \cite{Trott:2020ebl})
\begin{align}
    \left| \mathbf{1}^+_S \right\rangle & = \left| +\right\rangle\left| +\right\rangle + \left| -\right\rangle\left| -\right\rangle, \nn
    \left| \mathbf{1}^-_A \right\rangle & = \left| +\right\rangle\left| +\right\rangle - \left| -\right\rangle\left| -\right\rangle, \nn
    \left| \mathbf{2}^+,1 \right\rangle & =\left| +\right\rangle\left| -\right\rangle + \left| -\right\rangle\left| +\right\rangle, \nn
    \left| \mathbf{2}^+,2 \right\rangle & = \left| +\right\rangle\left| -\right\rangle - \left| -\right\rangle\left| +\right\rangle,\label{eq:SO2P}
\end{align}
with the superscript ${}^{+/-}$ denoting the even/odd parity of each irrep state. It is worth pointing out that in our setup the parity transformation leads to the following helicity changes: $h_1\to-h_2,h_2\to-h_1$. The reason for this strange helicity transformation under parity is due to the fact that we are only concerned with the CG coefficients, which do not carry the information of the external momentum (cf.~\eref{eq:mijCijm}). So when acting the parity we must change the labels of the two particles in order to hold the external momenta unchanged. On the other hand, the Higgs scalar is invariant under parity.

We now turn to charge conjugation. Although the transformation of gauge fields is straightforward, the charge conjugation of the Higgs field for $SU(2)$ has occasionally been misinterpreted. To avoid confusion, we explicitly review its charge-conjugation property (see, {\it e.g.}, \cite{Kondo:2022wcw,Durieux:2024zrg}). The Higgs doublet $H^a$ transforms in the fundamental representation $\mathbf{2}$ of $SU(2)$, which is a pseudo-real representation, while $H_a$ transforms in the anti-fundamental representation $\bar{\mathbf{2}}$. Under the conjugation transformation $C$, the Higgs doublet transforms as\,\footnote{Another definition of charge conjugation is given by $C_S$: $H^{a}\overset{C_{S}}{\longrightarrow}H_{a}^{\dagger}\overset{C_{S}}{\longrightarrow}H^{a}$. Our $C$ is sometimes denoted as $C_A$.}
\begin{equation}
    H^{a}\overset{C}{\longrightarrow}H^{a\dagger}=\epsilon^{ab}H_{b}^{\dagger}\overset{C}{\longrightarrow}-H^{a},
\end{equation}
where $\epsilon^{ab}$ is defined as $\epsilon^{11}=\epsilon^{22}=0,\epsilon^{12}=-\epsilon^{21}=1$. Notice that, in our convention, the charge conjugation $C$ acts within the same $\mathbf{2}$ or $\bar{\mathbf{2}}$ representation. 
When constructing the ERs, we will use the eigenstates of charge conjugation, which are given by
\begin{align}
&  H^{1}+iH_{2}^{\dagger}\overset{C}{\longrightarrow}-i\left(  H^{1}+iH_{2}^{\dagger}\right) \nn
&  H^{1}-iH_{2}^{\dagger}\overset{C}{\longrightarrow}i\left(  H^{1}-iH_{2}^{\dagger}\right) \nn
&  H^{2}+iH_{1}^{\dagger}\overset{C}{\longrightarrow}i\left(  H^{2}+iH_{1}^{\dagger}\right) \nn
&  H^{2}-iH_{1}^{\dagger}\overset{C}{\longrightarrow}-i\left(  H^{2}-iH_{1}^{\dagger}\right). \label{eq:CofH}
\end{align}
With these eigenstates, the charge conjugation of the direct product of two-particle states can be easily determined, as we will see in~\Cref{sec:aQGCER}.

Our goal is to construct the ERs of the amplitude cone, which in turn means to construct the group projectors $P^{ijkl}_{\bfm}=\sum_\alpha \langle\bfm,\ai\left| ij \right \rangle \left\langle kl\right|\bfm,\ai\rangle$. Since the combined effect of the $\langle\bfm,\ai|$ and $|\bfm,\ai\rangle$ factors in the projector always leads to a CP-even transformation, constructing a CP-even projector requires that $\left|ij \right\rangle$ and $\left|kl \right\rangle$ share the same CP transformation properties.

\subsection{Exchange symmetries of CG coefficients}
\label{sec:exchangeCG}

The spin of the intermediate state implies an additional symmetry associated with the exchange of the indices $i\leftrightarrow j$ in $C^{\bfm,\ai}_{i,j}$. At the level of the $2\to 2$ scattering amplitudes $\mathcal{M}_{ijkl}$, this symmetry corresponds to the simultaneous exchange $i\leftrightarrow j$ and $k\leftrightarrow l$. In parity-conserving theories, the amplitudes are invariant under this double exchange. In this subsection, we shall classify this symmetry according to the spin of the intermediate state, along with its CP symmetries.

To see this, let us examine the general 3-point amplitudes involving two massless and one massive particle, as the symmetry structure of these amplitudes is directly connected to our CG coefficients discussed in the last subsection.
Following \cite{arkani-hamed:2017jhn}, we have
\begin{equation}
    {A}_{ij\to X}=m_{ij} =\dfrac{g_{ab}}{M^{2\ell+h_i+h_j-1}}[12]^{\ell + h_i + h_j} \braket{1 \mathbf{X}}^{\ell + h_j - h_i}\braket{2 \mathbf{X}}^{\ell + h_i - h_j},
    \label{eq:amp}
\end{equation}
where $h_i$ and $h_j$ label the helicities of particles $i$ and $j$ with momentum $p_1$ and $p_2$ respectively, $\ell$ is the spin of $X$ (or the angular momentum of $ij$), $g_{ab}$ denotes the coupling constant with $a,b$ labeling the corresponding gauge indices, and $M$ denotes the mass of $X$. That is, here, we have decomposed a generic intermediate state (possibly a multi-particle state) into states with definite spins labeled by $X$. In our case, we are interested in amplitudes where particles $i$ and $j$ are bosons. The spin-statistics theorem then implies that under a simultaneous exchange of the particle species, momenta, and helicities, the amplitudes satisfy ${A}_{ij\to X} = {A}_{ji\to X}$. This leads to
\begin{equation}
    \label{eq:spin}
    g_{ab} = (-1)^{\ell + h_i + h_j} g_{ba}.
\end{equation}
This indicates that the exchange symmetry of the CG coefficients with respect to their gauge indices depends on the spin $\ell$ of $X$.

However, in parity-violating theories, $C^{\bfm,\ai}_{i,j}$ need not be symmetric or antisymmetric in exchanging $i$ and $j$, since a general construction may involve both P-even and P-odd sectors. In fact, the amplitude associated with the operator $O_{M,9}$ is generated by a process where $C^{\bfm,\ai}_{i,j}$ does have definite exchange symmetry. In such cases, $C^{\bfm,\ai}_{i,j}$ can be split into a symmetry component and an anti-symmetry component, with some degeneracy between the two.

\section{Extremal rays for aQGC positivity cone}\label{sec:aQGCER}

In the last section, we have identified and clarified the necessary ingredients for constructing the projectors of the amplitude cone. In the section, we shall use them to determine the ERs of the aQGC positivity cone $\mathcal{C}$ by performing the tensor product decomposition of the external states.

For clarity, we first make explicit our {\it notation} for the states and the symmetries. Our choices for the Higgs, $B$-boson and $W$-boson fields are $\phi_{\bfa}$ ($\bfa=1,2,3,4$) or ${H^a, H^\dagger_a}$ ($a=1,2$), $B_\mathbf{i}(\mathbf{i} = 1,2)$ and ${W^I_{\mathbf{i}}}$ ($I=1,2,3; \mathbf{i}=1,2$), respectively, where the index $a$ transforms in the fundamental $\mathbf{2}$ or anti-fundamental $\overline{\mathbf{2}}$ of $\mathrm{SU(2)}$, the index $I$ carries the adjoint $\mathbf{3}$ representation of $\mathrm{SU(2)}$, and $\mathbf{i}$ transforms under the real vector representation $\mathbf{2}$ of $\mathrm{SO(2)}$. In writing the CG coefficients, we will also use the index $\hat I = (0,I,\bfa)$, denoting the combined components of $B,~W$ and $H$. We will write the $\mathrm{SU(2)}$ and $\mathrm{U(1)}$ groups together as $\mathrm{SU(2)}_{\mathrm{U(1)}}$, {\it e.g.}, ${\bf 3}_0$ for the $W$-boson. When the meaning is clear, we will use the field labels to denote the external states, for example, $W^1_2 H^\dagger_2 \equiv \ket{W^1_2, H^\dagger_2}$.

\subsection{Direct construction}
\label{sec:directConstruction}

Let us recall that to find the ERs of the amplitude cone,  our first task is to find the group projectors $P_{\bfm}^{ijkl}\equiv\sum_{\alpha}C_{i,j}^{\bfm,\alpha}(C_{k,l}^{\bfm,\alpha})^{*}$ of the total symmetry group. These are potential ERs of the cone. The reason that they are potential ERs, rather than  ERs, is because the real ERs are obtained by a further projection on the amplitude space: $\bar P_\bfm^{ijkl}\equiv (P_\bfm^{ijkl}+P_\bfm^{i\bar{l}k\bar{j}})/2!$. To obtain all potential ERs, we must consider all possible electro-weak boson scatterings, {\it i.e.}, $i,j,k,l$ must run over all possible electro-weak bosons, and find the relevant CG coefficients (see Appendix~\ref{appd:CG} for the results).

$\bullet$ Let us start with a simple one: $WW\to X$ scattering. In this case, the hypercharge U(1) symmetry and the CP symmetries are trivial, so we only need to be concerned with the internal $SU(2)$ symmetry and little group $SO(2)$. Since the total symmetry group is the tensor product of these two, the full CG coefficient $C_{i,j}^{\bfm,\alpha}$ can be decomposed into $C_{i,j}^{\bfm,\alpha} = C_{I,J}^{\mathfrak{m},\mathtt{a}} C_{\bfi,\bfj}^{\mathbb{m},\text{a}}$, where $i=(I,\bfi),~j=(J,\bfj),~\bfm=(\mathfrak{m},\mathbb{m})$ and $\ai=(\mathtt{a},\text{a})$.
The $SO(2)$ CG coefficients have been essentially obtained in \Cref{sec:CP}. Explicitly, they are given by
\begin{equation}
    C^{\mathbb{1}_S}_{\bfi \bfj} =\delta_{\bfi \bfj},~~~~
    C^{\mathbb{1} _A}_{\bfi \bfj} =\epsilon_{\bfi \bfj},~~~~
    C^{\mathbb{2},\text{1}}_{\bfi \bfj} =
    \begin{pmatrix}
        1&0\\
        0&-1
    \end{pmatrix},~~
    C^{\mathbb{2},\text{2} }_{\bfi \bfj} =
    \begin{pmatrix}
        0&1\\
        1&0
    \end{pmatrix}
\label{eq:so2CGC}
\end{equation}
Also, the $SU(2)$ part is mostly a matter of reading the table of the standard $SU(2)$ CG coefficients. However, we are using a self-conjugate representation for the $W$ bosons: we choose $W^1, W^2, W^3$ as the external states, while the standard table of CG coefficients usually tabulates states associated with $W^+, W^0, W^-$. So an additional transformation between the bases is needed. To be concrete, since $W^\pm= \mp(W^1\mp i W^2)/\sqrt{2},~W^0=W^3$, the transformation is given by
\begin{equation}
 W^{I=1,2,3}=\sum_{\tilde I=+,0,-} S_{I\tilde I}W^{\tilde I},~~~~~   
    S_{I\tilde I}=\frac{1}{\sqrt{2}}
    \begin{pmatrix}
        -1 & 0 & 1\\
        -i & 0 & -i\\
         0 &\sqrt{2}& 0
    \end{pmatrix}.
\end{equation}
The states of the invariant subspace are expanded by the tensor product states with the CG coefficients
\begin{equation}
    \ket{\mathfrak{m},\mathfrak{a}}= \sum_{\tilde I,\tilde J} C^{\mathfrak{m}, \mathfrak{a}}_{\tilde I\tilde J}\ket{W^{\tilde I}}\ket{W^{\tilde J}}
   =\sum_{I,J} C^{\mathfrak{m},\mathfrak{a}}_{IJ}\ket{W^I}\ket{W^J},~~~~  C_{I,J}^{\mathfrak{m},\mathfrak{a}} =  \sum_{\tilde I,\tilde J} C_{\tilde I,\tilde J}^{\mathfrak{m},\mathfrak{a}}S^{-1}_{\tilde I I} S^{-1}_{\tilde J J}
\end{equation}
where $C_{\tilde I,\tilde J}^{\mathfrak{m},\mathfrak{a}}$'s are the CG coefficients readable from a standard $SU(2)$ table.

Having obtained the CG coefficients of the $WW\to X$ scattering, we further determine the parity of its total angular momentum (or the spin parity of $X$, {\it i.e.}, $\ell {\rm mod} 2$), which is useful for diagnosing whether two irreps are degenerate. To this end, note that the exchange symmetry of the $SU(2)$ CG coefficients can be inferred from Eq.~\eqref{eq:spin}, with the extra knowledge that in the current case $h_i+h_j=-2,0,2$: 
\be
C_{I,J}^{\mathfrak{m},\mathtt{a}}=(-1)^{\ell} C_{J,I}^{\mathfrak{m},\mathtt{a}} .
\label{eq:CIJspin}
\ee
where $\ell$ is the total angular momentum of $WW$. The $SU(2)$ irreps of the $WW$ tensor product decomposition is given by ${\bf3}\otimes {\bf3}={\bf1}_S\oplus{\bf3}_A\oplus{\bf5}$. As we will see, the irreps ${\bf1}_S$ and ${\bf3}_A$ can also arise from other scattering processes, leading to degenerate ERs with arbitrary parameters and rendering the amplitude cone non-polyhedral. To determine when the degeneracies arise, we use \eref{eq:CIJspin} to find out the spin parity of $X$. For the symmetric irrep ${\bf1}_S$, we have $\ell=\text{even}$, while $\ell=\text{odd}$ for anti-symmetric irrep ${\bf3}_A$. In other words, the positivity cone and its ERs also predict the spin parity of potential UV states, along side with other UV quantum numbers.

$\bullet$ For the $WB\to X$ scattering and the $BB\to X$ scattering, the $SO(2)$ CG coefficients are the same as the $WW\to X$ scattering, as they are also scatterings between spin-1 particles. For the $SU(2)$ group and the $U(1)$ group, there is no need to perform tensor product decomposition, as $B$ is a $SU(2)$ singlet and $W$ and $B$ both have 0 hypercharge. The relevant CG coefficients can be found in Appendix~\ref{appd:CG}. 

However, the $WB\to X$ scattering does contain a component that is degenerate with the $WW\to X$ scattering. To find the degenerate component, note that \eref{eq:CIJspin} still leads to
$C_{0,I}^{\mathfrak{m},\mathtt{a}}=(-1)^{\ell} C_{I,0}^{\mathfrak{m},\mathtt{a}}$,
where $0$ denotes the $B$ boson. Thus, after decomposing the $WB$ state into a symmetry part ($\ell=\text{even}$) plus an anti-symmetry part ($\ell=\text{odd}$), we see that the anti-symmetry part, which is in the ${\bf3}_A$ irrep, is degenerate with the ${\bf3}_A$ in the $WW\to X$ scattering if its $X$ spin parity is the same as that of the $WW\to X$ scattering. That is, we can construct an ER with the following augmented $SU(2)$ CG coefficients 
\be
\left(
\begin{array}{c|c|c}
0 & 0 &0 \\
\hline
0 & ~C_{I,J}^{\mathfrak{3}_A,\mathtt{a}}~&0  \\
\hline
0&0&0
\end{array}
\right)
+
x \left(
\begin{array}{c|c|c}
0 & C_{I,0}^{\mathfrak{3},\mathtt{a}} &0 \\
\hline
-C_{0,J}^{\mathfrak{3},\mathtt{a}} & ~0~&0  \\
\hline
0&0&0
\end{array}
\right)
\Rightarrow ~C_{\hat I,\hat J}^{\mathfrak{3},\mathtt{a}} =
\left(
\begin{array}{c|c|c}
0 & x C_{I,0}^{\mathfrak{3},\mathtt{a}} &0 \\
\hline
-x C_{0,J}^{\mathfrak{3},\mathtt{a}} & ~C_{I,J}^{\mathfrak{3}_A,\mathtt{a}}~&0  \\
\hline
0&0&0
\end{array}
\right),~~~{\rm UV~spin~}\ell={\rm odd}
\label{eq:SU2_3AC}
\ee
where $x$ is an arbitrary real parameter, encoding the potential degeneracy. Similarly, the $WW\to X$ scattering does contain a component, ${\bf1}_S$, that is degenerate with the $BB\to X$ scattering: 
\be
C_{\hat I,\hat J}^{\mathfrak{1}_S,\mathtt{a}} =
\left(
\begin{array}{c|c|c}
x C_{0,0}^{\mathfrak{1}_S,\mathtt{a}} & 0&0  \\
\hline
0 & ~C_{I,J}^{\mathfrak{1}_S,\mathtt{a}}~&0  \\
\hline
0&0&0
\end{array}
\right), ~~~{\rm UV~spin~}\ell={\rm even}
\ee
where $x$ is again an arbitrary real parameter, in principle independent of the $x$ parameter in \eref{eq:SU2_3AC}.

$\bullet$ For $HH\to X$ scattering, where the $SO(2)$ little group scalings are trivial, we work with real scalar fields in the following form
\begin{equation}
H^a =
    \begin{pmatrix}
      \phi_{\bf2} + i \phi_{\bf1} \\
      \phi_{\bf4} - i\phi_{\bf3}
    \end{pmatrix} ,
\end{equation} 
which transforms in the fundamental representation ${\bf2}$ of $SU(2)$. Thus, for the $SU(2)$ sector, it is again a matter of reading the standard $SU(2)$ table of the CG coefficients for $H^a$ and $H_a^\dagger$, plus a straightforward transformation of them to the real scalar basis. Again, the parity of the $\ell$ spin dictates the exchange symmetry of the CG coefficients $C_{\bfa,\bfb}^{\mathfrak{m},\mathtt{a}}=(-1)^{\ell} C_{\bfb,\bfa}^{\mathfrak{m},\mathtt{a}}$
(inferred from \eref{eq:CIJspin}). Thus, upon including the $HH\to X$ scattering, which also contains $SU(2)$ a ${\bf1}_S$ and ${\bf3}_A$ component, we see that the $C_{\hat I,\hat J}^{\mathfrak{1}_S,\mathtt{a}}$ and  $C_{\hat I,\hat J}^{\mathfrak{3}_A,\mathtt{a}}$ contain additional degeneracy and must be updated to be
\be
C_{\hat I,\hat J}^{\mathfrak{1}_S,\mathtt{a}} =
\left(
\begin{array}{c|c|c}
x C_{0,0}^{\mathfrak{1}_S,\mathtt{a}} & 0&0  \\
\hline
0 & C_{I,J}^{\mathfrak{1}_S,\mathtt{a}} &0 \\
\hline
0&0& y C_{\bfa,\bfb}^{\mathfrak{1}_S,\mathtt{a}}
\end{array}
\right),~~~~~
C_{\hat I,\hat J}^{\mathfrak{3}_A,\mathtt{a}} =
\left(
\begin{array}{c|c|c}
0 & x C_{I,0}^{\mathfrak{3},\mathtt{a}}&0  \\
\hline
-x C_{0,J}^{\mathfrak{3},\mathtt{a}} & C_{I,J}^{\mathfrak{3}_A,\mathtt{a}}&0  \\
\hline
0 & 0 & y C_{\bfa,\bfb}^{\mathfrak{3}_A,\mathtt{a}}
\end{array}
\right)
\ee
where $y$ is another independent real parameter, and the UV spin parity of $C_{\hat I,\hat J}^{\mathfrak{1}_S,\mathtt{a}}$ is even and the UV spin parity of $C_{\hat I,\hat J}^{\mathfrak{3}_A,\mathtt{a}}$ is odd.

$\bullet$ Now, we consider $WH\to X$ scattering. In this case, the $SO(2)$ little group part is trivial, so we again focus on the internal group part. Since we choose to use the real scalar $\phi_\bfa$ basis for the Higgs, let us first find out how these states transform under a $SU(2)$ transformation. Note that in the $SU(2)$ fundamental representation, the Higgs state $H^a$ transforms via the Wigner D-matrix
\begin{equation}
    D^{1/2}(\alpha,\beta,\gamma)H^a
   =H'^a,~~D^{(1/2)}(\alpha,\beta,\gamma)=
    \begin{pmatrix}
        e^{-\frac{i}{2}\left(  \alpha+\gamma\right)  }\cos\frac{\beta}{2} &
        -e^{-\frac{i}{2}\left(  \alpha-\gamma\right)  }\sin\frac{\beta}{2}\\
        e^{\frac{i}{2}\left(  \alpha-\gamma\right)  }\sin\frac{\beta}{2} & e^{\frac
        {i}{2}\left(  \alpha+\gamma\right)  }\cos\frac{\beta}{2}
    \end{pmatrix} .
\end{equation}
This can be reshaped into the following representation for the $\phi_\bfa$ states
\begin{equation}
    R_{\phi}(\alpha,\beta,\gamma)
    \begin{pmatrix}
    \phi_{\bf1}\\
    \phi_{\bf2}\\
    \phi_{\bf3}\\
    \phi_{\bf4}
    \end{pmatrix}=
    \begin{pmatrix}
    \phi'_{\bf1}\\
    \phi'_{\bf2}\\
    \phi'_{\bf3}\\
    \phi'_{\bf4}
    \end{pmatrix}
    \label{eq:Rphiabc}
\end{equation}
where 
\begin{equation}
R_{\phi}(\alpha,\beta,\gamma)=
\begin{pmatrix}
\cos\frac{\beta}{2}\cos\frac{\alpha+\gamma}{2} & -\cos\frac{\beta}{2}\sin
\frac{\alpha+\gamma}{2} & \sin\frac{\beta}{2}\cos\frac{\alpha-\gamma}{2} &
\sin\frac{\beta}{2}\sin\frac{\alpha-\gamma}{2}\\
\cos\frac{\beta}{2}\sin\frac{\alpha+\gamma}{2} & \cos\frac{\beta}{2}\cos
\frac{\alpha+\gamma}{2} & \sin\frac{\beta}{2}\sin\frac{\alpha-\gamma}{2} &
-\sin\frac{\beta}{2}\cos\frac{\alpha-\gamma}{2}\\
-\sin\frac{\beta}{2}\cos\frac{\alpha-\gamma}{2} & -\sin\frac{\beta}{2}%
\sin\frac{\alpha-\gamma}{2} & \cos\frac{\beta}{2}\cos\frac{\alpha+\gamma}{2} &
-\cos\frac{\beta}{2}\sin\frac{\alpha+\gamma}{2}\\
-\sin\frac{\beta}{2}\sin\frac{\alpha-\gamma}{2} & \sin\frac{\beta}{2}\cos
\frac{\alpha-\gamma}{2} & \cos\frac{\beta}{2}\sin\frac{\alpha+\gamma}{2} &
\cos\frac{\beta}{2}\cos\frac{\alpha+\gamma}{2}
\end{pmatrix}.\label{eq:Rphi}
\end{equation}
The 4D real $SU(2)$ representation $R_\phi(\ai,\bi,\gi)$ is related to the 2D complex representation $D^{(1/2)}(\alpha,\beta,\gamma)$ via the $U$ matrix
\begin{equation}
U=\frac{1}{\sqrt{2}}
\begin{pmatrix}
1 & -i & 0 & 0\\
0 & 0 & -1 & -i
\end{pmatrix},~~~UU^\dagger=\mathds{1}_{2\times 2}
\end{equation}
via
\begin{equation}
    UR_{\phi}(\alpha,\beta,\gamma)U^\dagger=D^{(1/2)}(\alpha,\beta,\gamma) .
\end{equation}
The representation $R_{\phi}(\alpha,\beta,\gamma)$ can be generated by the following generators $i\Gamma^I,I=1,2,3$, where 
these Gamma matrices are defined by
\begin{equation}
\Gamma^1 =\!
    \begin{pmatrix}
        0 & 0 & 0 & -1\\
        0 & 0 &-1 & 0\\
        0 & 1 & 0 & 0\\
        1 & 0 & 0 & 0
    \end{pmatrix}\!,
    ~
\Gamma^2 =\!
\begin{pmatrix}
    0 & 0 & 1 & 0\\
    0 & 0 & 0 & -1\\
    -1& 0 & 0 & 0\\
    0 & 1 & 0 & 0    
\end{pmatrix}\!,~
\Gamma^3 =\!
\begin{pmatrix}
    0 & -1 & 0 & 0\\
    1 & 0 & 0 & 0\\
    0 & 0 & 0 & -1\\
    0 & 0 & 1 & 0  
\end{pmatrix}\!,
~
\Gamma^4 =\!
\begin{pmatrix}
    0 & -1 & 0 & 0\\
    1 & 0 & 0 & 0\\
    0 & 0 & 0 & 1\\
    0 & 0 & -1 & 0      
\end{pmatrix}
\end{equation}
We have additionally defined, $i\Gamma^4$, which is the hypercharge generator. Thus, the hypercharge transformation can be constructed as $R_Y=\exp(-\Gamma^4)$. Hypercharge conservation then implies that incoming two-particle states are eigenstates of $R_Y\otimes R_Y$.

Next, we can compute the CG coefficients of $WH\to X$ scattering, following a procedure analogous to the $WW\to X$ case. Explicitly, note that a $SU(2)$ irrep can be decomposed as follows
\begin{equation}
    \ket{\mathfrak{m},\mathfrak{a}}= \sum_{\tilde I,\tilde a} C^{\mathfrak{m}, \mathfrak{a}}_{\tilde I\tilde a}\ket{W^{\tilde I}}\ket{H^{\tilde a}}
    =\sum_{\tilde I,\tilde a} C^{\mathfrak{m}, \mathfrak{a}}_{\tilde I\tilde a}S^{-1}_{\tilde I I}U_{\tilde a \bfa}\ket{W^{I}}\ket{\phi_{\bfa}}
    =\sum_{I,\bfa}C^{\mathfrak{m}, \mathfrak{a}}_{I \bfa}\ket{W^{I}}\ket{\phi_{\bfa}}
    \label{eq:maWphi00}
\end{equation}
from which we can infer that 
\be
C_{I,\bfa}^{\mathfrak{m},\mathfrak{a}} =  \sum_{\tilde I,\tilde a} C^{\mathfrak{m}, \mathfrak{a}}_{\tilde I\tilde a}S^{-1}_{\tilde I I}U_{\tilde a \bfa}
\ee
where $C_{I,\bfa}^{\mathfrak{m},\mathfrak{a}}$'s are the CG coefficients we are after and $C^{\mathfrak{m}, \mathfrak{a}}_{\tilde I\tilde a}$'s are the CG coefficients readable from a standard $SU(2)$ table. For example, the 2 components of the $\bf 2$ irrep can be expanded as
\begin{equation}
\left\vert \mathbf{2}_{1}\right\rangle   =
\begin{tabular}
[c]{ccccc}
& $\phi_{1}$ & $\phi_{2}$ & $\phi_{3}$ & $\phi_{4}$\\\cline{2-5}
$W_{1}$ & \multicolumn{1}{|c}{$0$} & \multicolumn{1}{|c}{$0$} &
\multicolumn{1}{|c}{$\frac{i}{\sqrt{6}}$} & \multicolumn{1}{|c|}{$\frac
{1}{\sqrt{6}}$}\\\cline{2-5}
$W_{2}$ & \multicolumn{1}{|c}{$0$} & \multicolumn{1}{|c}{$0$} &
\multicolumn{1}{|c}{$-\frac{1}{\sqrt{6}}$} & \multicolumn{1}{|c|}{$\frac
{i}{\sqrt{6}}$}\\\cline{2-5}
$W_{3}$ & \multicolumn{1}{|c}{$-\frac{i}{\sqrt{6}}$} &
\multicolumn{1}{|c}{$\frac{1}{\sqrt{6}}$} & \multicolumn{1}{|c}{$0$} &
\multicolumn{1}{|c|}{$0$}\\\cline{2-5}
\end{tabular}~~~~~~
\left\vert \mathbf{2}_{2}\right\rangle   =
\begin{tabular}
[c]{ccccc}
& $\phi_{1}$ & $\phi_{2}$ & $\phi_{3}$ & $\phi_{4}$\\\cline{2-5}
$W_{1}$ & \multicolumn{1}{|c}{$-\frac{i}{\sqrt{6}}$} &
\multicolumn{1}{|c}{$\frac{1}{\sqrt{6}}$} & \multicolumn{1}{|c}{$0$} &
\multicolumn{1}{|c|}{$0$}\\\cline{2-5}
$W_{2}$ & \multicolumn{1}{|c}{$-\frac{1}{\sqrt{6}}$} &
\multicolumn{1}{|c}{$-\frac{i}{\sqrt{6}}$} & \multicolumn{1}{|c}{$0$} &
\multicolumn{1}{|c|}{$0$}\\\cline{2-5}
$W_{3}$ & \multicolumn{1}{|c}{$0$} & \multicolumn{1}{|c}{$0$} &
\multicolumn{1}{|c}{$-\frac{i}{\sqrt{6}}$} & \multicolumn{1}{|c|}{$-\frac
{1}{\sqrt{6}}$}\\\cline{2-5}
\end{tabular}
\label{eq:2122phiW}
\end{equation}
and the $C_{I,\bfa}^{\mathfrak{m},\mathfrak{a}}$ coefficients are given by  
\begin{align}
C^{\mathbf{2},1}  & =\frac{1}{\sqrt{3}}
\begin{pmatrix}
0 & 0 & 0 & 1\\
0 & 0 & -1 & 0\\
0 & 1 & 0 & 0
\end{pmatrix}
,\quad C^{\mathbf{2},2}=\frac{1}{\sqrt{3}}
\begin{pmatrix}
0 & 0 & 1 & 0\\
0 & 0 & 0 & 1\\
-1 & 0 & 0 & 0
\end{pmatrix}
\nonumber\\
C^{\mathbf{2},3}  & =\frac{1}{\sqrt{3}}
\begin{pmatrix}
0 & 1 & 0 & 0\\
-1 & 0 & 0 & 0\\
0 & 0 & 0 & -1
\end{pmatrix}
,\quad C^{\mathbf{2},4}=\frac{1}{\sqrt{3}}
\begin{pmatrix}
-1 & 0 & 0 & 0\\
0 & -1 & 0 & 0\\
0 & 0 & -1 & 0
\end{pmatrix}.\label{eq:WH2cgreal}
\end{align}
However, these are not the final (internal-group) CG coefficients that we use to construct the ERs, as we must additionally impose the conservation of the CP and hypercharges in the scattering. A linear combination of these four CG coefficients will satisfy the CP and hypercharge symmetries.

To this end, notice that the ${\bf 2}$ irrep states can be rewritten as
\begin{align*}
    \ket{\mathbf{2}_1} &= (W^1+iW^2)H^\dagger_2 + W^3 H^\dagger_1,\\
    \ket{\mathbf{2}_2} &= (W^1-iW^2)H^\dagger_1 - W^3 H^\dagger_2
\end{align*}
which actually automatically respect hypercharge conservation, as can be easily verified by acting on them with the hypercharge transform matrix $R_Y$. However, they do not respect the charge conjugation and parity symmetry. To get  the C-eigenstates $\ket{\mathbf{m}^{C}}$, according to \eref{eq:CofH}, we can linearly add the conjugate representation to construct the following states
\begin{align}
    \ket{{1}^{+i}} &= (W^1+iW^2)(H_2^\dagger+iH^1)+W^3(H_1^\dagger-iH^2),\nn
    \ket{{2}^{-i}} &= (W^1+iW^2)(H_2^\dagger-iH^1)+W^3(H_1^\dagger+iH^2),\nn
    \ket{{3}^{-i}} &= (W^1-iW^2)(H_1^\dagger+iH^2)-W^3(H_2^\dagger-iH^1),\nn
    \ket{{4}^{+i}} &= (W^1-iW^2)(H_1^\dagger-iH^2)-W^3(H_2^\dagger+iH^1).
\end{align}
where, for example, ${}^{+i}$ in $\mathbf{1}^{+i}$ denotes the eigenvalue $\blue{\eta_C}$ of the charge conjugation operator. Building upon this and further taking the tensor product with the $SO(2)$ irrep states $\ket{\mathbb{2},\text{a}}$ (cf.~\eref{eq:Pxy}, which are one-particle states, additionally adding the Higgs states to form two-particle states), we find that the irreducible two-particle states are given by,
\begin{equation}
\begin{aligned}
&\ket{1_{\rm WH}}=\ket{{1}^{+i}}\ket{\mathbb{2},1},\\
&\ket{2_{\rm WH}}=\ket{{2}^{-i}}\ket{\mathbb{2},2},\\
&\ket{3_{\rm WH}}=\ket{{3}^{-i}}\ket{\mathbb{2},2},\\
&\ket{4_{\rm WH}}=\ket{{4}^{+i}}\ket{\mathbb{2},1},\\
\end{aligned}
    ~~~~~~~~~~~
\begin{aligned}
&\ket{5_{\rm WH}}=\ket{{1}^{+i}}\ket{\mathbb{2},2},\\
&\ket{6_{\rm WH}}=\ket{{2}^{-i}}\ket{\mathbb{2},1},\\
&\ket{7_{\rm WH}}=\ket{{3}^{-i}}\ket{\mathbb{2},1},\\
&\ket{8_{\rm WH}}=\ket{{4}^{+i}}\ket{\mathbb{2},2}.
\end{aligned}
\end{equation}
\noindent However, $\{\ket{1_{\rm WH}},\ket{4_{\rm WH}},\ket{5_{\rm WH}},\ket{8_{\rm WH}}\}$ and $\{\ket{2_{\rm WH}},\ket{3_{\rm WH}},\ket{6_{\rm WH}},\ket{7_{\rm WH}}\}$ span two different invariant subspaces under the transformations of $SU(2)\times SO(2)$, so a final valid state must be a linear combination of states from the two subspaces. Using the transformation properties of $SU(2)\times SO(2)$ as well as the hypercharge conservation, we find that the valid states are given by
\begin{align}
    \ket{C^1}&=\ket{1^+_{\rm WH}} +i c\ket{2^+_{\rm WH}},\nn
    \ket{C^2}&=\ket{4^+_{\rm WH}} +i c \ket{3^+_{\rm WH}},\nn
    \ket{C^3}&=i\ket{5^+_{\rm WH}}+ c \ket{6^+_{\rm WH}},\nn
    \ket{C^4}&=i\ket{8^+_{\rm WH}}+ c \ket{7^+_{\rm WH}},
    \label{eq:c1c2c3c4}
\end{align}
where $c$ is a real parameter controlling the mixing between states of different parity.
Here $\ket{k^{\pm}_{\rm WH}}=\ket{k_{\rm WH}} \mp i\eta_C \ket{k_{\rm WH}}^T$, where $\ket{k_{\rm WH}}^T=C^{aI}_{0 \mathbf{i}}\ket{H^aW_{\mathbf{i}}^I}$ denotes the transposed state, in contrast to $\ket{k_{\rm WH}}=C^{Ia}_{\mathbf{i} 0}\ket{W_{\mathbf{i}}^I H^a}$, and the signs $\pm$ label the corresponding CP properties.
Thus, these states, together with their conjugates, are the final irrep states we use for the $WH\to X$ scattering, and following the procedure of \eref{eq:maWphi00} or \eref{eq:2122phiW} we can read out the CG coefficients from these irreps.

Note that because of the combination of two states in \eref{eq:c1c2c3c4}, CP-even but P-odd processes or ERs are allowed in the $WH$ scattering. To see this, let us consider, for example, the following projector $P^{ijkl}_{C^1} = \langle ij\ket{C^1}\bra{C^1}kl\rangle$, where
\begin{align}
    \ket{C^1}\bra{C^1}&=(\ket{1^+_{\rm WH}}+i c \ket{2^+_{\rm WH}})(\bra{1^+_{\rm WH}}-ic \bra{2^+_{\rm WH}})\nn
     &=\ket{1^+_{\rm WH}}\bra{1^+_{\rm WH}}-ic \ket{1^+_{\rm WH}}\bra{2^+_{\rm WH}} +ic \ket{2^+_{\rm WH}}\bra{1^+_{\rm WH}}+c^2\ket{2^+_{\rm WH}}\bra{2^+_{\rm WH}}.
\end{align}
In this case, $\ket{1^+_{\rm WH}}\bra{2^+_{\rm WH}}$ and $\ket{2^+_{\rm WH}}\bra{1^+_{\rm WH}}$, with the initial and final states carrying opposite parity properties, are CP-even but P-odd. Note that this projector includes the information of the $O_{M9}$ operator.

$\bullet$ Finally, for the $BH\to X$ scattering, the CG coefficients of the irrep $\bf 2$ can be straightforwardly obtained in the real basis $\{\ket{B\phi^\bfa}\}$, as the $B$ boson is not charged under the $SU(2)$. The $\{\ket{B\phi^\bfa}\}$ states are transformed in the same way as \eref{eq:Rphiabc}. Thus, the $BH\to X$ scattering contains a component that is degenerate with the $WH\to X$ scattering. Viewing $B$ as the 0-th component before $W^{1,2,3}$, we can augment the CG coefficient matrix of
 \eqref{eq:WH2cgreal} to
\begin{align}
C^{\mathbf{2},1}  & =\frac{1}{\sqrt{3}}
\begin{pmatrix}
x & 0 & 0 & 0\\
0 & 0 & 0 & 1\\
0 & 0 & -1 & 0\\
0 & 1 & 0 & 0
\end{pmatrix}
,\quad C^{\mathbf{2},2}=\frac{1}{\sqrt{3}}
\begin{pmatrix}
0 & x & 0 & 0\\
0 & 0 & 1 & 0\\
0 & 0 & 0 & 1\\
-1 & 0 & 0 & 0
\end{pmatrix}
\nonumber\\
C^{\mathbf{2},3}  & =\frac{1}{\sqrt{3}}
\begin{pmatrix}
0 & 0 & -x & 0\\
0 & 1 & 0 & 0\\
-1 & 0 & 0 & 0\\
0 & 0 & 0 & -1
\end{pmatrix}
,\quad C^{\mathbf{2},4}=\frac{1}{\sqrt{3}}
\begin{pmatrix}
0 & 0 & 0  & x\\
-1 & 0 & 0 & 0\\
0 & -1 & 0 & 0\\
0 & 0 & -1 & 0
\end{pmatrix}.
\end{align}
where $x$ is an arbitrary real parameter encoding the degeneracy from the $BH$ scattering.

\subsection{Construction via Casimir}\label{sec:Casimirconstruction}

Let us first explicitly spell out the total symmetry group and the representations of the relevant external particles in our problem; see Table~\ref{tab:rep}. Note that, to construct the group projector $P_{\bfm}^{ijkl}\equiv\sum_{\alpha}C_{i,j}^{\bfm,\alpha}(C_{k,l}^{\bfm,\alpha})^{*}$, we shall include both the continuous/Lie group symmetries as well as the discrete/finite group ones. For charge conjugation $C$, $H$ and $H^\dagger$ together form a 4D representation---in the following method we work with complex Higgs fields, rather than real scalars as in the direct construction of the previous subsection. For parity $P$, it is necessary to consider the $i,j$ particles as a whole, because parity changes the external momenta. So the task is reduced to compute the CG coefficients $C_{i,j}^{\bfm,\alpha}=\langle \bfm,\ai |i,j\rangle$.

\begin{table}[htbp]
\centering
\begin{tabular}{|c|c|c|c|c|}
\hline
 &  $W$& $B$ & $H$ & $H^\dagger$\\
\hline
$\mathrm{SU(2)_{U(1)}}$ & $\mathbf{3}_0$ & $\mathbf{1}_0$ & $\mathbf{2}_{1/2}$ & $\overline{\mathbf{2}}_{-1/2}$\\
\hline
$\mathrm{SO(2)}$& $\mathbf{2}$ & $\mathbf{2}$ & $\mathbf{1}$ & $\mathbf{1}$\\
\hline
\end{tabular}
\caption{Representations of the external particles for the relevant continuous internal and spacetime symmetries. There are also the discrete symmetries $C$ and $P$, which are not listed here.
\label{tab:rep}}
\end{table}

Note that the symmetries of the external states contain both a Lie group part $G=SU(2)\times U(1)\times SO(2)$ and a finite(discrete) group part $S$:
\be
\label{eq:ij2GS}
|i,j\rangle = |i(G,S),j(G,S) \rangle,
\ee
and they transform under different group actions, $R^{(G)}$ and $R^{(S)}$, which will be treated slightly differently.

For the Lie group sector, the CG coefficients can be obtained by solving an eigenvalue problem involving the quadratic Casimirs. To see how it works, we start with
\bal
| \bfm,\ai \rangle &= \sum_{i,j} | i,j\rangle  \langle i,j | \bfm,\ai \rangle =    \sum_{i,j} \(C_{i,j}^{\bfm,\alpha}\)^* | i,j\rangle 
\eal
Under a group action, both sides will transform according to their respective representations:
\be
R^{(G_\mathbf{m})}_{\alpha \beta} | {\bfm}, {\beta} \rangle
= \sum_{i,j,p,q} \(C_{i,j}^{\bfm,\alpha}\)^* R^{(G)}_{ip}R^{(G)}_{j q} | p,q\rangle .
\ee
Contracting both sides with $\langle i,j|$, we have
\be
R^{(G_\mathbf{m})}_{\alpha \beta} \(C_{i,j}^{\bfm,\beta}\)^*
= \sum_{p,q} \(C_{p,q}^{\bfm,\alpha}\)^* R^{(G)}_{pi}R^{(G)}_{q j}.
\label{eq:RCCRR}
\ee
We now split $R^{(G)}$ (as well as $R^{(G_\bfm)}$) into two consecutive actions $R^{(G)}=R^{(G)}_1 R^{(G)}_2$ and consider infinitesimal group actions near the identity $R^{(G)}_1=I +i\varepsilon^{(1)}_r T_r$ and $R^{(G)}_2=I +i\varepsilon^{(2)}_r T_r$, where $T_r$ are the generators of the Lie algebra. Then, matching the ${\cal O}(\varepsilon^{(1)}\varepsilon^{(2)})$ of \eref{eq:RCCRR} gives rise to 
\bal
(T_r T_s)^{({\bfm})}_{{\alpha}{ \beta}} \(C_{i,j}^{\bfm,\beta}\)^*
 &= 
\sum_{p,q}
\(C_{p,q}^{\bfm,\alpha}\)^*
\left[(T_r \otimes I + I \otimes T_r
)(T_s \otimes I + I \otimes T_s
)\right]_{p q,ij} \\
 &\equiv\sum_{p,q,p',q'} \(C_{p,q}^{\bfm,\alpha}\)^*
\left[(T_r)_{p {p'}} I_{q{q'}} + I_{p {p'}}  (T_r)_{q {q'}} \right]
 \\ &~~~~~~~~~~~~~~~~~~~~~~~~ \times \left[(T_s)_{{p'} i} I_{{q'} j} + I_{{p'} i} (T_s)_{{q'} j}
\right]
\eal
Acting the Killing form\,\footnote{For the abelian group, which has no Killing form, we define the ``quadratic Casimir'' simply as the square of the linear Casimir.} on both sides, we get an equation involving the quadratic Casimir $C_2$:
\bal
C_2(T)^{(\bfm)}_{\alpha \beta}\(C_{i,j}^{\bfm,\beta}\)^* &= 
\sum_{p,q}
\(C_{p,q}^{\bfm,\alpha}\)^*
C_2(T\otimes I + I \otimes T)_{p q,i j} 
\\
C_2(T)^{(\bfm)}\(C_{i,j}^{\bfm,\ai}\)^* &= 
\sum_{p,q}
\(C_{p,q}^{\bfm,\alpha}\)^*
C_2(T\otimes I + I \otimes T)_{p q,i j}
\\
C_2(T)^{(\bfm)} C_{i,j}^{\bfm,\ai} &= 
\sum_{p,q} C_2(T\otimes I + I\otimes T)_{i j, p q}
C_{p,q}^{\bfm,\alpha}
\label{eq:casimirEigen}
\eal
where in the second equation above we have used the fact that the quadratic Casimir for a given irrep is proportional to the identity matrix $C_2(T)^{(\bfm)}_{\ai \beta} = C_2(T)^{(\bfm)}\dd_{\ai \beta}$. From \eref{eq:casimirEigen}, for a given irrep ($\bfm$ and $\ai$), we see that the problem of finding the CG coefficients can be viewed as an eigenvalue problem of computing the eigenvector $C_{i,j}^{\bfm,\ai}$, if we view $i,j$ (or $p,q$) as one index. 

To obtain the explicit form of the eigenvalue problem, we simply need to identify the algebra generators $T_r$ in the given representation. For our current problem, the continuous symmetry group is $SU(2)\times U(1)\times SO(2)$. We shall choose the basis of the representation space in the following order 
\be
\label{eq:Vspacedef}
V = \mathrm{Span}\{B_1,B_2,W^1_1,W^1_2,W^2_1,W^2_2,W^3_1,W^3_2,H^1,\\H^2,H_1^{\dagger},H_2^{\dagger}\} .
\ee
In this basis, the Lie algebra generators, which are now $12\times12$ matrices, are given by
\begin{equation}
\begin{aligned}
    J^i_{\mathrm{SU}(2)} &= 0_{2 \times 2} \oplus 
     \left[J^i_{\mathbf{3}} \oplus 
     J^i_{\mathbf{3}}\right]^p \oplus J^i_{\mathbf{2}} \oplus J^i_{\Bar{\mathbf{2}}} \\
    Y_{U(1)} &= 0_{8\times 8} \oplus\dfrac{1}{2} \oplus \dfrac{1}{2} \oplus-\dfrac{1}{2} \oplus -\dfrac{1}{2}\\
    J^{z}_{\mathrm{SO(2)}} & = J^{z}_{\mathbf{2},B} \oplus J^z_{\mathbf{2},W} \oplus J^z_{\mathbf{2},W}\oplus J^z_{\mathbf{2},W} \oplus 0_{4 \times 4} 
\end{aligned}
\end{equation}
where $J^i_{\bf 2}= \f12\sigma^i$ and $(J^i_{\mathbf{3}})_{jk}= - i \epsilon_{ijk}$, and $p = (W^1,W^2,W^3,W^1,W^2,W^3) \to (W^1,W^1,W^2,\\ W^2,W^3,W^3)$ indicates a permutation of basis in order to agree with the basis $V$. The quadratic Casimir $C_2(T\otimes I + I \otimes T)$ for $SU(2)\times U(1)\times SO(2)$ is given by  a linear combination of 
\begin{equation}
\label{eq:3casimirs}
(\mathbf{J}_{\mathrm{SU(2)}}\otimes I + I \otimes \mathbf{J}_{\mathrm{SU(2)}} )^2,~~ (Y_{\mathrm{U(1)}}\otimes I + I \otimes Y_{\mathrm{U(1)}})^2 ,~~  (J^z_{\mathrm{SO(2)}} \otimes I + I \otimes J^z_{\mathrm{SO(2)}})^2 .
\end{equation}
This means that the $C_{i,j}^{\bfm,\ai}$ we seek is a simultaneous eigenvector for the above three Casimirs.

For the discrete/CP symmetry sector, parity acts on the $i,j$ particles as a single entity. The CG coefficients $C_{i,j}^{\bfm,\alpha}$ for this sector is merely notational. In this case, the equivalent of \eref{eq:RCCRR} becomes
\be
R^{(S_\bfm)}_{\alpha \beta} \(C_{i,j}^{\bfm,\beta}\)^* = \sum_{p,q} \(C_{p,q}^{\bfm,\alpha}\)^*R^{(S)}_{{pq},{ij}} ,
\ee
where the right hand side now only has one $R$ because we regard the discrete group part as furnishing a single representation. In our convention, the CP symmetry forms $S = Z_4$  (cf.~Section \ref{sec:CP}). This is an abelian finite group, which only has 1D irreps, so the left hand side can be written as $R^{(S_\bfm)}_{\alpha \beta} = R^{(S_\bfm)} \delta_{\alpha \beta}$, leading to
\be
\label{eq:CPeigen0}
R^{(S_\bfm)} C_{i,j}^{\bfm,\alpha} = \sum_{p,q} R^{(S)}_{{ij},{pq}} C_{p,q}^{\bfm,\alpha}.
\ee
Thus, it is again an eigenvector problem to compute $C_{p,q}^{\bfm,\alpha}$. However, different from the Lie group sector, it is now straightforward to construct the representation matrix $R^{(S)}_{ij,pq}$. Still choosing \eref{eq:Vspacedef} as the basis, from Section \ref{sec:CP}, we find that the $R^{(S)}_{ij,pq}$ matrix is given by
\be
R^{(S)}_{{ij},{pq}} = C_{ij,p'q'} P_{p'q',p,q},~~~C = C_0 \otimes C_0, \quad P_{pq,ij} = (P_0)_{pj}(P_0)_{qi} ,
\ee
where we have defined
\bal
C_{0} & = \(-I\)_{2\times 2} \oplus I_{6 \times 6} \oplus C_H , \\
P_{0} & =  P_B \oplus P_W \oplus P_W \oplus P_W \oplus I_{4 \times 4} ,
\\
(C_H)_{ip} &= (-1)^{i+1}\dd_{i,5-p},~~~ P_B = P_W = {\rm diag}(1,-1) .
\eal
For $U(1)$, the quadratic operator $Y_{\rm tot}^{2}$, with
$Y_{\rm tot}=Y_{\mathrm{U(1)}}\otimes I+
I\otimes Y_{\mathrm{U(1)}}$, does not distinguish the inequivalent
irreps with charges $\pm q$ ($q\neq0$).
We nevertheless use its square because we work with CP eigenstates:
CP reverses hypercharge, so
\begin{equation}
 (CP)Y_{\rm tot}( CP)^{-1}=-Y_{\rm tot},
\qquad
[CP,Y_{\rm tot}^{2}]=0.
\end{equation}
For EFTs in which CP symmetry is not imposed, charge-resolved irreps
should instead be obtained by additionally diagonalizing
$Y_{\rm tot}$ within the degenerate eigenspaces.

With both the Lie group and discrete group eigenvector equations established, for a given irrep $(\bfm,\ai)$, we shall seek an eigenvector $C_{i,j}^{\bfm,\alpha}$ that simultaneously solves these eigenvector equations, {\it i.e.}, \eref{eq:casimirEigen} for each of the 3 quadratic Casimirs in \eref{eq:3casimirs} as well as \eref{eq:CPeigen0}, subject to the additional condition \eref{eq:spin}. The resulting CG coefficients are listed in Appendix~\ref{appd:CG}.

For some irreps, the constrained eigen problem above is not uniquely determined---there can be up to 2 free (real) parameters in the obtained CG coefficients. In terms of the group tensor-product decomposition, this means that they are multiplicities for some irreps (see Table \ref{tab:repofX} or Appendix~\ref{appd:CG}), which can also be seen via the direct construction method of \Cref{sec:directConstruction}. In the table, we have additionally denoted the UV spin parity $(J \ \mathrm{mod \ 2})$ for the irreps. Strictly speaking, the UV spin parity is not part of the irrep $m$, but it can be predicted in our formalism and help to break the degeneracy. The existence of these degeneracies means that the amplitude cone a non-polyhedral cone, which complicates the task of fully determining its structure.

\begin{table}[htbp]
\centering
\begin{tabular}{c|l l l l}
\hline
multiplicity & \multicolumn{4}{c}{$(SO(2)_{|h|},SU(2)_{U(1)})^{\rm CP}_{(\ell \ \mathrm{mod \ 2})}$ }\\
\hline
\multirow{3}*{$ 1$} & $(\mathbf{1}_0, \mathbf{1}_0)^+_{1}$ & $(\mathbf{1}_0, \mathbf{1}_1)^\pm_{1}$&  $(\mathbf{1}_0, \mathbf{3}_0)^+_{0}$ & $(\mathbf{1}_0, \mathbf{3}_1)^\pm_{0}$
\\
~  &
$(\mathbf{1}_0, \mathbf{5}_0)^+_{0}$ & $(\mathbf{1}_0, \mathbf{5}_0)^-_{0}$& $(\mathbf{2}_1, \mathbf{4}_{1/2})^{\pm i}$ & $(\mathbf{2}_2, \mathbf{5}_0)^+_{0}$
\\
~ & $(\mathbf{2}_2, \mathbf{3}_0)^-_{0}$ & $(\mathbf{2}_2, \mathbf{3}_0)^+_{1}$ & $(\mathbf{2}_2, \mathbf{3}_0)^-_{1}$ & 
\\
\hline
\multirow{2}*{$2$} & $(\mathbf{1}_0, \mathbf{1}_0)^-_{0}$ & $(\mathbf{1}_0, \mathbf{3}_0)^-_{0}$ & $(\mathbf{1}_0, \mathbf{3}_0)^+_{1}$ &
$(\mathbf{2}_2, \mathbf{1}_0)^+_{0}$
\\
~ & $(\mathbf{2}_1,\mathbf{2}_{1/2})^{\pm i}$ &
  & 
\\
\hline
$3$ & $(\mathbf{1}_0, \mathbf{1}_0)^+_{0}$ & $(\mathbf{1}_0, \mathbf{3}_0)^-_{1}$
\\
\hline
\end{tabular}
\caption{Multiplicities of the irreps $(SO(2)_{|h|},SU(2)_{U(1)})^{\rm CP}_{(\ell \ \mathrm{mod \ 2})}$. For example,  an irrep with multiplicity 2 means that the corresponding CG coefficient vector $C^{\bfm,\ai}_{i,j}$ contains $2-1=1$ free parameter. The $SO(2)$ irreps are denoted by their dimensions together with their helicities $h$. The CP symmetry has 4 eigenvalues: $\pm 1$ and $\pm i$. ($\ell$ mod 2) denotes the UV spin parity (see Section \ref{sec:exchangeCG}).
\label{tab:repofX}}
\end{table}

\subsection{Extremal rays}
\label{sec:ERs}

Having obtained the CG coefficients via two different methods, we are now ready to enumerate the ERs of the positivity cone. For a given irrep $\bfm$, potential ERs are given by $\bar P_{\bfm}^{ijkl}= \f{1}{2}\sum_{\alpha}[C_{i,j}^{\bfm,\alpha}(C_{k,l}^{\bfm,\alpha})^{*} +C_{i,\bar{l}}^{\bfm,\alpha}(C_{k,\bar{j}}^{\bfm,\alpha})^{*}]$. Remember that these are only potential ERs because it is essentially a projection to a subspace, going from $P_{\bfm}^{ijkl}= \sum_{\alpha}C_{i,j}^{\bfm,\alpha}(C_{k,l}^{\bfm,\alpha})^{*}$ to $\bar P_{\bfm}^{ijkl}$, and as a result, some of $\bar P_{\bfm}^{ijkl}$ can be written as conical hulls of the rest. However, as we will see shortly, most of the potential ERs are real ERs.

Let us first enumerate the potential ERs. To represent them, we need to determine the dimension of the vector space spanned by these projectors. This can be achieved by analyzing the linear dependence relations among the $\bar P_{\bfm}^{ijkl}$'s
after factoring out the degeneracy parameters $x$, $y$, and $u$. We find that the 
$\bar P_{\bfm}^{ijkl}$'s span a 25D vector space. We shall choose the basis of this space to be closely related to the aQGC operators: The first 22 basis vectors are given by the amplitudes corresponding to the individual $22$ aQGC operators $\{M^{\mathcal{I}}_{ijkl},\mathcal{I}=1,2\dots,22\}$, and the rest 3 basis vectors are chosen to be components of $\bar P_{\bfm}^{ijkl}$'s that contain continuous parameters. Specifically, the rest 3 basis vectors are chosen to be: The vector containing only the constant and $x$ components of $({\bf1}_0,{\bf3}_0)_1^-$, and the vector containing only the $x^2$ components of $({\bf 1}_0,{\bf3}_0)_1^+$. 
Most of the 3 basis vectors have nontrivial overlaps with $\{M^{\mathcal{I}}_{ijkl},\mathcal{I}=1,2\dots,22\}$, except for the vector containing the $x$ components of $({\bf1}_0,{\bf3}_0)_1^-$ (corresponding to the $WB\to WW$ scattering), which is orthogonal to the space spanned by the aQGC operators. Therefore, we shall consider the 25D space that is related to the aQGC amplitude cone. 
Having defined this basis, the potential ERs can be expressed as
{\footnotesize
\begin{align}
    &\vec{e}_1= \left(0, 0, 0, 0, 0, 0, 0, 0, 0, 2, 0, -2, 0, 0, 0, 0, 0, 0, 0, 0, 0, 0, 0, 0, 0\right),\nn
    &\vec{e}_2= \left(0, 0, 0, 0, 0, 0, 0, 0, 0, 0, 0, 0, 0, 0, 0, 0, 0, -2 , -2 , 8 , 0, 0, 0, 0, 0\right),\nn
    &\vec{e}_3= \left(0,0,0,0,0,0,0,0,0,0,0,0,0,0,0,0,0,0,-2,8,0,0,0,\f{1}{2},0\right),\nn
    &\vec{e}_4= \left(0,0,0,0,0,0,0,0,0,0,0,0,-\f{1}{4},-\f{1}{12},1,-\f{1}{3},0,0,0,0,0,0,0,0,0\right),\nn
    &\vec{e}_5= \left(0,0,0,0,0,0,0,0,0,0,0,0,-\f{1}{12},\f{1}{4},0,0,0,0,0,0,0,0,0,0,0\right),\nn
    &\vec{e}_6= \left(0,0,0,0,0,0,0,0,0,0,0,0,\f{1}{12},-\f{1}{4},1,-\f{1}{3},0,0,0,0,0,0,\f{5}{12},0,0\right),\nn
    &\vec{e}_7= \left(0,0,0,0,0,0,0,0,0,0,0,0,-1,1,4,-4,0,0,0,0,0,0,1,0,0\right),\nn
    &\vec{e}_8= \left(0,0,0,0,0,0,0,0,0,0,0,0,0,0,0,0,-16 ,-4 ,4 ,16 ,0,0,0,1,0\right),\nn
    &\vec{e}_9= \left(0,0,0,0,0,0,0,0,0,2,0,0,0,0,0,0,0,0,0,0,0,0,0,0,0\right),\nn
    &\vec{e}_{10}= \left(0,0,0,0,0,0,0,0,0,0,-2,2,0,0,0,0,0,0,0,0,0,0,0,0,0\right),\nn
    &\vec{e}_{11}(x)= \left(0,0,0,0,0,0,0,0,0,0,0,0,0,-\f{1}{2},0,1,4 x,0,-2 x,0,-2 x^2,4 x^2,0,0,0\right),\nn
    &\vec{e}_{12}(x)= \left(0,0,0,0,0,0,0,0,0,0,0,0,-\f{1}{4},0,0,1,4 x,-x,0,0,-x^2,4 x^2,-\f{1}{2},-\f{x}{2},0\right),\nn
    &\vec{e}_{13}(x)= \left(-\f{8x^2}{3},-\f{8}{3}\left(3 x^2+1\right),0,0,0,0,-\f{64}{3}\left(x^2-1\right),0,\f{64x}{3},0,0,0,0,0,0,0,0,0,0,0,0,0,0,0,0\right),\nn
    &\vec{e}_{14}(x)= \left(0,0,0,0,0,0,0,0,0,0,0,0,0,0,4,-4,0,0,0,0,0,0,1,x^2,-x\right),\nn
    &\vec{e}_{15}(x)= \left(0,0,0,0,-16 x,0,0,0,0,0,-2 x^2,4 x^2,0,0,0,0,0,0,4,0,0,0,0,0,0\right),\nn
    &\vec{e}_{16}(x,y)= \left(0,-8 y,0,0,0,0,-64 y,-64 x y,0,2 y^2,-4 y^2,2 y^2,0,0,0,0,-16 x^2,0,0,16 x^2,0,0,1,x^2,x\right),\nn
    &\vec{e}_{17}(x,y)= \left(2 y,0,4 x y,0,0,0,0,0,0,0,2 y^2,0,\f{1}{2},0,0,0,0,2 x,0,0,2 x^2,0,0,0,0\right),\nn    
    &\vec{e}_{18}(x,y,u)= \left(-x^2,-4,-2 u^2,-4 (y^2+u^2),8 x u,-16 (y+x u),16 (-1+x^2),16 (-x y+u),-16 x,0,\dots,0\right).\label{eq:potential ERs}
\end{align}
}
\noindent where it is understood that the real parameters $x$, $y$, and $u$ in different potential ERs are unrelated and can be chosen independently. Because of these real parameters, the amplitude cone is non-polyhedral and has curved surfaces as its boundaries, featuring an infinite number of ERs.

To determine whether a potential ER is really an ER, we can check whether it can be expressed as a conical hull of the rest. For the discrete potential ERs, we find that 
\be
\label{eq:nonERs}
\vec{e}_6,~~\vec{e}_{9} ~~\text{are not real ERs.}
\ee
For the continuous ERs ({\it i.e.}, $\vec{e}_{11}$ to $\vec{e}_{18}$), in principle, it is possible that parts of or the whole potential ER is redundant and contained within the positivity cone. However, as we see in the next section, numerically, we find that it seems that these continuous potential ERs all live on the boundaries of the positivity cone, though not always extremal.

\section{Optimal positivity bounds for aQGC coefficients}\label{sec:bounds}

The amplitude/positivity cone is determined by its ERs---once the ERs are known it is easy to check whether a given EFT is within the cone or not. The positivity bounds are essentially the inequalities describing the boundaries of the amplitude cone. Mathematically, there are two equivalent representations of a convex cone: the inequality representation (or H-representation) and the ER representation (or V-representation), see Appendix~\ref{appd:convgeo} for slightly more details. The positivity inequalities are actually the ERs of the dual cone of the amplitude cone. For a polyhedral cone, vertex enumeration can be used to essentially construct the ERs of the dual cone from the ERs of the primal cone, giving rise to optimal, analytical positivity bounds. In our case, however, some ERs are characterized by one or two free parameters, preventing us from getting optimal bounds analytically. 
In this section, we will present the strategy for obtaining some partial analytical bounds and computing the full optimal numerical bounds.

\subsection{Linear analytical bounds}\label{sec:linearPosiBound}

It is instructive to first derive some partial analytical positivity bounds on the Wilson coefficients, which are easiest to use and may provide useful intuition. We will focus the simplest bounds that are linear in the Wilson coefficients. The conversion from the ER representation to the inequality representation can be achieved through vertex enumeration. This is a complete solution for a polyhedral cone, but for a non-polyhedral cone, like our case, approximations are needed. 

Generally, vertex enumeration works as follows. Given the ERs $e_i^{n}$ of a convex polyhedral cone that live in a $d$-dimensional space, one can construct the ERs of the dual cone via 
\begin{equation}
    E^{n}=\pm \epsilon^{n_{1} n_{2}\dots n_{d-1} n}e_{i_{1}}^{n_{1}}e_{i_{2}}^{n_{2}}\cdots e_{i_{d-1}}^{n_{d-1}}\label{eq:erdual}
\end{equation}
where $\epsilon^{n_1\dots n_d}$ is the Levi-Civita symbol and $i_1,\dots,i_{d-1}$ label distinct ERs, as those listed in \eref{eq:potential ERs}. This is because an ER in the dual cone is a 1D intersection of the dual cone facets: $\vec{E}\cdot \vec{e}_{i_1}=\vec{E}\cdot \vec{e}_{i_2}=...=\vec{E}\cdot \vec{e}_{i_{d-1}}=0$. That is, an ER in the dual cone lives in the null space of $d-1$ given $e_i$'s, thus giving rise to \eref{eq:erdual}. There is a sign ambiguity in \eref{eq:erdual}, which corresponds to different orderings of $\vec{e}_{i_1}, \vec{e}_{i_2}, ..., \vec{e}_{i_{d-1}}$. This can be fixed by imposing the defining property of the dual cone
\begin{equation}
    E^{n}e_{i}^{n}=\epsilon^{n_{1}n_{2}\dots n_{d-1}n}e_{i_{1}}^{n_{1}}e_{i_{2}}^{n_{2}}\cdots e_{i_{d-1}}^{n_{d-1}}e_{i}^{n}\geq0,~~\forall\vec{e}_{i}\in \mathcal{C}.
\end{equation}

In the presence of free parameters in the ERs $\vec{e}_i(z)$, as in our case (where we have up to three real parameters $z=(x,y,u)$), a dual-cone ER must additionally be orthogonal to the neighboring ERs of $\vec{e}_i(z)$. To take into account of the infinitesimally neighboring ERs, we can include derivatives of $\vec{e}_i(z)$ with respect to the free parameters when constructing the dual-cone ERs. Incorporating this consideration, for cases involving continuous parameters, we can build dual-cone ERs as follows 
\begin{equation}
    E^{n}_{\rm cont}(z_1,z_2,\dots)=\pm \epsilon^{n_{1}n_{2}\dots n_{d-1}n}e_{i_{1}}^{n_{1}}(z_1)e_{i_{1}}'^{n_{2}}(z_1) e_{i_{2}}^{n_{3}}(z_2) e_{i_{2}}'^{n_{4}}(z_2)\cdots,\label{eq:combinationE}
\end{equation}
where primes denote derivatives with respect to the corresponding free parameters. Note that there are also cases where $\vec{e}_{i}$ ({\it i.e.}, $\vec{e}_{16}$, $\vec{e}_{17}$, and $\vec{e}_{18}$) contains two or three free parameters, which should be dealt with partial derivatives. 

Denoting all dual-cone ERs collectively as $\vec{E}(z_1,z_2,\dots)$, the positivity bounds in our case are then given by
\begin{equation}
    \vec{E}(z_1,z_2,\dots)\cdot\vec{F}\geq 0, ~~~~{\bf subject~to}~~\vec{E}(z_1,z_2,\dots)\cdot \vec{e}_i\geq 0,~\forall\vec{e}_i\in \mathcal{C}.\label{eq:posibounds}
\end{equation}
where $\vec{F}=(F_{M,\mathcal{I}_1}, F_{S,\mathcal{I}_2}, F_{T,\mathcal{I}_3},0,0,0)$ and $\mathcal{I}_1,~\mathcal{I}_2$ and $\mathcal{I}_3$ label different aQGC Wilson coefficients.

The above construction amounts to a brute-force (continuous) generalization of (discrete) vertex enumeration. \eref{eq:posibounds} defines a positive semi-definite program with high-degree polynomials of the parameters $(z_1,z_2,\dots)$. Optimizing over $(z_1,z_2,\dots)$, we can extract positivity bounds on the Wilson coefficients $\vec{F}$. In practice, however, solving this problem analytically becomes intractable for generic $(z_1,z_2,\dots)$. Following~\cite{Yamashita:2020gtt}, we adopt the following approximations when constructing $\vec{E}$:
\begin{itemize}
    \item Each $\vec{e}_i(z)$ is restricted to taking values $\vec{e}_i(0)$, $\vec{e}_i(\infty)$ or $\vec{e}_i(\pm z_0)$, and likewise for its derivatives;
    \item All these $\vec{e}_i(z)$'s appearing in the combination~\eref{eq:combinationE} share the same parameter $z_0$.
\end{itemize}
Note that this approximation is performed when constructing the dual cone, which restricts the size of the dual cone, corresponding to produce a conservative primal amplitude cone.
For ERs depending on a single parameter, it is straightforward to apply the above approximations across different ERs. For ERs depending on multiple free parameters, generic cases are still difficult to solve. To make the problem analytically tractable, we make a further approximation by restricting ourselves to two cases: $y=u=0$ and $y=x=u$. Remember that the parameters $y$ and $u$ encode the degeneracy of ERs from the sector of Higgs scattering. By choosing $y=u=0$, the scenario is reduced to the case of transversal vector boson scattering. We would like to emphasize that these approximations are adapted to obtain the analytical bounds, and will not be used in computing the numerical bounds in the next subsection.

\begin{figure}
    \centering
    \includegraphics[width=0.6\linewidth]{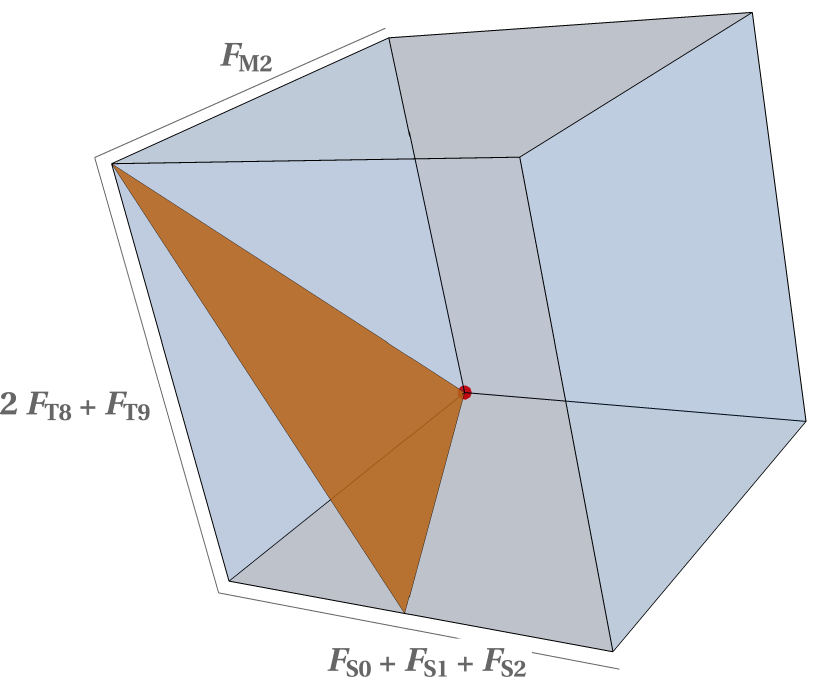}
    \caption{Example of extra linear bounds from mixing different types of operators. The light blue region represents the allowed parameter space carved out by $F_{S0}+F_{S1}+F_{S2}\geq0$ and $2F_{T8}+F_{T9}\geq0$. The mixing introduces another linear bound $2\left(  F_{S0}+F_{S1}+F_{S2}\right)  +2F_{T8}+F_{T9}-2F_{M2} \geq0$ (space above the brown plane), which chops off the triangular cone below the brown plane.}
    \label{fig:FMFSFT}
\end{figure}

By choosing the $x$ and $y$ in appropriate ERs to be zero, we can reproduce the previous analytic linear positivity bounds for transversal vector boson scattering \cite{Yamashita:2020gtt}
\begin{align}
    &F_{T9} \geq 0\nn
    &F_{T2} + 2 F_{T3} \geq 0,\nn
    &F_{T7} \geq 0,\nn
    &4 F_{T0} + 4 F_{T1} + 3 F_{T2} + 5 F_{T3} \geq 0,\nn
    &8 F_{T0} + 4 F_{T1} + 3 F_{T2} + 2 F_{T3} \geq 0,\nn
    &12 F_{T0} + 4 F_{T1} + 5 F_{T2} + 3 F_{T3} \geq 0,\nn
    &4 F_{T1} + F_{T2} + 2 F_{T3} \geq 0,\nn
    &F_{T2}\geq 0,\nn
    &2 F_{T8} + F_{T9} \geq 0,\nn
    &2 F_{T4} + 4F_{T6} + F_{T7} \geq 0.
\end{align}
Slightly less transparent but still straightforward to extract from our ERs are the analytic linear positivity bounds for Higgs scattering, which are given by~\cite{Zhang:2020jyn} (see \cite{Remmen:2024hry} for an extension of the extremal bounds on Higgs scattering in HEFT),
\begin{align}
    &F_{S0}+F_{S1}+F_{S2}\geq0 \nn
    &F_{S0}+F_{S2}\geq0 \nn
    &F_{S0}\geq0
\end{align}
Now, with our setup, we can obtain linear analytical bounds on the M-type aQGC operators. Recall that the presence of continuous parameters is due to degeneracy arising from the superposition of different particle states with the same group structure. As discussed in \Cref{sec:aQGC}, M-type aQGC operators involve interactions between the Higgs scalar and the electroweak gauge bosons $W$ and $B$. 
To derive bounds on these coefficients, we can neglect the degeneracy due to ERs that involve vector boson scattering only, but we must keep all contributions from the Higgs sector.
The new linear analytical bounds on the M-type of aQGC coefficients are:
\begin{align}
    &F_{M3}   \leq0\nn
    &F_{M7}-8F_{M1}-2F_{M9}   \geq0\nn
    &F_{M7}+2F_{M9}-8F_{M1}   \geq0
\end{align}
Additionally, we can also derive linear analytical bounds for mixtures of different types of aQGC operators, for example, by choosing $x=y=u$ in the continuous ERs, {\it i.e.,} using a subset of the ERs $\vec{e}_{16}$, $\vec{e}_{17}$, and $\vec{e}_{18}$:
\begin{align}
    &2\left(  F_{S0}+F_{S1}+F_{S2}\right)  +2F_{T8}+F_{T9}-2F_{M2} \geq0\nn
    &F_{M7}-8F_{M1}+2F_{M8}+32F_{S0}+4F_{T7}-4F_{M3}\geq0\nn
    &F_{M7}-8F_{M1}+2F_{M8}+16(F_{S0}+F_{S2})+4F_{T7}-4F_{M3}\geq0
\end{align}
These bounds indicate that mixing different types of operators results in stronger constraints. For example, consider a case where all the aQGCs vanish except for $F_{S0}+F_{S1}+F_{S2}$, $2F_{T8}+F_{T9}$ and $F_{M2}$. Then, the positivity bound for considering S-type operators only is $F_{S0}+F_{S1}+F_{S2}\geq0$, while considering only T-type operators leads to $2F_{T8}+F_{T9}\geq0$. However, mixing different types of operators leads to an additional bound $2\left(  F_{S0}+F_{S1}+F_{S2}\right)  +2F_{T8}+F_{T9}-2F_{M2}\geq0$, further reducing the allowed parameter space, as illustrated by Figure~\ref{fig:FMFSFT}. It is worth pointing out that the bound $F_{S0}+F_{S1}+F_{S2}\geq0$ (or $2F_{T8}+F_{T9}\geq0$) is the optimal bound when agnostic about the remaining operators, which pictorially corresponds to project the bounds to the $F_{S0}+F_{S1}+F_{S2}$ (or $2F_{T8}+F_{T9}$) axis.

\subsection{Optimal positivity bounds}\label{sec:NumBound}

As mentioned, our positivity cone for the aQGC coefficients is non-polyhedral, so generic optimal bounds are obtained numerically via discretizing the continuous parameters. To verify whether a set of aQGC coefficients are ultimately consistent with the fundamental principles of quantum field theory, one should use the algorithms presented in this subsection. We shall supplement this subsection with a Python package {\tt SMEFTaQGC} (attached to the arXiv submission), which implements an algorithm to verify whether a given set of aQGC coefficients are within the optimal positivity bounds, and an algorithm to search for optimal bounds for an arbitrary linear combination of aQGC coefficients.

To find the optimal numerical bounds, we shall choose the following discretization scheme for the continuous parameters:  map $z$ to a finite range via the tangent function and then discretize evenly on this range, 
\begin{equation}
\label{eq:disScheme}
z_{i}=\left\{\begin{array}{ll}
\tan \frac{i \pi}{2 N}, & \text { for }-N+1 \leq i \leq N-1 \\
\infty, & i=N  .
\end{array}\right.
\end{equation}
The procedure for obtaining the ERs in the limit $x_N, y_N \to \infty$ requires some clarification. For an ER with one continuous parameter, we can simply define $\vec{e}_i(x)$ at infinity as 
\be
\vec{e}_i(\infty)=\lim_{x\to\infty} x^{-2}\vec{e}_i(x) .
\ee
For an ER with two continuous parameters, we can extract the effective ERs at infinity as follows. Suppose an ER has the form $\vec{e}(x,y)=(0, a_1,a_2 x,a_3 y, a_4 x y, a_5 x^2, a_6y^2)$, where $a_n$ are constants. As $x,y\to \infty$, we have
\be
\vec{e}(x,y)\to (0, 0, 0, 0, a_4 x y, a_5 x^2, a_6y^2)= y^2(0, 0, 0, 0, a_4 \tilde x, a_5 \tilde x^2, a_6) = y^2 \tilde{\vec{e}}(\tilde x)
\ee
where we have defined $\tilde x=x/y$, which can be finite, and positively re-scaled ER $\tilde{\vec{e}}(\tilde x)$. For a convex cone, a positively re-scaled ER is as good as the original ER. Thus, a two-parameter ER at infinity effectively gives rise to a new ER with one parameter, for which we again should discrete according to \eref{eq:disScheme}.

Each single-parameter ER contributes $2N$ numerical ERs while each two-parameter ER contributes $(2N-1)^2+2N$ numerical ERs, with the last $2N$ arising from the ER evaluated at infinity. For convenience, all numerical ERs are collectively denoted by $\vec{\tt e}_{i}$ and discretized positivity cone is referred to as ${\tt C}$.

\subsubsection{Positivity check} 
\label{sec:positivityCheck}

With the explicit forms of the discretized ERs, it is rather efficient to check whether a given set of Wilson coefficients $\vec{F}=(F_{M,\mathcal{I}_1}, F_{S,\mathcal{I}_2}, F_{T,\mathcal{I}_3},0,0,0)$ lies within the positivity cone that defines the consistent physical parameter space satisfying the S-matrix axioms. It can be determined by the following linear program: To check whether there exists a set of positive real numbers (or decision variables) $w_i$ such that $\sum_{i} w_{i} \vec{\tt e}_{i}=\vec{F}$, where $\vec{\tt e}_{i}$'s are the discretized ERs of the amplitude cone listed in \Cref{sec:ERs}. Or, in the standard optimization language, we may write 
\begin{equation}
\begin{array}{ll}
\min_{w_i} & 0 \\
\text {such that} & \sum_{i} w_{i} \vec{\tt e}_i=\vec{F},~~~ w_{i} \geq 0 ,
\end{array}\label{eq:inoutcone}    
\end{equation}
with the objective function being simply zero. Since the objective function is a constant, the search can be terminated once a set of feasible $w_i$ is found. This problem can be efficiently solved using standard algorithms such as the simplex method or the interior-point method. In the accompanying Python package {\tt SMEFTaQGC}, this functionality is implemented in a function named {\tt CheckPositivity}, which invokes the {\tt linprog} routine in {\tt scipy}.

For example, this verification might be useful when constructing phenomenological EFT models to fit the experimental data. One should make sure that the proposed aQGC coefficients lie within the positivity cone. 

If the initial proposed phenomenological EFT model lies outside the positivity cone, {\it i.e.}, if {\tt CheckPositivity} returns a negative result, we have implemented another function {\tt FeasibleDirection}, which provides a feasible direction that allows model builders to adjust the aQGC coefficients so as to satisfy the positivity bounds. For this purpose, we  define a ``most interior'' ray within the positivity cone. There are, of course, ambiguities in defining the most interior ray geometrically, and physically the underlying theory may not correspond to the most interior ray. Nevertheless, it may provide some useful intuition on how to modify the model. We shall choose the most interior ray\,\footnote{Another choice would be the average of the normalized ERs: $\vec{F}_0=\sum_i \hat{\vec{{\tt e}}}_{i}/\text{(number of ERs)}$.}
to be $\vec{F}_0=\sum_i\alpha_i \hat{\vec{\tt e}}_{i}$ that maximizes the minimal coefficient among all of the $\alpha_i$'s. Here, the decision variables $\alpha_i$ are subject to $\sum_i\alpha_i=1$, and the normalized ($\hat{\vec{{\tt e}}}_{i}=\vec{{\tt e}}_{i}/|\vec{{\tt e}}_{i}|$) ERs are constrained to correspond to a valid set of aQGC coefficients, $\sum_i \alpha_i \hat{\tt e}^n_i=0,~n=23,24,25$, reflecting the fact that a physical coefficient vector has vanishing entries in its last three components $\vec{F}=(F_{M,\mathcal{I}_1}, F_{S,\mathcal{I}_2}, F_{T,\mathcal{I}_3},0,0,0)$. This optimization can be implemented as a linear program with the objective of maximizing a semi-positive $\lambda$ that is subject to $\lambda \le \alpha_i$ for all $i$.

Then, we can find a feasible direction $\vec{d}_0=\vec{F}_0-\vec{F}$, where $\vec{F}$ is the initially-chosen, inconsistent set of aQGCs. More generally, we can consider a direction close to $\vec{d}_0$: $\vec{d}=\vec{d}_0+(\text{small deviations})$, which will still lead us toward the positivity cone.

For a given feasible direction $\vec{d}$, it intersects with the positivity cone at two points, corresponding to the minimal distance required to reach the cone and the maximal distance for which the trajectory remains inside the cone along this direction. These two distances can also be computed with linear programming, by minimizing or maximizing $\mu$ with respect to semi-positive $w_i$, subject to $\sum_{i} w_{i} \vec{\tt e}_{i}=\vec{F}+\mu \vec{d}$. This functionality is implemented by the {\tt FeasibleBoundary} command of our package. For a point outside the positivity cone, {\tt FeasibleBoundary} returns two positive values, corresponding to the minimal and maximal distances.

Another scenario that we can use {\tt FeasibleBoundary} is as follows. For an arbitrary point $\vec{F}$ inside the positivity cone and a given direction $\vec{d}$, the command determines the boundary reached along this direction. In this case, {\tt FeasibleBoundary} again outputs two values, but one of them is zero and the other gives the distance from $\vec{F}$ to this boundary along the $\vec{d}$ direction.

\subsubsection{Eliminate redundant potential ERs}

After discretizing the continuous ERs $\vec{\tt e}_{i}$, the genuine ERs of the positivity cone can be identified from the potential ERs listed in \eref{eq:potential ERs}. This can be done by numerically testing whether each potential ER lies within the cone generated by the remaining potential ERs. More precisely, suppose there are $m$ numerical potential ERs indexed by the set $\mathcal{I}=\{1,2,\dots,m\}$. For a given potential ER $\vec{\tt e}_{k}$, we define the reduced index set $\mathcal{I}'=\{1,\dots,k-1,k+1,\dots,m\}$ and then test whether $\vec{\tt e}_{k}$ belongs to the cone 
\begin{equation}
    \mathcal{C}_{m-1}=\text{cone}(\{\vec{\tt e}_{i},i\in \mathcal{I}'\}).
\end{equation}
This procedure amounts to solving the following linear programming problem, similar to \eref{eq:inoutcone}:
\begin{equation}
\begin{array}{ll}
\min_{w_i} & 0 \\
\text {such that} & \sum_{i\in\mathcal{I}'} w_{i} \vec{\tt e}_{i}=\vec{\tt e}_{k}, ~~w_{i} \geq 0.
\end{array}    
\end{equation}
If this optimization problem is feasible, the vector $\vec{\tt e}_{k}$ is not a genuine ER and can be dropped without affecting the positivity cone. Checking this for potential ERs, we find that for isolated ERs, $\vec{e}_6$ and $\vec{e}_{9}$ are not real ERs, while among the continuous families, $\vec{e}_{13}(x)$ is non-genuine for $0.2786\leq|x|\leq 3.7207$. All other continuous ERs are confirmed to be genuine ERs, at least, within the evenly-discretized interval of $x,y\in[-63.65,63.65]$. It may sound surprising that only some portions of the continuous ERs appears to be non-genuine, but one should keep in mind that all these continuous ERs lie only within two- or three-dimensional subspaces of the high-dimensional space in which the positivity cone resides.

\begin{figure}
    \centering
    \includegraphics[width=0.6\linewidth]{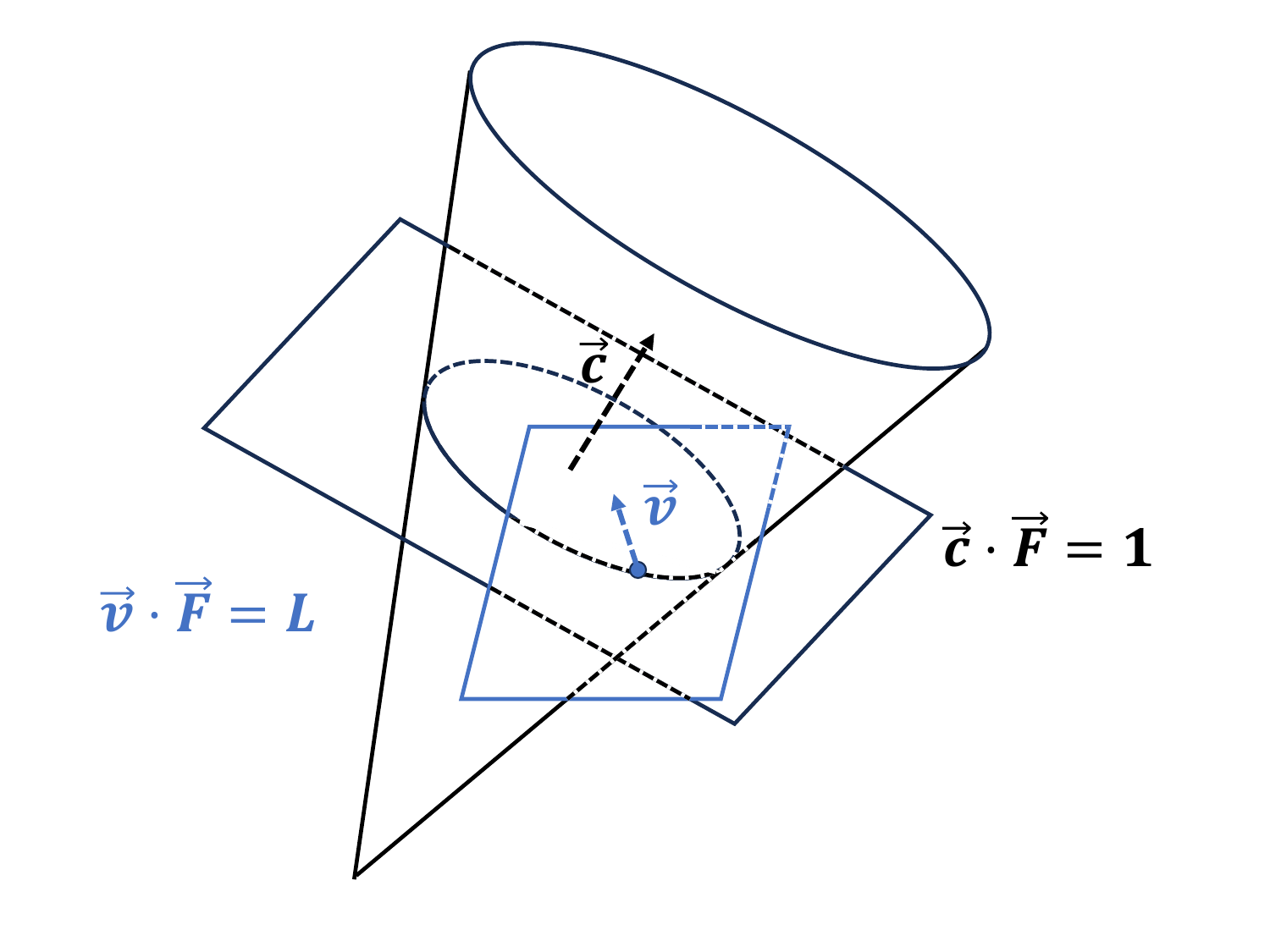}
    \caption{Optimal bound on a hyperplane section. The hyperplane section is defined by $\vec{c}\cdot\vec{F}=1$ ($\vec{c}$ being the normal vector to the hyperplane), and the objective function is $L=\vec{v}\cdot\vec{F}$ ($\vec{v}$ being the normal vector to the blue hyperplane). Varying $\vec{F}$ within the section is equivalent to parallel-translating the hyperplane $\vec{v}\cdot\vec{F}=L$ along the $\vec{v}$ direction. The optimal value of $L$ is attained on the boundary of the cone on the hyperplane section.}
    \label{fig:conecross-sec}
\end{figure}

\subsubsection{Optimal bounds on linear aQGC combinations}

A low-dimensional convex cone can be charted by probing its boundary from an interior point using linear programming. However, this intuitive approach becomes intractable for high-dimensional cones, such as in our case. In general, the full positivity bounds, which provide an inequality representation of the positivity cone, can be obtained numerically from the ER representation. For complicated cones like ours, however, computing this inequality representation is numerically challenging, and the resulting bounds are often too numerous to be practically useful.

Nevertheless, for a given hyperplane section of the positivity cone, specified by $\vec{c}\cdot \vec{F}=1$,  the optimal bound for any linear combination of the aQGC coefficients $L(\vec{F})=\vec{v}\cdot\vec{F}=\sum_i v_iF_i$ can be easily obtained by solving the following linear program ($\vec{c}$ and $\vec{v}$ being constant):
\begin{equation}
    \begin{array}{ll}
    \min/\max_{w_i,\vec{F}} & L(\vec{F})=\vec{v}\cdot \vec{F} \\
    \text {such that} &\vec{c}\cdot \vec{F}=1,\\
                      & \sum_{i} w_{i} \vec{\tt e}_{i}=\vec{F},~~w_{i} \geq 0 ,
    \end{array}\label{eq:optimalLP}
\end{equation}
where $\vec{F}$ is the vector of the aQGC coefficients, taking values on the hyperplane section $\vec{c}\cdot \vec{F}=1$. Here,  $\vec{c}$ is a normal vector to the hyperplane section and must lie inside the dual cone, and $\vec{v}$ specifies the linear combination of Wilson coefficients and is the normal vector to the blue hypersurface schematically depicted in Figure~\ref{fig:conecross-sec}. Pictorially, as shown in Figure~\ref{fig:conecross-sec}, the linear program \eqref{eq:optimalLP} aims to find the blue hyperplane that intersects the boundary of the positivity cone on the black hyperplane.

Note that an optimal bound obtained via \eref{eq:optimalLP} can be interpreted as a linear homogeneous positivity bound. To see this, suppose that the linear program gives the following two-sided bound: 
\be
L_{\min}\leq \vec{v}\cdot \vec{F}\leq L_{\max} ,
\label{eq:twosidedBounds}
\ee 
where $L_{\min}$ and $L_{\max}$ denote the minimum and maximum values of $L(\vec{F})$ respectively.
Remember that the $\vec{F}$ for this bound is subject to $\vec{c}\cdot\vec{F}=1$. So, multiplying this bound by $\kappa \vec{c}\cdot\vec{F}$ ($\kappa$ being positive), we get
\begin{align}
 L_{\min}\vec{c}\cdot \vec{F'} \leq \vec{v}\cdot \vec{F'}\leq L_{\max}\vec{c}\cdot \vec{F'}\label{eq:HomoOptB}
\end{align}
where we have defined $\vec{F'}=\kappa \vec{F}$. Since $\kappa$ is an arbitrary positive number, $\vec{F'}$ is a generic vector of aQGC coefficients, not subject to the constraint $\vec{c}\cdot\vec{F'}=1$, and \eref{eq:HomoOptB} is an optimal linear positivity bound. 

The linear program \eqref{eq:optimalLP} is implemented in the {\tt SMEFTaQGC} package as {\tt OptimalBound}. It takes $\vec{c}$ and $\vec{v}$ as inputs, and outputs $L_{\min}$ and $L_{\max}$, which allow us to get the bound \eref{eq:HomoOptB}. A notable observation is that, by choosing appropriate $\vec{v}$ and $\vec{c}$, \eref{eq:HomoOptB} can lead to a linear bound involving a mixture of different types of aQGCs. 

Let us present an example to illustrate this. In general, $\vec{c}$ is a given vector within the dual cone: $\vec{c}\cdot \vec{\tt e}_{i}\geq0$ for all $i$. Here, however, we adopt an alternative approach to construct an {\it ad hoc} vector $\vec{c}$ to maximize its minimal projection onto the numerical ERs, ensuring that it lies deep inside the dual cone. This leads to the following optimization problem:
\begin{equation}
    \begin{array}{ll}
    \max_{\vec{c},t} & t \\
    \text {such that} &\vec{c}\cdot \vec{F}_0=1,\\
                      & \vec{c}\cdot \vec{\tt e}_{i}\geq t,~~\forall i\\
                      &t\geq 0.
    \end{array}    \label{eq:LPc}
\end{equation}
where $\vec{F}_0$ is a reference point in the interior of the positivity cone. 
While different choices of the reference vector $\vec{F}_0$ lead to different vectors $\vec{c}$, all such choices are equally suitable for our purposes. 
For example, we can choose the reference point to be on the ``most interior ray'' $\vec{F}_0$ constructed in \Cref{sec:positivityCheck}. 
Solving linear program \eqref{eq:LPc}, we find that the vector $\vec{c}$ is
\begin{align}
   &\vec{c} \simeq  (0, -0.971, 0, -0.486, -0.008, -0.005, 0.121, 0.005, -0.002, 7.769, 2.590, 5.179, 10.417,\nn
   & 24.190, 16.130, 18.991, 2.138, -0.387, 1.295, 0.874, 2.590, 2.590, 16.623, 25.401, 0.200),
    \label{eq:adhoc_c}
\end{align}
which encompasses all S-, M-, and T-type aQGCs.
Now, let us choose the vector $\vec{v}$ as $(0,0,0,0,0,0,0,0,0,1,1,1,0,0,0,0,0,0,
0,0,0,0,0,0,0)$, corresponding to choosing the linear objective function in \ref{eq:optimalLP} as $L(\vec{F})=F_{S,0}+F_{S,1}+F_{S,2}$. Solving \ref{eq:optimalLP}
then gives an optimal bound is $0\leq F_{S,0}+F_{S,1}+F_{S,2}\leq 0.386 (\vec{c}\cdot \vec{F})$, where $\vec{c}$ is given by \eref{eq:adhoc_c}. As expected, different types of aQGCs contribute to this upper bound, resulting in a tighter constraint.

In \Cref{sec:linearPosiBound}, we have obtained a series of linear analytical bounds. With the current numerical setup, we can verify whether these analytical bounds are optimal, {\it i.e.}, whether they are ERs of the dual cone. Indeed, we find many of those linear analytical bounds are optimal.

It is worth noting that not every choice of $\vec{c}$ leads to a two-sided bound of the form \eqref{eq:twosidedBounds}, since, depending on its orientation, the intersection between the hyperplane $\vec{c}\cdot \vec{F}=1$ and the positivity cone is not guaranteed to be a bounded subspace. That means that some combinations of aQGCs are unbounded by the positivity cone.

\begin{figure}
    \centering
    \includegraphics[width=0.6\linewidth]{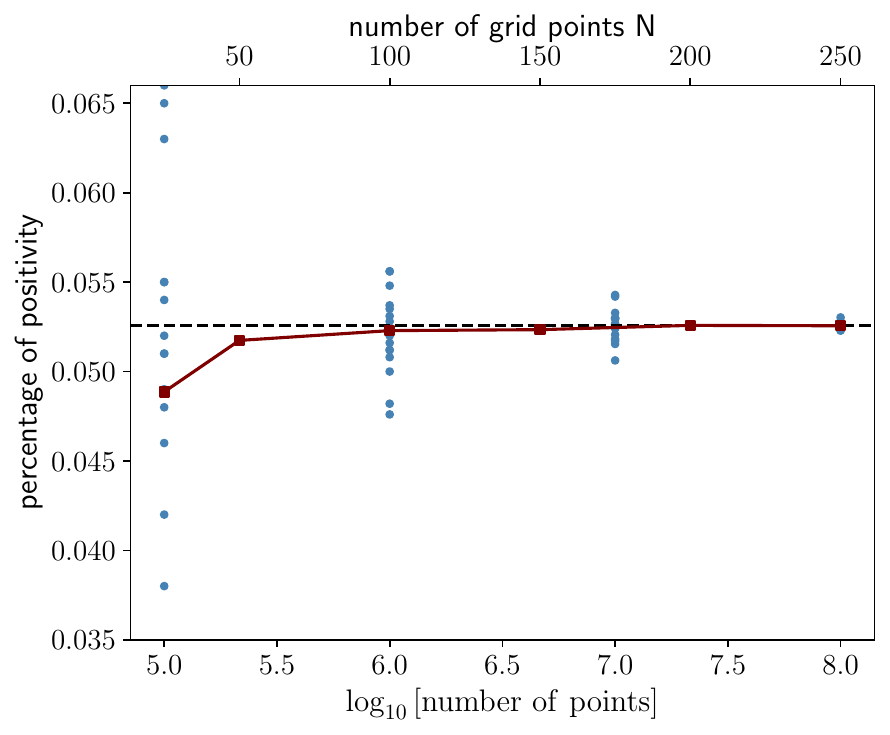}
    \caption{Percentage of positivity and convergence of the discretization scheme. The vertical axis shows the fraction of the positivity cone, measured by its solid angle in the higher-dimensional aQGC sphere. Each blue point is a sampling with the number of sampling points denoted on the lower horizontal axis ($10^5$, $10^6$, $10^7$ or $10^8$ points), after discretizing the continuous parameters with $N=200$ according to \eref{eq:disScheme}. The dashed line $(0.05255\%)$ denotes the averaged percentage of the positivity region for $15$ samplings with $10^8$ points. Each maroon point denotes the average of $15$ samplings, each with $10^7$ points, evaluated with different discretizations $N=25,50,100,150,200,250$.}
    \label{fig:solidangle}
\end{figure}

\subsubsection{Percentage of positivity}

One measure of the strength of the positivity bounds is the solid-angle percentage of the positivity cone on the high-dimensional sphere of aQGC coefficients. This percentage of positivity can be computed by uniformly sampling on the 22-dimensional aQGC sphere. Uniform random sampling on a high-dimensional sphere can be achieved using the normalized Gaussian method~\cite{muller1959note,marsaglia1972choosing}, in which one first draws each component of a vector independently from a standard normal distribution and then rescales the vector to unit length. Using the linear program \eref{eq:inoutcone} and $\mathcal{N}=10^7$ sampling points, we count the number $\mathcal{N}_{\tt C}$ of points that lie within the positivity cone. The solid angle $\Omega_{\rm positivity}$ is then defined as the fraction of points that fall inside the cone, 
\begin{equation}
    \Omega_{\rm positivity}=\frac{\mathcal{N}_{{\tt C}} }{\mathcal{N}}.
\end{equation}
Doing this for the discretized positivity cone ${\tt C}$, we find that the percentage of positivity is merely 
\begin{equation}
    \Omega_{\rm positivity}=(0.05255\pm0.000186)\% ,
\end{equation} 
indicated by the black dashed line in Figure~\ref{fig:solidangle}. Figure~\ref{fig:solidangle} also shows the convergence of positivity percentages for different numbers of sampling points $\mathcal{N}$. The Monte Carlo sampling error is estimated by the square root of the variance $\Delta=\sqrt{{(1-\Omega_{\rm positivity})\Omega_{\rm positivity}}/{15\mathcal{N}}}\simeq0.000186\%$, which comes from viewing $\Omega_{\rm positivity}$ as a probability in a Gaussian random sampling. In Figure~\ref{fig:solidangle}, the red line shows the convergence of the results with respect to choosing different grid points $N$ (cf. \eref{eq:disScheme}) to discretize the continuous ERs.

\section{Conclusion}\label{sec:conclusion}

Among the processes measured at the LHC, vector boson scattering provides a powerful avenue for exploring potential effects of physics beyond the Standard Model. In particular, probing deviations in QGCs is a central objective of the LHC electroweak program. In this work, we have derived a set of theoretical bounds on the 22 dimension-8 anomalous QGC couplings within the SMEFT framework, thereby constraining possible departures from the Standard Model. These so-called positivity bounds follow from fundamental principles of quantum field theory or the S-matrix, such as Lorentz invariance, unitarity, and analyticity, and are agnostic about the UV completion. Consequently, these bounds are robust and highly model-independent. In theories with multiple degrees of freedom and rich internal symmetries, optimal ($s^2$ order) positivity bounds, available from the axiomatic principles of QFT, can be extracted using the convex-geometry approach by identifying the extremal rays.

Generalizing previous work on extremal positivity bounds in VBS~\cite{Yamashita:2020gtt}, we incorporate the longitudinal modes of vector bosons and the Higgs mode, and consider the complete set of $22$ aQGC operators. Given the substantial number of low-energy modes involved, a more systematic approach is developed to deal with the extremal positivity problem. After deriving the CG coefficients by direct construction, we classify the intermediate/UV states according to their CP properties and the spin–parity. Upon identifying the CP eigenstates, parity-violating and parity-conserving operators can be treated uniformly within the CP-conserving framework. Furthermore, these results are independently confirmed using a novel method based on evaluating the eigenvalues of the Casimir operators.

For the convex cone generated by the complete set of aQGCs consistent with the S-matrix axioms, we have identified all isolated and continuous ERs, which are then used to derive various positivity bounds on these couplings. More concretely, after obtaining the ERs of the amplitude cone, we have derived linear analytical bounds on the full set of aQGC coefficients,  including constraints on parity-violating operators $O_{M,8}$ and $O_{M,9}$ and on mixed types of coefficients.

On the numerical side, after discretizing the continuous ERs, we construct efficient algorithms to determine whether a given set of aQGC coefficients lies within the positivity cone and to extract optimal positivity bounds along specified directions. When a proposed coefficient vector falls outside the positivity cone, feasible directions are also provided to modifying the proposed vector. These procedures are implemented using linear programming within the convex-geometry framework and are made available through a Python package {\tt SMEFTaQGC}, which accompanies this work.

Positivity bounds dramatically reduce the allowed region of parameter space in the SMEFT, and their constraining power increases as the number of modes in the problem grows. The naive SMEFT parameter space may be viewed as a high-dimensional sphere, within which the positivity cone forms a convex subregion and the SM is at the vertex of the positivity cone. For the full aQGC coefficient space, we find that the positivity cone occupies only a very small solid angle,
\be
\Omega_{\rm positivity}=(0.05255\pm0.000186)\% ,
\ee
of the total naive SMEFT space.

In this paper, we have computed the positivity bounds using the primal amplitude cone, in which the ERs of the amplitude cone are analytically enumerated and then converted into bounds on the primal cone via vertex enumeration. As proposed in \cite{Li:2021lpe}, an alternative approach is to derive the ERs of the dual cone to the amplitude cone using semi-definite programming, which is by design more numerical in nature. The ERs of the dual cone correspond directly to the positivity bounds on the amplitude cone. We leave a systematic implementation of this approach for future work~\cite{Chen:2026xx}.

\acknowledgments

We would like to thank Joe Davighi, Dong-Yu Hong, Ken Mimasu, Shi-Lin Wan and Zhuo-Hui Wang for helpful discussions. SYZ acknowledges support from the National Natural Science Foundation of China under grant No.~12475074 and No.~12247103.
~\\

\appendix

\section{Basic concepts of convex cones}\label{appd:convgeo}

In this appendix, we briefly review several basic concepts about convex cones that are relevant to understand our analysis of the amplitude cone and its dual.

\begin{itemize}
    \item {\it Conical hull}: Given a set $S$ in a linear space, the conical hull of $S$ is the set of all non-negative linear combinations of elements of $S$,
    \begin{equation}
    \mathrm{cone}(S)\equiv \left\{ {\textstyle \sum_{i}} \lambda_i s_i~\big|~ s_i\in S,\; \lambda_i\ge 0 \right\}.
    \end{equation}
    \item {\it Convex cone}: 
    A convex cone $C$ is a set closed under addition and multiplication by non-negative scalars:
    $x,y\in C,\;\lambda\ge 0~\Rightarrow~ x+y\in C,\;\lambda x\in C$.
    Every convex cone can be written as the conical hull of its generating set. A cone is called {\it salient} (or pointed) if it contains no nontrivial straight line, \ie, if $x\in C$ and $-x\in C$ imply $x=0$.
    \item {\it Dual cones}:
    Given a convex cone $C$, its \emph{dual cone} is defined as
    \begin{equation}
     C^* \equiv \{\, y \mid y\cdot x \ge 0,\ \forall\, x\in C \,\}.\label{eq:dualcone}
    \end{equation}
    The dual cone is always convex and salient.

\end{itemize}

Having introduced the basic notions of convex cones and their duals, we now turn to several structural features that play a central role in our analysis, such as faces, facets, and extremal rays.

\begin{itemize}

\item
\emph{Face}: A face $F$ of a convex cone is a flat piece of its boundary. More precisely, a face is a convex subset of the boundary formed by intersecting the cone with a supporting hyperplane. A supporting hyperplane of a convex cone is a (codimension-one) hyperplane that touches the cone such that the entire cone lies on one side of it.

\item 
\emph{Facet}: A face of codimension one is called a facet. Geometrically, it is a flat boundary ``wall'' obtained where a supporting hyperplane touches the cone, corresponding to a single linear inequality.

\item 
\emph{Extremal rays}: An extremal ray is a one-dimensional face of a cone. Geometrically, it is a boundary direction that cannot be written as a sum of two other independent directions in the cone. Extremal rays play a central role because a salient cone is fully determined by its extremal rays: Every element of the cone can be written as a non-negative combination of them.

\end{itemize}

Faces describe the nested boundary structure of a cone. For example, for a (high-dimensional) polyhedral cone, the faces are the facets (codimension-one flat faces), the lower-dimensional faces obtained as intersections of several facets, the (one-dimensional) extremal rays, and the apex. 
While facets correspond to the cone's bounding inequalities, extremal rays give the minimal generating directions of the cone. Thus, a convex cone admits two complementary descriptions: the inequality representation (or H(alfspace)-representation) and the extremal ray representation (or V(ertex)-representation).

Finally, all positive semi-definite matrices 
\be
\mathrm{PSD}_n = \{ M\succeq 0\}
\ee
form a salient convex cone. A standard result states that its extremal rays consist precisely of the rank-$1$ positive semi-definite matrices: 
\be
M = u u^\dagger,~ u\neq 0 .
\ee
This is because any $M\in\mathrm{PSD}_n$ admits a  positive semi-definite decomposition into such rank-$1$ generators $M = \sum_{i} \lambda_i\, u_i u_i^\dagger,~ \lambda_i\ge 0$.

\section{CG coefficients $C^{\bfm,\ai}_{i,j}$}\label{appd:CG}

{
\setlength{\tabcolsep}{2pt}
\begin{longtable}{|c|c|c|c|c||c||c||c|c|}
    \caption{CG-coefficients for constructing ERs of the aQGC positivity cone}\label{tab:cgcoe} \\
    \hline
   \diagbox{$|\mathbf{m},\alpha\rangle$}{$|i,j\rangle$} & $H^aH^b$ & $H^\dagger_aH^b$ & $H^aH^\dagger_b$ & $H^\dagger_a H^\dagger_b$ &  $B_{\mathbf{i}}B_{\mathbf{j}}$ &$W_{\mathbf{i}}^IW_{\mathbf{j}}^J$ & $B_{\mathbf{i}}W_{\mathbf{j}}^J$ & $W^I_{\mathbf{i}}B_{\mathbf{j}}$\\
    \hline
    $X$ in $(\mathbf{1}_0,\mathbf{1}_0)^+_0$ & & $\delta^b{}_a$ & $\delta^a{}_b$ & & $\delta_{\mathbf{ij}}$ & $\delta_{\mathbf{ij}}\delta^{IJ}$ & & \\
    \hline
    $X$ in $(\mathbf{1}_0,\mathbf{1}_0)^+_1$ & & $\delta^b{}_a$ & $-\delta^a{}_b$ &  & & & &\\
    \hline
        $X$ in $(\mathbf{1}_0,\mathbf{1}_0)^-_0$ & &  &  &  & $\epsilon_{\mathbf{ij}}$&$\epsilon_{\mathbf{ij}}\delta^{IJ}$ & &\\
    \hline
    $X^\pm$ in $(\mathbf{1}_0,\mathbf{1}_1)^\pm_1$ & $\epsilon^{ab}$ & & & $\pm \epsilon_{ab}$  & & & &\\
    \hline
    $X^K$ in $(\mathbf{1}_0,\mathbf{3}_0)^+_0$ & &  & &  & & &$\epsilon_{\mathbf{ij}}\delta^{JK}$ & $\epsilon_{\mathbf{ij}}\delta^{IK}$\\
    \hline
    $X^K$ in $(\mathbf{1}_0,\mathbf{3}_0)^+_1$ & &  &  &  & &$\epsilon_{\mathbf{ij}}\epsilon^{IJK}$ &$\delta_{\mathbf{ij}}\delta^{JK}$ &$-\delta_{\mathbf{ij}}\delta^{IK}$\\
    \hline
    $X^K$ in $(\mathbf{1}_0,\mathbf{3}_0)^-_0$ & & $(\tau^I)^b{}_a$ & $(\tau^I)^a{}_b$ &  & & & $\delta_{\mathbf{ij}}\delta^{JK}$ &$\delta_{\mathbf{ij}}\delta^{IK}$\\
    \hline
    $X^K$ in $(\mathbf{1}_0,\mathbf{3}_0)^-_1$ & & $(\tau^I)^b{}_a$ & $-(\tau^I)^a{}_b$ &  & &$\delta_{\mathbf{ij}}\epsilon^{IJK}$ & $\epsilon_{\mathbf{ij}}\delta^{JK}$ &$-\epsilon_{\mathbf{ij}}\delta^{IK}$\\
    \hline
    $X^{\pm,I}$ in $(\mathbf{1}_0,\mathbf{3}_1)^\pm_0$ & $(\tau^I\epsilon)^{ab}$ & & & $\mp (\epsilon\tau^I)_{ab}$  & & & &\\
    \hline
    $X^A$ in $(\mathbf{1}_0,\mathbf{5}_0)^+_0$ &  & & &  & &$\delta_{\mathbf{ij}}C^{A;IJ}_{\mathbf{5}}$ & &\\
    \hline
    $X^A$ in $(\mathbf{1}_0,\mathbf{5}_0)^-_0$ &  & & &  & &$\epsilon_{\mathbf{ij}}C^{A;IJ}_{\mathbf{5}}$ & &\\
    \hline
    $X_{\mathbf{k}}$ in $(\mathbf{2}_2,\mathbf{1}_0)^+_0$ & & & &  & $C^{\mathbf{2}}_{\mathbf{k;ij}}$& $C^{\mathbf{2}}_{\mathbf{k;ij}}\delta^{IJ}$& &\\
    \hline
    $X^K_{\mathbf{k}}$ in $(\mathbf{2}_2,\mathbf{3}_0)^-_0$ &  & & &  & & & $C^{\mathbf{2}}_{\mathbf{k;ij}}\delta^{JK}$ &$C^{\mathbf{2}}_{\mathbf{k;ij}}\delta^{IK}$\\
    \hline
   $X^K_{\mathbf{k}}$ in $(\mathbf{2}_2,\mathbf{3}_0)^+_1$ &  & & &  & & & $C^{\mathbf{2}}_{\mathbf{k;ij}}\delta^{JK}$ &$-C^{\mathbf{2}}_{\mathbf{k;ij}}\delta^{IK}$\\
    \hline
   $X^K_{\mathbf{k}}$ in $(\mathbf{2}_2,\mathbf{3}_0)^-_1$ &  & & &  & &$C^{\mathbf{2}}_{\mathbf{k;ij}}\epsilon^{IJK}$ & &\\
    \hline
   $X^A_{\mathbf{k}}$ in $(\mathbf{2}_2,\mathbf{5}_0)^+_0$ &  & & &  & & $C^{\mathbf{2}}_{\mathbf{k;ij}}C_{\mathbf{5}}^{K;IJ}$& &\\
    \hline
   
\end{longtable}
\begin{longtable}{|c|c|c||c|c|}
\hline
    \diagbox{$|\mathbf{m},\alpha\rangle$}{$|i,j\rangle$} & $W^I_{\mathbf{i}}H^b$ & $W^I_{\mathbf{i}}H^\dagger_b$ & $B_\mathbf{i}H^b$ & $B_\mathbf{i}H^\dagger_b$ \\
    \hline
    $X^{\pm i,c}_{\mathbf{k}}$ in $(\mathbf{2}_{1},\mathbf{2}_{1/2})^{\pm i}$ & $(\tau^I)^b{}_c\delta_{\mathbf{ik}}$ & $\pm i(\epsilon\tau^I)_{bc}\delta_{\mathbf{ik}}$ & $\delta^b{}_c\delta_{\mathbf{ik}}$ & $\pm i\epsilon_{bc}\delta_{\mathbf{ik}}$ \\
    \hline
    $X^{\pm i,\alpha}_{\mathbf{k}}$ in $(\mathbf{2}_{1},\mathbf{4}_{1/2})^{\pm i}$ & 
    $C^{bI}_{\mathbf{4},\alpha}\delta_{\mathbf{ik}}$ & $\pm i(\epsilon_{bc} C^{cI}_{\mathbf{4},\alpha})\delta_{\mathbf{ik}}$ & &
    \\
    \hline
\end{longtable}

}

For an easy reference, in Table~\ref{tab:cgcoe}, we explicitly list all the CG-coefficients $C^{\bfm,\ai}_{i,j}= \langle \mathbf{m},\alpha|i,j\rangle$ required to construct the ERs of the aQGC positivity cone, which are mostly written in a compact tensor notation. In the table, the columns and rows enumerate product states $|i,j\rangle$ and irrep states $|\mathbf{m},\alpha\rangle$ respectively,
and zeros are omitted to avoid clutter. $C^{\mathbf{2}}_{\mathbf{k;ij}}$ denotes the CG-coefficients of the $\mathbf{2}$ irrep in $SO(2)$ decomposition $\mathbf{2}\otimes\mathbf{2} = \mathbf{1}\oplus\mathbf{1}\oplus\mathbf{2}$, while $C^{A;IJ}_{\mathbf{5}}$  and $C^{aI}_{\mathbf{4},\alpha}$ denotes the CG-coefficients of the $\mathbf{5}$ and $\mathbf{4}$ irrep in $SU(2)$ decomposition $\mathbf{3}\otimes\mathbf{3} = \mathbf{1} \oplus \mathbf{3} \oplus \mathbf{5}$ and $\mathbf{2}\otimes\mathbf{3} = \mathbf{2}\oplus\mathbf{4}$. $\ti^I$ are Pauli matrices, and the various $\delta$'s and $\epsilon$'s with appropriate indices are Kronecker deltas and 2D Levi-Civita symbols. See the beginning of Section \ref{sec:aQGCER} for the meaning of the remaining notation.

\section{Positivity bounds including dim-6 aQGCs}\label{appd:dim6part}

For reasons outlined below \eref{eq:smeftLagGen}, we have neglected the dim-6 contributions in the main text when expressing the positivity bounds. However, should the dim-6 aQGC contributions be sizable in the electroweak scattering amplitudes, it is actually rather easy to take these contributions into account, as we will do in this appendix.

Note that the dim-$6$ aQGC operators \cite{Degrande:2012wf} are given by
\begin{equation}
\begin{aligned}
O_{WWW} &  =4\operatorname{Tr}[\hat{W}_{\mu \nu}\hat{W}^{\nu\rho}\hat{W}_{\rho}^{\mu}],\\
O_{HW} &  =\left(  D^{\mu}H\right)  ^{\dagger}\hat{W}_{\mu \nu}\left(D^{\nu}H\right),\\
O_{HB} &  =2\left(  D^{\mu}H\right)  ^{\dagger}\hat{B}_{\mu \nu}\left(  D^{\nu
}H\right)  ,
\end{aligned}\label{eq:dim6QGC}
\end{equation}
where $\hat{W}^{\mu\nu}$ and $\hat{B}^{\mu\nu}$ are defined in \eref{eq:WBfield}.
The dim-6 contribution to the Lagrangian is given by $ \mathcal{L}^{(6)}=(f_{WWW}O_{WWW}+f_{HW}O_{HW}+f_{HB}O_{HB})/\Lambda^2$.
The corresponding rescaled coefficients of these operators will be written as $c_{WWW}\equiv g^3f_{WWW},c_{HW}\equiv gf_{HW}$ and $c_{HB}\equiv g'f_{HB}$, respectively.
Note that the group-theoretical construction of the ERs of the amplitude cone and the subsequent derivation of the dual cone remain unchanged when the dim-6 operators are included. The only change is that the dim-6 operators will shift the dim-8-only amplitude vector $\vec{F}$ considered in the main text by quadratic dim-6 contributions, which can be expanded in the basis introduced in \Cref{sec:ERs} as follows: 
\begin{align}
&(F_{M0},F_{M1}+6c_{HW}c_{WWW},...,F_{M7}+48c_{HW}c_{WWW},...,\nn
    &F_{S0}-\frac{c_{HW}^2+4c_{HB}^2}{8},F_{S1}+\frac{c_{HW}^2}{4}, F_{S2}-\frac{c_{HW}^2-4c_{HB}^2}{8},F_{T,\mathcal{I}_3},-9c_{WWW}^2,0,0).
\end{align}
where $\mathcal{I}_3$ runs from $0$ to $9$.
For example, for the linear bounds listed in \Cref{sec:linearPosiBound}, we now instead have
\begin{align}
    &F_{S0}\geq\frac{c_{HW}^2+4c_{HB}^2}{8},\nn
    &F_{S0}+F_{S2}\geq \frac{c_{HW}^2}{4},\nn
    &F_{S0}+F_{S1}+F_{S2}\geq 0,\nn
    &4F_{T1}+F_{T2}+2F_{T3}\geq 36c_{WWW}^2,\nn
    &4F_{T0}+4F_{T1}+3F_{T2}+5F_{T3}\geq72c^2_{WWW},\nn
    &F_{T2}+2F_{T3}\geq36c_{WWW}^2,\nn
    &F_{M7}-8F_{M1}+2F_{M8}+32F_{S0}+4F_{T7}-4F_{M3}\geq16c_{HB}^2+4c_{HW}^2,\nn
    &F_{M7}-8F_{M1}+2F_{M8}+16(F_{S0}+F_{S2})+4F_{T7}-4F_{M3}\geq4c_{HW}^2
\end{align}
We see that the structure of these bounds are of the form $\sum (\text{dim-8})> \sum(\text{dim-6})^2$, which are stronger bounds compared to the dim-8-only bounds, as $\sum (\text{dim-8})> \sum(\text{dim-6})^2$ mathematically implies $\sum (\text{dim-8})>0$. In the {\tt SMEFTaQGC} package, to include dim-6 effects, we can turn on the Boolean flag {\tt dim6} by setting it to be \texttt{True}.

\bibliographystyle{JHEP}
\bibliography{biblio.bib}

@article{Durieux:2024zrg,
    author = "Durieux, Gauthier and Remmen, Grant N. and Rodd, Nicholas L. and {\'E}boli, O. J. P. and Gonzalez-Garcia, Maria C. and Kondo, Dan and Murayama, Hitoshi and Okabe, Risshin",
    title = "{LHC EFT WG note: Basis for anomalous quartic gauge couplings}",
    eprint = "2411.02483",
    archivePrefix = "arXiv",
    primaryClass = "hep-ph",
    reportNumber = "CERN-LHCEFTWG-2024-002, CERN-LPCC-2024-002",
    doi = "10.21468/SciPostPhysCommRep.6",
    month = "11",
    year = "2024"
}

@article{Yamashita:2020gtt,
    author = "Yamashita, Kimiko and Zhang, Cen and Zhou, Shuang-Yong",
    title = "{Elastic positivity vs extremal positivity bounds in SMEFT: a case study in transversal electroweak gauge-boson scatterings}",
    eprint = "2009.04490",
    archivePrefix = "arXiv",
    primaryClass = "hep-ph",
    reportNumber = "USTC-ICTS/PCFT-20-29",
    doi = "10.1007/JHEP01(2021)095",
    journal = "JHEP",
    volume = "01",
    pages = "095",
    year = "2021"
}

@article{Kondo:2022wcw,
    author = "Kondo, Dan and Murayama, Hitoshi and Okabe, Risshin",
    title = "{23, 381, 6242, 103268, 1743183, {\textellipsis} : Hilbert series for CP-violating operators in SMEFT}",
    eprint = "2212.02413",
    archivePrefix = "arXiv",
    primaryClass = "hep-ph",
    doi = "10.1007/JHEP03(2023)107",
    journal = "JHEP",
    volume = "03",
    pages = "107",
    year = "2023"
}

@article{Zhang:2020jyn,
    author = "Zhang, Cen and Zhou, Shuang-Yong",
    title = "{Convex Geometry Perspective on the (Standard Model) Effective Field Theory Space}",
    eprint = "2005.03047",
    archivePrefix = "arXiv",
    primaryClass = "hep-ph",
    reportNumber = "USTC-ICTS/PCFT-20-14",
    doi = "10.1103/PhysRevLett.125.201601",
    journal = "Phys. Rev. Lett.",
    volume = "125",
    number = "20",
    pages = "201601",
    year = "2020"
}

@article{Remmen:2019cyz,
    author = "Remmen, Grant N. and Rodd, Nicholas L.",
    title = "{Consistency of the Standard Model Effective Field Theory}",
    eprint = "1908.09845",
    archivePrefix = "arXiv",
    primaryClass = "hep-ph",
    doi = "10.1007/JHEP12(2019)032",
    journal = "JHEP",
    volume = "12",
    pages = "032",
    year = "2019"
}

@article{Murphy:2020rsh,
    author = "Murphy, Christopher W.",
    title = "{Dimension-8 operators in the Standard Model Effective Field Theory}",
    eprint = "2005.00059",
    archivePrefix = "arXiv",
    primaryClass = "hep-ph",
    doi = "10.1007/JHEP10(2020)174",
    journal = "JHEP",
    volume = "10",
    pages = "174",
    year = "2020"
}

@article{Li:2020gnx,
    author = "Li, Hao-Lin and Ren, Zhe and Shu, Jing and Xiao, Ming-Lei and Yu, Jiang-Hao and Zheng, Yu-Hui",
    title = "{Complete set of dimension-eight operators in the standard model effective field theory}",
    eprint = "2005.00008",
    archivePrefix = "arXiv",
    primaryClass = "hep-ph",
    doi = "10.1103/PhysRevD.104.015026",
    journal = "Phys. Rev. D",
    volume = "104",
    number = "1",
    pages = "015026",
    year = "2021"
}

@article{Almeida:2020ylr,
    author = "Almeida, Eduardo da Silva and {\'E}boli, O. J. P. and Gonzalez{\textendash}Garcia, M. C.",
    title = "{Unitarity constraints on anomalous quartic couplings}",
    eprint = "2004.05174",
    archivePrefix = "arXiv",
    primaryClass = "hep-ph",
    reportNumber = "YITP-SB-2020-8",
    doi = "10.1103/PhysRevD.101.113003",
    journal = "Phys. Rev. D",
    volume = "101",
    number = "11",
    pages = "113003",
    year = "2020"
}

@article{Trott:2020ebl,
    author = "Trott, Timothy",
    title = "{Causality, unitarity and symmetry in effective field theory}",
    eprint = "2011.10058",
    archivePrefix = "arXiv",
    primaryClass = "hep-ph",
    doi = "10.1007/JHEP07(2021)143",
    journal = "JHEP",
    volume = "07",
    pages = "143",
    year = "2021"
}

@article{Bellazzini:2020cot,
    author = "Bellazzini, Brando and Elias Mir{\'o}, Joan and Rattazzi, Riccardo and Riembau, Marc and Riva, Francesco",
    title = "{Positive moments for scattering amplitudes}",
    eprint = "2011.00037",
    archivePrefix = "arXiv",
    primaryClass = "hep-th",
    doi = "10.1103/PhysRevD.104.036006",
    journal = "Phys. Rev. D",
    volume = "104",
    number = "3",
    pages = "036006",
    year = "2021"
}

@article{Arkani-Hamed:2017jhn,
    author = "Arkani-Hamed, Nima and Huang, Tzu-Chen and Huang, Yu-tin",
    title = "{Scattering amplitudes for all masses and spins}",
    eprint = "1709.04891",
    archivePrefix = "arXiv",
    primaryClass = "hep-th",
    reportNumber = "NCTS-TH/1714, NCTS-TH-1714",
    doi = "10.1007/JHEP11(2021)070",
    journal = "JHEP",
    volume = "11",
    pages = "070",
    year = "2021"
}

@article{Zhang:2021eeo,
    author = "Zhang, Cen",
    title = "{SMEFTs living on the edge: determining the UV theories from positivity and extremality}",
    eprint = "2112.11665",
    archivePrefix = "arXiv",
    primaryClass = "hep-ph",
    doi = "10.1007/JHEP12(2022)096",
    journal = "JHEP",
    volume = "12",
    pages = "096",
    year = "2022"
}

@article{marsaglia1972choosing,
  title={Choosing a point from the surface of a sphere},
  author={Marsaglia, George},
  journal={The Annals of Mathematical Statistics},
  volume={43},
  number={2},
  pages={645--646},
  year={1972},
  publisher={Institute of Mathematical Statistics}
}

@article{muller1959note,
  title={A note on a method for generating points uniformly on n-dimensional spheres},
  author={Muller, Mervin E},
  journal={Communications of the ACM},
  volume={2},
  number={4},
  pages={19--20},
  year={1959},
  publisher={ACM New York, NY, USA}
}

@article{Bi:2019phv,
    author = "Bi, Qi and Zhang, Cen and Zhou, Shuang-Yong",
    title = "{Positivity constraints on aQGC: carving out the physical parameter space}",
    eprint = "1902.08977",
    archivePrefix = "arXiv",
    primaryClass = "hep-ph",
    reportNumber = "USTC-ICTS-19-01",
    doi = "10.1007/JHEP06(2019)137",
    journal = "JHEP",
    volume = "06",
    pages = "137",
    year = "2019"
}

@article{Zhang:2018shp,
    author = "Zhang, Cen and Zhou, Shuang-Yong",
    title = "{Positivity bounds on vector boson scattering at the LHC}",
    eprint = "1808.00010",
    archivePrefix = "arXiv",
    primaryClass = "hep-ph",
    reportNumber = "USTC-ICTS-18-13",
    doi = "10.1103/PhysRevD.100.095003",
    journal = "Phys. Rev. D",
    volume = "100",
    number = "9",
    pages = "095003",
    year = "2019"
}

@article{Li:2021lpe,
    author = "Li, Xu and Xu, Hao and Yang, Chengjie and Zhang, Cen and Zhou, Shuang-Yong",
    title = "{Positivity in Multifield Effective Field Theories}",
    eprint = "2101.01191",
    archivePrefix = "arXiv",
    primaryClass = "hep-ph",
    reportNumber = "USTC-ICTS/PCFT-21-01",
    doi = "10.1103/PhysRevLett.127.121601",
    journal = "Phys. Rev. Lett.",
    volume = "127",
    number = "12",
    pages = "121601",
    year = "2021"
}

@article{Martin:1965jj,
    author = "Martin, Andre",
    title = "{Extension of the axiomatic analyticity domain of scattering amplitudes by unitarity. 1.}",
    doi = "10.1007/BF02720568",
    journal = "Nuovo Cim. A",
    volume = "42",
    pages = "930--953",
    year = "1965"
}

@inproceedings{deRham:2022hpx,
    author = "de Rham, Claudia and Kundu, Sandipan and Reece, Matthew and Tolley, Andrew J. and Zhou, Shuang-Yong",
    title = "{Snowmass White Paper: UV Constraints on IR Physics}",
    booktitle = "{Snowmass 2021}",
    eprint = "2203.06805",
    archivePrefix = "arXiv",
    primaryClass = "hep-th",
    month = "3",
    year = "2022"
}

@article{Chen:2026xx,
    author = "Chen, Zhuo-Yan and Chang, Fu-Ming and Zhou, Shuang-Yong",
    title = "{Full positivity bounds for anomalous quartic gauge couplings in SMEFT: Dual cone, in preparation}",
    eprint = "xxxx.xxxxx",
    archivePrefix = "arXiv",
    primaryClass = "hep-th",
    year = "2026"
}

@article{Adams:2006sv,
    author = "Adams, Allan and Arkani-Hamed, Nima and Dubovsky, Sergei and Nicolis, Alberto and Rattazzi, Riccardo",
    title = "{Causality, analyticity and an IR obstruction to UV completion}",
    eprint = "hep-th/0602178",
    archivePrefix = "arXiv",
    reportNumber = "CERN-PH-TH-2006-033, HUTP-06-A0005",
    doi = "10.1088/1126-6708/2006/10/014",
    journal = "JHEP",
    volume = "10",
    pages = "014",
    year = "2006"
}

@article{deRham:2017avq,
    author = "de Rham, Claudia and Melville, Scott and Tolley, Andrew J. and Zhou, Shuang-Yong",
    title = "{Positivity bounds for scalar field theories}",
    eprint = "1702.06134",
    archivePrefix = "arXiv",
    primaryClass = "hep-th",
    doi = "10.1103/PhysRevD.96.081702",
    journal = "Phys. Rev. D",
    volume = "96",
    number = "8",
    pages = "081702",
    year = "2017"
}

@article{deRham:2017zjm,
    author = "de Rham, Claudia and Melville, Scott and Tolley, Andrew J. and Zhou, Shuang-Yong",
    title = "{UV complete me: Positivity Bounds for Particles with Spin}",
    eprint = "1706.02712",
    archivePrefix = "arXiv",
    primaryClass = "hep-th",
    doi = "10.1007/JHEP03(2018)011",
    journal = "JHEP",
    volume = "03",
    pages = "011",
    year = "2018"
}

@article{Pham:1985cr,
    author = "Pham, T. N. and Truong, Tran N.",
    title = "{Evaluation of the Derivative Quartic Terms of the Meson Chiral Lagrangian From Forward Dispersion Relation}",
    reportNumber = "Print-85-0588 (ECOLE POLY)",
    doi = "10.1103/PhysRevD.31.3027",
    journal = "Phys. Rev. D",
    volume = "31",
    pages = "3027",
    year = "1985"
}

@article{Pennington:1994kc,
    author = "Pennington, M. R. and Portoles, J.",
    title = "{The Chiral Lagrangian parameters, l1, l2, are determined by the rho resonance}",
    eprint = "hep-ph/9409426",
    archivePrefix = "arXiv",
    reportNumber = "DTP-94-54",
    doi = "10.1016/0370-2693(94)01551-M",
    journal = "Phys. Lett. B",
    volume = "344",
    pages = "399--406",
    year = "1995"
}

@article{Ananthanarayan:1994hf,
    author = "Ananthanarayan, B. and Toublan, D. and Wanders, G.",
    title = "{Consistency of the chiral pion pion scattering amplitudes with axiomatic constraints}",
    eprint = "hep-ph/9410302",
    archivePrefix = "arXiv",
    reportNumber = "UNIL-TP-4-94",
    doi = "10.1103/PhysRevD.51.1093",
    journal = "Phys. Rev. D",
    volume = "51",
    pages = "1093--1100",
    year = "1995"
}

@article{Comellas:1995hq,
    author = "Comellas, Jordi and Latorre, Jose Ignacio and Taron, Josep",
    title = "{Constraints on chiral perturbation theory parameters from QCD inequalities}",
    eprint = "hep-ph/9507258",
    archivePrefix = "arXiv",
    reportNumber = "UB-ECM-PF-95-14",
    doi = "10.1016/0370-2693(95)01110-C",
    journal = "Phys. Lett. B",
    volume = "360",
    pages = "109--116",
    year = "1995"
}

@article{Manohar:2008tc,
    author = "Manohar, Aneesh V. and Mateu, Vicent",
    title = "{Dispersion Relation Bounds for pi pi Scattering}",
    eprint = "0801.3222",
    archivePrefix = "arXiv",
    primaryClass = "hep-ph",
    reportNumber = "IFIC-08-01, FTUV-07-0121",
    doi = "10.1103/PhysRevD.77.094019",
    journal = "Phys. Rev. D",
    volume = "77",
    pages = "094019",
    year = "2008"
}

@article{Low:2009di,
    author = "Low, Ian and Rattazzi, Riccardo and Vichi, Alessandro",
    title = "{Theoretical Constraints on the Higgs Effective Couplings}",
    eprint = "0907.5413",
    archivePrefix = "arXiv",
    primaryClass = "hep-ph",
    doi = "10.1007/JHEP04(2010)126",
    journal = "JHEP",
    volume = "04",
    pages = "126",
    year = "2010"
}

@article{Sanz-Cillero:2013ipa,
    author = "Sanz-Cillero, Juan Jose and Yao, De-Liang and Zheng, Han-Qing",
    title = "{Positivity constraints on the low-energy constants of the chiral pion-nucleon Lagrangian}",
    eprint = "1312.0664",
    archivePrefix = "arXiv",
    primaryClass = "hep-ph",
    reportNumber = "FTUAM-13-37, IFT-UAM-CSIC-13-127",
    doi = "10.1140/epjc/s10052-014-2763-0",
    journal = "Eur. Phys. J. C",
    volume = "74",
    pages = "2763",
    year = "2014"
}

@article{Bellazzini:2016xrt,
    author = "Bellazzini, Brando",
    title = "{Softness and amplitudes{\textquoteright} positivity for spinning particles}",
    eprint = "1605.06111",
    archivePrefix = "arXiv",
    primaryClass = "hep-th",
    reportNumber = "SACLAY-T16-038",
    doi = "10.1007/JHEP02(2017)034",
    journal = "JHEP",
    volume = "02",
    pages = "034",
    year = "2017"
}

@article{deRham:2018qqo,
    author = "de Rham, Claudia and Melville, Scott and Tolley, Andrew J. and Zhou, Shuang-Yong",
    title = "{Positivity Bounds for Massive Spin-1 and Spin-2 Fields}",
    eprint = "1804.10624",
    archivePrefix = "arXiv",
    primaryClass = "hep-th",
    reportNumber = "Imperial/TP/2018/CdR/01, IMPERIAL-TP-2018-CDR-01",
    doi = "10.1007/JHEP03(2019)182",
    journal = "JHEP",
    volume = "03",
    pages = "182",
    year = "2019"
}

@article{deRham:2017imi,
    author = "de Rham, Claudia and Melville, Scott and Tolley, Andrew J. and Zhou, Shuang-Yong",
    title = "{Massive Galileon Positivity Bounds}",
    eprint = "1702.08577",
    archivePrefix = "arXiv",
    primaryClass = "hep-th",
    doi = "10.1007/JHEP09(2017)072",
    journal = "JHEP",
    volume = "09",
    pages = "072",
    year = "2017"
}

@article{Bellazzini:2015cra,
    author = "Bellazzini, Brando and Cheung, Clifford and Remmen, Grant N.",
    title = "{Quantum Gravity Constraints from Unitarity and Analyticity}",
    eprint = "1509.00851",
    archivePrefix = "arXiv",
    primaryClass = "hep-th",
    reportNumber = "CALT-TH-2015-044, SACLAY-T15-161",
    doi = "10.1103/PhysRevD.93.064076",
    journal = "Phys. Rev. D",
    volume = "93",
    number = "6",
    pages = "064076",
    year = "2016"
}

@article{Cheung:2016yqr,
    author = "Cheung, Clifford and Remmen, Grant N.",
    title = "{Positive Signs in Massive Gravity}",
    eprint = "1601.04068",
    archivePrefix = "arXiv",
    primaryClass = "hep-th",
    reportNumber = "CALT-TH-2015-062",
    doi = "10.1007/JHEP04(2016)002",
    journal = "JHEP",
    volume = "04",
    pages = "002",
    year = "2016"
}

@article{Bonifacio:2016wcb,
    author = "Bonifacio, James and Hinterbichler, Kurt and Rosen, Rachel A.",
    title = "{Positivity constraints for pseudolinear massive spin-2 and vector Galileons}",
    eprint = "1607.06084",
    archivePrefix = "arXiv",
    primaryClass = "hep-th",
    doi = "10.1103/PhysRevD.94.104001",
    journal = "Phys. Rev. D",
    volume = "94",
    number = "10",
    pages = "104001",
    year = "2016"
}

@article{Du:2016tgp,
    author = "Du, Meng-Lin and Guo, Feng-Kun and Mei{\ss}ner, Ulf-G. and Yao, De-Liang",
    title = "{Aspects of the low-energy constants in the chiral Lagrangian for charmed mesons}",
    eprint = "1610.02963",
    archivePrefix = "arXiv",
    primaryClass = "hep-ph",
    doi = "10.1103/PhysRevD.94.094037",
    journal = "Phys. Rev. D",
    volume = "94",
    number = "9",
    pages = "094037",
    year = "2016"
}

@article{Bellazzini:2017fep,
    author = "Bellazzini, Brando and Riva, Francesco and Serra, Javi and Sgarlata, Francesco",
    title = "{Beyond Positivity Bounds and the Fate of Massive Gravity}",
    eprint = "1710.02539",
    archivePrefix = "arXiv",
    primaryClass = "hep-th",
    reportNumber = "CERN-TH-2017-201",
    doi = "10.1103/PhysRevLett.120.161101",
    journal = "Phys. Rev. Lett.",
    volume = "120",
    number = "16",
    pages = "161101",
    year = "2018"
}

@article{Hinterbichler:2017qyt,
    author = "Hinterbichler, Kurt and Joyce, Austin and Rosen, Rachel A.",
    title = "{Massive Spin-2 Scattering and Asymptotic Superluminality}",
    eprint = "1708.05716",
    archivePrefix = "arXiv",
    primaryClass = "hep-th",
    doi = "10.1007/JHEP03(2018)051",
    journal = "JHEP",
    volume = "03",
    pages = "051",
    year = "2018"
}

@article{Bellazzini:2017bkb,
    author = "Bellazzini, Brando and Riva, Francesco and Serra, Javi and Sgarlata, Francesco",
    title = "{The other effective fermion compositeness}",
    eprint = "1706.03070",
    archivePrefix = "arXiv",
    primaryClass = "hep-ph",
    reportNumber = "CERN-TH-2017-119",
    doi = "10.1007/JHEP11(2017)020",
    journal = "JHEP",
    volume = "11",
    pages = "020",
    year = "2017"
}

@article{Bonifacio:2018vzv,
    author = "Bonifacio, James and Hinterbichler, Kurt",
    title = "{Bounds on Amplitudes in Effective Theories with Massive Spinning Particles}",
    eprint = "1804.08686",
    archivePrefix = "arXiv",
    primaryClass = "hep-th",
    doi = "10.1103/PhysRevD.98.045003",
    journal = "Phys. Rev. D",
    volume = "98",
    number = "4",
    pages = "045003",
    year = "2018"
}

@article{Bellazzini:2019xts,
    author = "Bellazzini, Brando and Lewandowski, Matthew and Serra, Javi",
    title = "{Positivity of Amplitudes, Weak Gravity Conjecture, and Modified Gravity}",
    eprint = "1902.03250",
    archivePrefix = "arXiv",
    primaryClass = "hep-th",
    doi = "10.1103/PhysRevLett.123.251103",
    journal = "Phys. Rev. Lett.",
    volume = "123",
    number = "25",
    pages = "251103",
    year = "2019"
}

@article{Melville:2019wyy,
    author = "Melville, Scott and Noller, Johannes",
    title = "{Positivity in the Sky: Constraining dark energy and modified gravity from the UV}",
    eprint = "1904.05874",
    archivePrefix = "arXiv",
    primaryClass = "astro-ph.CO",
    doi = "10.1103/PhysRevD.101.021502",
    journal = "Phys. Rev. D",
    volume = "101",
    number = "2",
    pages = "021502",
    year = "2020",
    note = "[Erratum: Phys.Rev.D 102, 049902 (2020)]"
}

@article{Melville:2019tdc,
    author = "Melville, Scott and Roest, Diederik and Stefanyszyn, David",
    title = "{UV Constraints on Massive Spinning Particles: Lessons from the Gravitino}",
    eprint = "1911.03126",
    archivePrefix = "arXiv",
    primaryClass = "hep-th",
    doi = "10.1007/JHEP02(2020)185",
    journal = "JHEP",
    volume = "02",
    pages = "185",
    year = "2020"
}

@article{deRham:2019ctd,
    author = "de Rham, Claudia and Tolley, Andrew J.",
    title = "{Speed of gravity}",
    eprint = "1909.00881",
    archivePrefix = "arXiv",
    primaryClass = "hep-th",
    reportNumber = "Imperial/TP/2019/CdR/04",
    doi = "10.1103/PhysRevD.101.063518",
    journal = "Phys. Rev. D",
    volume = "101",
    number = "6",
    pages = "063518",
    year = "2020"
}

@article{Alberte:2019xfh,
    author = "Alberte, Lasma and de Rham, Claudia and Momeni, Arshia and Rumbutis, Justinas and Tolley, Andrew J.",
    title = "{Positivity Constraints on Interacting Spin-2 Fields}",
    eprint = "1910.11799",
    archivePrefix = "arXiv",
    primaryClass = "hep-th",
    reportNumber = "Imperial/TP/2019/LA/03",
    doi = "10.1007/JHEP03(2020)097",
    journal = "JHEP",
    volume = "03",
    pages = "097",
    year = "2020"
}

@article{Alberte:2019zhd,
    author = "Alberte, Lasma and de Rham, Claudia and Momeni, Arshia and Rumbutis, Justinas and Tolley, Andrew J.",
    title = "{Positivity Constraints on Interacting Pseudo-Linear Spin-2 Fields}",
    eprint = "1912.10018",
    archivePrefix = "arXiv",
    primaryClass = "hep-th",
    reportNumber = "Imperial/TP/2019/LA/04",
    doi = "10.1007/JHEP07(2020)121",
    journal = "JHEP",
    volume = "07",
    pages = "121",
    year = "2020"
}

@article{Ye:2019oxx,
    author = "Ye, Gen and Piao, Yun-Song",
    title = "{Positivity in the effective field theory of cosmological perturbations}",
    eprint = "1908.08644",
    archivePrefix = "arXiv",
    primaryClass = "hep-th",
    doi = "10.1140/epjc/s10052-020-7973-z",
    journal = "Eur. Phys. J. C",
    volume = "80",
    number = "5",
    pages = "421",
    year = "2020"
}

@article{Wang:2020jxr,
    author = "Wang, Yu-Jia and Guo, Feng-Kun and Zhang, Cen and Zhou, Shuang-Yong",
    title = "{Generalized positivity bounds on chiral perturbation theory}",
    eprint = "2004.03992",
    archivePrefix = "arXiv",
    primaryClass = "hep-ph",
    doi = "10.1007/JHEP07(2020)214",
    journal = "JHEP",
    volume = "07",
    pages = "214",
    year = "2020"
}

@article{Alberte:2020jsk,
    author = "Alberte, Lasma and de Rham, Claudia and Jaitly, Sumer and Tolley, Andrew J.",
    title = "{Positivity Bounds and the Massless Spin-2 Pole}",
    eprint = "2007.12667",
    archivePrefix = "arXiv",
    primaryClass = "hep-th",
    reportNumber = "Imperial/TP/2020/LA/02",
    doi = "10.1103/PhysRevD.102.125023",
    journal = "Phys. Rev. D",
    volume = "102",
    number = "12",
    pages = "125023",
    year = "2020"
}

@article{Huang:2020nqy,
    author = "Huang, Yu-tin and Liu, Jin-Yu and Rodina, Laurentiu and Wang, Yihong",
    title = "{Carving out the Space of Open-String S-matrix}",
    eprint = "2008.02293",
    archivePrefix = "arXiv",
    primaryClass = "hep-th",
    reportNumber = "NCTS-TH/2011",
    doi = "10.1007/JHEP04(2021)195",
    journal = "JHEP",
    volume = "04",
    pages = "195",
    year = "2021"
}

@article{Tokuda:2020mlf,
    author = "Tokuda, Junsei and Aoki, Katsuki and Hirano, Shin'ichi",
    title = "{Gravitational positivity bounds}",
    eprint = "2007.15009",
    archivePrefix = "arXiv",
    primaryClass = "hep-th",
    reportNumber = "KOBE-COSMO-20-13, YITP-20-98",
    doi = "10.1007/JHEP11(2020)054",
    journal = "JHEP",
    volume = "11",
    pages = "054",
    year = "2020"
}

@article{Bellazzini:2018paj,
    author = "Bellazzini, Brando and Riva, Francesco",
    title = "{New phenomenological and theoretical perspective on anomalous ZZ and Z{\ensuremath{\gamma}} processes}",
    eprint = "1806.09640",
    archivePrefix = "arXiv",
    primaryClass = "hep-ph",
    doi = "10.1103/PhysRevD.98.095021",
    journal = "Phys. Rev. D",
    volume = "98",
    number = "9",
    pages = "095021",
    year = "2018"
}

@article{Remmen:2020vts,
    author = "Remmen, Grant N. and Rodd, Nicholas L.",
    title = "{Flavor Constraints from Unitarity and Analyticity}",
    eprint = "2004.02885",
    archivePrefix = "arXiv",
    primaryClass = "hep-ph",
    doi = "10.1103/PhysRevLett.127.149901",
    journal = "Phys. Rev. Lett.",
    volume = "125",
    number = "8",
    pages = "081601",
    year = "2020",
    note = "[Erratum: Phys.Rev.Lett. 127, 149901 (2021)]"
}

@article{Chen:2023bhu,
    author = "Chen, Qing and Mimasu, Ken and Wu, Tong Arthur and Zhang, Guo-Dong and Zhou, Shuang-Yong",
    title = "{Capping the positivity cone: dimension-8 Higgs operators in the SMEFT}",
    eprint = "2309.15922",
    archivePrefix = "arXiv",
    primaryClass = "hep-ph",
    doi = "10.1007/JHEP03(2024)180",
    journal = "JHEP",
    volume = "03",
    pages = "180",
    year = "2024"
}

@inproceedings{Hong:2024fbl,
    author = "Hong, Dong-Yu and Wang, Zhuo-Hui and Zhou, Shuang-Yong",
    title = "{On Capped Higgs Positivity Cone}",
    eprint = "2404.04479",
    archivePrefix = "arXiv",
    primaryClass = "hep-ph",
    month = "4",
    year = "2024"
}

@article{Weinberg:1978kz,
    author = "Weinberg, Steven",
    editor = "Deser, S.",
    title = "{Phenomenological Lagrangians}",
    reportNumber = "HUTP-78-A051A",
    doi = "10.1016/0378-4371(79)90223-1",
    journal = "Physica A",
    volume = "96",
    number = "1-2",
    pages = "327--340",
    year = "1979"
}

@article{Buchmuller:1985jz,
    author = "Buchmuller, W. and Wyler, D.",
    title = "{Effective Lagrangian Analysis of New Interactions and Flavor Conservation}",
    reportNumber = "CERN-TH-4254/85",
    doi = "10.1016/0550-3213(86)90262-2",
    journal = "Nucl. Phys. B",
    volume = "268",
    pages = "621--653",
    year = "1986"
}

@article{Leung:1984ni,
    author = "Leung, Chung Ngoc and Love, S. T. and Rao, S.",
    title = "{Low-Energy Manifestations of a New Interaction Scale: Operator Analysis}",
    reportNumber = "FERMILAB-PUB-84-074-T",
    doi = "10.1007/BF01588041",
    journal = "Z. Phys. C",
    volume = "31",
    pages = "433",
    year = "1986"
}

@article{deRham:2025vaq,
    author = "de Rham, Claudia and Tolley, Andrew J. and Wang, Zhuo-Hui and Zhou, Shuang-Yong",
    title = "{Primal S-matrix bootstrap with dispersion relations}",
    eprint = "2506.22546",
    archivePrefix = "arXiv",
    primaryClass = "hep-th",
    reportNumber = "Imperial/TP/2025/cdr/3, USTC-ICTS/PCFT-25-24",
    doi = "10.1007/JHEP01(2026)027",
    journal = "JHEP",
    volume = "01",
    pages = "027",
    year = "2026"
}

@article{Wan:2024eto,
    author = "Wan, Shi-Lin and Zhou, Shuang-Yong",
    title = "{Matrix moment approach to positivity bounds and UV reconstruction from IR}",
    eprint = "2411.11964",
    archivePrefix = "arXiv",
    primaryClass = "hep-th",
    doi = "10.1007/JHEP02(2025)168",
    journal = "JHEP",
    volume = "02",
    pages = "168",
    year = "2025"
}

@article{Xu:2024iao,
    author = "Xu, Hao and Hong, Dong-Yu and Wang, Zhuo-Hui and Zhou, Shuang-Yong",
    title = "{Positivity bounds on parity-violating scalar-tensor EFTs}",
    eprint = "2410.09794",
    archivePrefix = "arXiv",
    primaryClass = "hep-th",
    doi = "10.1088/1475-7516/2025/01/102",
    journal = "JCAP",
    volume = "01",
    pages = "102",
    year = "2025"
}

@article{Du:2021byy,
    author = "Du, Zong-Zhe and Zhang, Cen and Zhou, Shuang-Yong",
    title = "{Triple crossing positivity bounds for multi-field theories}",
    eprint = "2111.01169",
    archivePrefix = "arXiv",
    primaryClass = "hep-th",
    reportNumber = "USTC-ICTS/PCFT-21-42",
    doi = "10.1007/JHEP12(2021)115",
    journal = "JHEP",
    volume = "12",
    pages = "115",
    year = "2021"
}

@article{Eboli:2016kko,
    author = "{\'E}boli, O. J. P. and Gonzalez-Garcia, M. C.",
    title = "{Classifying the bosonic quartic couplings}",
    eprint = "1604.03555",
    archivePrefix = "arXiv",
    primaryClass = "hep-ph",
    reportNumber = "YITP-SB-16-09",
    doi = "10.1103/PhysRevD.93.093013",
    journal = "Phys. Rev. D",
    volume = "93",
    number = "9",
    pages = "093013",
    year = "2016"
}

@article{Green:2016trm,
    author = "Green, Daniel R. and Meade, Patrick and Pleier, Marc-Andre",
    title = "{Multiboson interactions at the LHC}",
    eprint = "1610.07572",
    archivePrefix = "arXiv",
    primaryClass = "hep-ex",
    reportNumber = "FERMILAB-PUB-16-507-CMS, BNL-114542-2017-JA",
    doi = "10.1103/RevModPhys.89.035008",
    journal = "Rev. Mod. Phys.",
    volume = "89",
    number = "3",
    pages = "035008",
    year = "2017"
}

@article{Chang:2013aya,
    author = "Chang, Jung and Cheung, Kingman and Lu, Chih-Ting and Yuan, Tzu-Chiang",
    title = "{WW scattering in the era of post-Higgs-boson discovery}",
    eprint = "1303.6335",
    archivePrefix = "arXiv",
    primaryClass = "hep-ph",
    doi = "10.1103/PhysRevD.87.093005",
    journal = "Phys. Rev. D",
    volume = "87",
    pages = "093005",
    year = "2013"
}

@article{Li:2022aby,
    author = "Li, Xu",
    title = "{Positivity bounds at one-loop level: the Higgs sector}",
    eprint = "2212.12227",
    archivePrefix = "arXiv",
    primaryClass = "hep-ph",
    doi = "10.1007/JHEP05(2023)230",
    journal = "JHEP",
    volume = "05",
    pages = "230",
    year = "2023"
}

@article{Remmen:2020uze,
    author = "Remmen, Grant N. and Rodd, Nicholas L.",
    title = "{Signs, spin, SMEFT: Sum rules at dimension six}",
    eprint = "2010.04723",
    archivePrefix = "arXiv",
    primaryClass = "hep-ph",
    doi = "10.1103/PhysRevD.105.036006",
    journal = "Phys. Rev. D",
    volume = "105",
    number = "3",
    pages = "036006",
    year = "2022"
}

@article{Gu:2020thj,
    author = "Gu, Jiayin and Wang, Lian-Tao",
    title = "{Sum Rules in the Standard Model Effective Field Theory from Helicity Amplitudes}",
    eprint = "2008.07551",
    archivePrefix = "arXiv",
    primaryClass = "hep-ph",
    reportNumber = "MITP/20-040",
    doi = "10.1007/JHEP03(2021)149",
    journal = "JHEP",
    volume = "03",
    pages = "149",
    year = "2021"
}

@article{Fuks:2020ujk,
    author = "Fuks, Benjamin and Liu, Yiming and Zhang, Cen and Zhou, Shuang-Yong",
    title = "{Positivity in electron-positron scattering: testing the axiomatic quantum field theory principles and probing the existence of UV states}",
    eprint = "2009.02212",
    archivePrefix = "arXiv",
    primaryClass = "hep-ph",
    reportNumber = "USTC-ICTS/PCFT-20-25",
    doi = "10.1088/1674-1137/abcd8c",
    journal = "Chin. Phys. C",
    volume = "45",
    number = "2",
    pages = "023108",
    year = "2021"
}

@article{Gu:2020ldn,
    author = "Gu, Jiayin and Wang, Lian-Tao and Zhang, Cen",
    title = "{Unambiguously Testing Positivity at Lepton Colliders}",
    eprint = "2011.03055",
    archivePrefix = "arXiv",
    primaryClass = "hep-ph",
    reportNumber = "MITP/20-063",
    doi = "10.1103/PhysRevLett.129.011805",
    journal = "Phys. Rev. Lett.",
    volume = "129",
    number = "1",
    pages = "011805",
    year = "2022"
}

@article{Bonnefoy:2020yee,
    author = "Bonnefoy, Quentin and Gendy, Emanuele and Grojean, Christophe",
    title = "{Positivity bounds on Minimal Flavor Violation}",
    eprint = "2011.12855",
    archivePrefix = "arXiv",
    primaryClass = "hep-ph",
    reportNumber = "DESY-20-201, DESY 20-201",
    doi = "10.1007/JHEP04(2021)115",
    journal = "JHEP",
    volume = "04",
    pages = "115",
    year = "2021"
}

@article{Davighi:2021osh,
    author = "Davighi, Joe and Melville, Scott and You, Tevong",
    title = "{Natural selection rules: new positivity bounds for massive spinning particles}",
    eprint = "2108.06334",
    archivePrefix = "arXiv",
    primaryClass = "hep-th",
    reportNumber = "CERN-TH-2021-122",
    doi = "10.1007/JHEP02(2022)167",
    journal = "JHEP",
    volume = "02",
    pages = "167",
    year = "2022"
}

@article{Chala:2021wpj,
    author = "Chala, Mikael and Santiago, Jose",
    title = "{Positivity bounds in the standard model effective field theory beyond tree level}",
    eprint = "2110.01624",
    archivePrefix = "arXiv",
    primaryClass = "hep-ph",
    doi = "10.1103/PhysRevD.105.L111901",
    journal = "Phys. Rev. D",
    volume = "105",
    number = "11",
    pages = "L111901",
    year = "2022"
}

@article{Ghosh:2022qqq,
    author = "Ghosh, Diptimoy and Sharma, Rajat and Ullah, Farman",
    title = "{Amplitude{\textquoteright}s positivity vs. subluminality: causality and unitarity constraints on dimension 6 {\&} 8 gluonic operators in the SMEFT}",
    eprint = "2211.01322",
    archivePrefix = "arXiv",
    primaryClass = "hep-ph",
    doi = "10.1007/JHEP02(2023)199",
    journal = "JHEP",
    volume = "02",
    pages = "199",
    year = "2023"
}

@article{Remmen:2022orj,
    author = "Remmen, Grant N. and Rodd, Nicholas L.",
    title = "{Spinning sum rules for the dimension-six SMEFT}",
    eprint = "2206.13524",
    archivePrefix = "arXiv",
    primaryClass = "hep-ph",
    reportNumber = "CERN-TH-2022-105",
    doi = "10.1007/JHEP09(2022)030",
    journal = "JHEP",
    volume = "09",
    pages = "030",
    year = "2022"
}

@article{Li:2022tcz,
    author = "Li, Xu and Zhou, Shun",
    title = "{Origin of neutrino masses on the convex cone of positivity bounds}",
    eprint = "2202.12907",
    archivePrefix = "arXiv",
    primaryClass = "hep-ph",
    doi = "10.1103/PhysRevD.107.L031902",
    journal = "Phys. Rev. D",
    volume = "107",
    number = "3",
    pages = "L031902",
    year = "2023"
}

@article{Li:2022rag,
    author = "Li, Xu and Mimasu, Ken and Yamashita, Kimiko and Yang, Chengjie and Zhang, Cen and Zhou, Shuang-Yong",
    title = "{Moments for positivity: using Drell-Yan data to test positivity bounds and reverse-engineer new physics}",
    eprint = "2204.13121",
    archivePrefix = "arXiv",
    primaryClass = "hep-ph",
    reportNumber = "USTC-ICTS/PCFT-22-12, KCL-PH-TH/2022-09",
    doi = "10.1007/JHEP10(2022)107",
    journal = "JHEP",
    volume = "10",
    pages = "107",
    year = "2022"
}

@article{Altmannshofer:2023bfk,
    author = "Altmannshofer, Wolfgang and Gori, Stefania and Lehmann, Benjamin V. and Zuo, Jianhong",
    title = "{UV physics from IR features: New prospects from top flavor violation}",
    eprint = "2303.00781",
    archivePrefix = "arXiv",
    primaryClass = "hep-ph",
    reportNumber = "MIT-CTP/5534",
    doi = "10.1103/PhysRevD.107.095025",
    journal = "Phys. Rev. D",
    volume = "107",
    number = "9",
    pages = "095025",
    year = "2023"
}

@article{Davighi:2023acq,
    author = "Davighi, Joe and Melville, Scott and Mimasu, Ken and You, Tevong",
    title = "{Positivity and the electroweak hierarchy}",
    eprint = "2308.06226",
    archivePrefix = "arXiv",
    primaryClass = "hep-ph",
    reportNumber = "KCL-PH-TH/2023-44",
    doi = "10.1103/PhysRevD.109.033009",
    journal = "Phys. Rev. D",
    volume = "109",
    number = "3",
    pages = "033009",
    year = "2024"
}

@article{Caron-Huot:2020cmc,
    author = "Caron-Huot, Simon and Van Duong, Vincent",
    title = "{Extremal Effective Field Theories}",
    eprint = "2011.02957",
    archivePrefix = "arXiv",
    primaryClass = "hep-th",
    doi = "10.1007/JHEP05(2021)280",
    journal = "JHEP",
    volume = "05",
    pages = "280",
    year = "2021"
}

@article{EliasMiro:2022xaa,
    author = "Elias Miro, Joan and Guerrieri, Andrea and Gumus, Mehmet Asim",
    title = "{Bridging positivity and S-matrix bootstrap bounds}",
    eprint = "2210.01502",
    archivePrefix = "arXiv",
    primaryClass = "hep-th",
    doi = "10.1007/JHEP05(2023)001",
    journal = "JHEP",
    volume = "05",
    pages = "001",
    year = "2023"
}

@article{Remmen:2024hry,
    author = "Remmen, Grant N. and Rodd, Nicholas L.",
    title = "{Positively Identifying HEFT or SMEFT}",
    eprint = "2412.07827",
    archivePrefix = "arXiv",
    primaryClass = "hep-ph",
    month = "12",
    year = "2024"
}

@inproceedings{deBlas:2022ofj,
    author = "de Blas, Jorge and Du, Yong and Grojean, Christophe and Gu, Jiayin and Miralles, Victor and Peskin, Michael E. and Tian, Junping and Vos, Marcel and Vryonidou, Eleni",
    title = "{Global SMEFT Fits at Future Colliders}",
    booktitle = "{Snowmass 2021}",
    eprint = "2206.08326",
    archivePrefix = "arXiv",
    primaryClass = "hep-ph",
    month = "6",
    year = "2022"
}

@article{Covarelli:2021gyz,
    author = "Covarelli, Roberto and Pellen, Mathieu and Zaro, Marco",
    title = "{Vector-Boson scattering at the LHC: Unraveling the electroweak sector}",
    eprint = "2102.10991",
    archivePrefix = "arXiv",
    primaryClass = "hep-ph",
    reportNumber = "FR-PHENO-2021-05, TIF-UNIMI-2021-2, VBSCAN-PUB-02-21",
    doi = "10.1142/S0217751X2130009X",
    journal = "Int. J. Mod. Phys. A",
    volume = "36",
    number = "16",
    pages = "2130009",
    year = "2021"
}

@article{Chala:2021pll,
    author = "Chala, Mikael and Guedes, Guilherme and Ramos, Maria and Santiago, Jose",
    title = "{Towards the renormalisation of the Standard Model effective field theory to dimension eight: Bosonic interactions I}",
    eprint = "2106.05291",
    archivePrefix = "arXiv",
    primaryClass = "hep-ph",
    doi = "10.21468/SciPostPhys.11.3.065",
    journal = "SciPost Phys.",
    volume = "11",
    pages = "065",
    year = "2021"
}

@article{Boughezal:2021tih,
    author = "Boughezal, Radja and Mereghetti, Emanuele and Petriello, Frank",
    title = "{Dilepton production in the SMEFT at O(1/{\ensuremath{\Lambda}}4)}",
    eprint = "2106.05337",
    archivePrefix = "arXiv",
    primaryClass = "hep-ph",
    reportNumber = "LA-UR-21-25379",
    doi = "10.1103/PhysRevD.104.095022",
    journal = "Phys. Rev. D",
    volume = "104",
    number = "9",
    pages = "095022",
    year = "2021"
}

@article{Gomez-Ambrosio:2018pnl,
    author = "Gomez-Ambrosio, Raquel",
    title = "{Studies of Dimension-Six EFT effects in Vector Boson Scattering}",
    eprint = "1809.04189",
    archivePrefix = "arXiv",
    primaryClass = "hep-ph",
    reportNumber = "IPPP/18/78, VBSCAN-PUB-05-18",
    doi = "10.1140/epjc/s10052-019-6893-2",
    journal = "Eur. Phys. J. C",
    volume = "79",
    number = "5",
    pages = "389",
    year = "2019"
}

@article{Chala:2023xjy,
    author = "Chala, Mikael and Li, Xu",
    title = "{Positivity restrictions on the mixing of dimension-eight SMEFT operators}",
    eprint = "2309.16611",
    archivePrefix = "arXiv",
    primaryClass = "hep-ph",
    doi = "10.1103/PhysRevD.109.065015",
    journal = "Phys. Rev. D",
    volume = "109",
    number = "6",
    pages = "065015",
    year = "2024"
}

@article{Yang:2023ncf,
    author = "Yang, Chengjie and Ren, Zhe and Yu, Jiang-Hao",
    title = "{Positivity from J-Basis operators in the standard model effective Field Theory}",
    eprint = "2312.04663",
    archivePrefix = "arXiv",
    primaryClass = "hep-ph",
    doi = "10.1007/JHEP05(2024)221",
    journal = "JHEP",
    volume = "05",
    pages = "221",
    year = "2024"
}

@article{Tolley:2020gtv,
    author = "Tolley, Andrew J. and Wang, Zi-Yue and Zhou, Shuang-Yong",
    title = "{New positivity bounds from full crossing symmetry}",
    eprint = "2011.02400",
    archivePrefix = "arXiv",
    primaryClass = "hep-th",
    doi = "10.1007/JHEP05(2021)255",
    journal = "JHEP",
    volume = "05",
    pages = "255",
    year = "2021"
}

@article{Chiang:2022ltp,
    author = "Chiang, Li-Yuan and Huang, Yu-tin and Rodina, Laurentiu and Weng, He-Chen",
    title = "{De-projecting the EFThedron}",
    eprint = "2204.07140",
    archivePrefix = "arXiv",
    primaryClass = "hep-th",
    doi = "10.1007/JHEP05(2024)102",
    journal = "JHEP",
    volume = "05",
    pages = "102",
    year = "2024"
}

@article{Arkani-Hamed:2020blm,
    author = "Arkani-Hamed, Nima and Huang, Tzu-Chen and Huang, Yu-tin",
    title = "{The EFT-Hedron}",
    eprint = "2012.15849",
    archivePrefix = "arXiv",
    primaryClass = "hep-th",
    reportNumber = "NCTS-TH/2014, CALT-TH 2020-061",
    doi = "10.1007/JHEP05(2021)259",
    journal = "JHEP",
    volume = "05",
    pages = "259",
    year = "2021"
}

@article{Sinha:2020win,
    author = "Sinha, Aninda and Zahed, Ahmadullah",
    title = "{Crossing Symmetric Dispersion Relations in Quantum Field Theories}",
    eprint = "2012.04877",
    archivePrefix = "arXiv",
    primaryClass = "hep-th",
    doi = "10.1103/PhysRevLett.126.181601",
    journal = "Phys. Rev. Lett.",
    volume = "126",
    number = "18",
    pages = "181601",
    year = "2021"
}

@article{Chiang:2021ziz,
    author = "Chiang, Li-Yuan and Huang, Yu-tin and Li, Wei and Rodina, Laurentiu and Weng, He-Chen",
    title = "{Into the EFThedron and UV constraints from IR consistency}",
    eprint = "2105.02862",
    archivePrefix = "arXiv",
    primaryClass = "hep-th",
    doi = "10.1007/JHEP03(2022)063",
    journal = "JHEP",
    volume = "03",
    pages = "063",
    year = "2022"
}

@article{Guerrieri:2021tak,
    author = "Guerrieri, Andrea and Sever, Amit",
    title = "{Rigorous Bounds on the Analytic S Matrix}",
    eprint = "2106.10257",
    archivePrefix = "arXiv",
    primaryClass = "hep-th",
    doi = "10.1103/PhysRevLett.127.251601",
    journal = "Phys. Rev. Lett.",
    volume = "127",
    number = "25",
    pages = "251601",
    year = "2021"
}

@article{Alberte:2021dnj,
    author = "Alberte, Lasma and de Rham, Claudia and Jaitly, Sumer and Tolley, Andrew J.",
    title = "{Reverse Bootstrapping: IR Lessons for UV Physics}",
    eprint = "2111.09226",
    archivePrefix = "arXiv",
    primaryClass = "hep-th",
    reportNumber = "Imperial/TP/2021/LA/1",
    doi = "10.1103/PhysRevLett.128.051602",
    journal = "Phys. Rev. Lett.",
    volume = "128",
    number = "5",
    pages = "051602",
    year = "2022"
}

@article{Caron-Huot:2021rmr,
    author = "Caron-Huot, Simon and Mazac, Dalimil and Rastelli, Leonardo and Simmons-Duffin, David",
    title = "{Sharp boundaries for the swampland}",
    eprint = "2102.08951",
    archivePrefix = "arXiv",
    primaryClass = "hep-th",
    doi = "10.1007/JHEP07(2021)110",
    journal = "JHEP",
    volume = "07",
    pages = "110",
    year = "2021"
}

@article{Fernandez:2022kzi,
    author = "Fernandez, Clara and Pomarol, Alex and Riva, Francesco and Sciotti, Francesco",
    title = "{Cornering large-N$_{c}$ QCD with positivity bounds}",
    eprint = "2211.12488",
    archivePrefix = "arXiv",
    primaryClass = "hep-th",
    doi = "10.1007/JHEP06(2023)094",
    journal = "JHEP",
    volume = "06",
    pages = "094",
    year = "2023"
}

@article{Albert:2023jtd,
    author = "Albert, Jan and Rastelli, Leonardo",
    title = "{Bootstrapping pions at large N. Part II. Background gauge fields and the chiral anomaly}",
    eprint = "2307.01246",
    archivePrefix = "arXiv",
    primaryClass = "hep-th",
    reportNumber = "YITP-SB-2023-15",
    doi = "10.1007/JHEP09(2024)039",
    journal = "JHEP",
    volume = "09",
    pages = "039",
    year = "2024"
}

@article{Ma:2023vgc,
    author = "Ma, Teng and Pomarol, Alex and Sciotti, Francesco",
    title = "{Bootstrapping the chiral anomaly at large N$_{c}$}",
    eprint = "2307.04729",
    archivePrefix = "arXiv",
    primaryClass = "hep-th",
    doi = "10.1007/JHEP11(2023)176",
    journal = "JHEP",
    volume = "11",
    pages = "176",
    year = "2023"
}

@article{Li:2023qzs,
    author = "Li, Yue-Zhou",
    title = "{Effective field theory bootstrap, large-N {\ensuremath{\chi}}PT and holographic QCD}",
    eprint = "2310.09698",
    archivePrefix = "arXiv",
    primaryClass = "hep-th",
    doi = "10.1007/JHEP01(2024)072",
    journal = "JHEP",
    volume = "01",
    pages = "072",
    year = "2024"
}

@article{Albert:2023seb,
    author = "Albert, Jan and Henriksson, Johan and Rastelli, Leonardo and Vichi, Alessandro",
    title = "{Bootstrapping mesons at large N: Regge trajectory from spin-two maximization}",
    eprint = "2312.15013",
    archivePrefix = "arXiv",
    primaryClass = "hep-th",
    reportNumber = "YITP-SB-2023-41",
    doi = "10.1007/JHEP09(2024)172",
    journal = "JHEP",
    volume = "09",
    pages = "172",
    year = "2024"
}

@article{Caron-Huot:2022ugt,
    author = "Caron-Huot, Simon and Li, Yue-Zhou and Parra-Martinez, Julio and Simmons-Duffin, David",
    title = "{Causality constraints on corrections to Einstein gravity}",
    eprint = "2201.06602",
    archivePrefix = "arXiv",
    primaryClass = "hep-th",
    reportNumber = "CALT-TH 2021-003",
    doi = "10.1007/JHEP05(2023)122",
    journal = "JHEP",
    volume = "05",
    pages = "122",
    year = "2023"
}

@article{Chiang:2022jep,
    author = "Chiang, Li-Yuan and Huang, Yu-tin and Li, Wei and Rodina, Laurentiu and Weng, He-Chen",
    title = "{(Non)-projective bounds on gravitational EFT}",
    eprint = "2201.07177",
    archivePrefix = "arXiv",
    primaryClass = "hep-th",
    month = "1",
    year = "2022"
}

@article{Henriksson:2022oeu,
    author = "Henriksson, Johan and McPeak, Brian and Russo, Francesco and Vichi, Alessandro",
    title = "{Bounding violations of the weak gravity conjecture}",
    eprint = "2203.08164",
    archivePrefix = "arXiv",
    primaryClass = "hep-th",
    doi = "10.1007/JHEP08(2022)184",
    journal = "JHEP",
    volume = "08",
    pages = "184",
    year = "2022"
}

@article{Hong:2023zgm,
    author = "Hong, Dong-Yu and Wang, Zhuo-Hui and Zhou, Shuang-Yong",
    title = "{Causality bounds on scalar-tensor EFTs}",
    eprint = "2304.01259",
    archivePrefix = "arXiv",
    primaryClass = "hep-th",
    doi = "10.1007/JHEP10(2023)135",
    journal = "JHEP",
    volume = "10",
    pages = "135",
    year = "2023"
}

@article{Bellazzini:2023nqj,
    author = "Bellazzini, Brando and Isabella, Giulia and Ricossa, Sergio and Riva, Francesco",
    title = "{Massive gravity is not positive}",
    eprint = "2304.02550",
    archivePrefix = "arXiv",
    primaryClass = "hep-th",
    doi = "10.1103/PhysRevD.109.024051",
    journal = "Phys. Rev. D",
    volume = "109",
    number = "2",
    pages = "024051",
    year = "2024"
}

@article{deRham:2025htd,
    author = "de Rham, Claudia and Jaitly, Sumer and Kaplanek, Greg",
    title = "{Mixed signals in the IR: Positivity bounds with indefinite species}",
    eprint = "2512.11980",
    archivePrefix = "arXiv",
    primaryClass = "hep-th",
    reportNumber = "Imperial-TP-cdr-05",
    month = "12",
    year = "2025"
}

@article{Liu:2025deo,
    author = "Liu, Zhen and Lyu, Kun-Feng and Wu, Tong Arthur",
    title = "{LHC Shines on Positivity}",
    eprint = "2512.04336",
    archivePrefix = "arXiv",
    primaryClass = "hep-ph",
    month = "12",
    year = "2025"
}

@article{Huang:2025icl,
    author = "Huang, Yu-tin and Ricossa, Sara and Riva, Francesco and Tsai, Jie-Da",
    title = "{The Rise of Linear Trajectories}",
    eprint = "2510.07991",
    archivePrefix = "arXiv",
    primaryClass = "hep-th",
    month = "10",
    year = "2025"
}

@article{Cheung:2025nhw,
    author = "Cheung, Clifford and Remmen, Grant N.",
    title = "{Multipositivity bounds for scattering amplitudes}",
    eprint = "2505.05553",
    archivePrefix = "arXiv",
    primaryClass = "hep-th",
    reportNumber = "CALT-TH 2025-010",
    doi = "10.1103/wt4x-2149",
    journal = "Phys. Rev. D",
    volume = "112",
    number = "1",
    pages = "016017",
    year = "2025"
}

@article{Brivio:2017vri,
    author = "Brivio, Ilaria and Trott, Michael",
    title = "{The Standard Model as an Effective Field Theory}",
    eprint = "1706.08945",
    archivePrefix = "arXiv",
    primaryClass = "hep-ph",
    doi = "10.1016/j.physrep.2018.11.002",
    journal = "Phys. Rept.",
    volume = "793",
    pages = "1--98",
    year = "2019"
}

@article{Contino:2013kra,
    author = "Contino, Roberto and Ghezzi, Margherita and Grojean, Christophe and Muhlleitner, Margarete and Spira, Michael",
    title = "{Effective Lagrangian for a light Higgs-like scalar}",
    eprint = "1303.3876",
    archivePrefix = "arXiv",
    primaryClass = "hep-ph",
    reportNumber = "CERN-PH-TH-2013-047, KA-TP-06-2013, PSI-PR-13-04",
    doi = "10.1007/JHEP07(2013)035",
    journal = "JHEP",
    volume = "07",
    pages = "035",
    year = "2013"
}

@article{ATLAS:2020nlt,
    author = "Aad, Georges and others",
    collaboration = "ATLAS",
    title = "{Observation of electroweak production of two jets and a Z-boson pair}",
    eprint = "2004.10612",
    archivePrefix = "arXiv",
    primaryClass = "hep-ex",
    reportNumber = "CERN-EP-2020-016",
    doi = "10.1038/s41567-022-01757-y",
    journal = "Nature Phys.",
    volume = "19",
    number = "2",
    pages = "237--253",
    year = "2023"
}

@article{He:1994br,
    author = "He, Hong-Jian and Kuang, Yu-Ping and Yuan, C. -P.",
    title = "{Equivalence theorem and probing the electroweak symmetry breaking sector}",
    eprint = "hep-ph/9410400",
    archivePrefix = "arXiv",
    reportNumber = "VPI-IHEP-94-04, TUIMP-TH-94-60, MSUHEP-40909",
    doi = "10.1103/PhysRevD.51.6463",
    journal = "Phys. Rev. D",
    volume = "51",
    pages = "6463--6473",
    year = "1995"
}

@article{Degrande:2012wf,
    author = "Degrande, Celine and Greiner, Nicolas and Kilian, Wolfgang and Mattelaer, Olivier and Mebane, Harrison and Stelzer, Tim and Willenbrock, Scott and Zhang, Cen",
    title = "{Effective Field Theory: A Modern Approach to Anomalous Couplings}",
    eprint = "1205.4231",
    archivePrefix = "arXiv",
    primaryClass = "hep-ph",
    reportNumber = "MPP-2011-149, SI-HEP-2011-17, CP3-12-25",
    doi = "10.1016/j.aop.2013.04.016",
    journal = "Annals Phys.",
    volume = "335",
    pages = "21--32",
    year = "2013"
}

@article{Bonnefoy:2025uzf,
    author = "Bonnefoy, Quentin and Cort{\'e}s, Vicente and Gendy, Emanuele and Grojean, Christophe and von Merkl, Karim Ritter and Pilatus, Paula Naomi",
    title = "{Geometry of effective field theory positivity cones}",
    eprint = "2508.18165",
    archivePrefix = "arXiv",
    primaryClass = "math-ph",
    doi = "10.1007/s00023-026-01687-y",
    journal = "Ann. Henri Poincaré ",
    month = "4",
    year = "2026"
}

@article{Chakraborty:2024ciu,
    author = "Chakraborty, Debsubhra and Chattopadhyay, Susobhan and Gupta, Rick S.",
    title = "{Complete set of positivity constraints on the HEFT at NLO}",
    eprint = "2412.14155",
    archivePrefix = "arXiv",
    primaryClass = "hep-ph",
    doi = "10.1103/7qgz-9ykc",
    journal = "Phys. Rev. D",
    volume = "113",
    number = "5",
    pages = "053007",
    year = "2026"
}

@article{Han:1998sg,
    author = "Han, Tao and Lykken, Joseph D. and Zhang, Ren-Jie",
    title = "{On Kaluza-Klein states from large extra dimensions}",
    eprint = "hep-ph/9811350",
    archivePrefix = "arXiv",
    reportNumber = "MADPH-98-1092, FERMILAB-PUB-98-364",
    doi = "10.1103/PhysRevD.59.105006",
    journal = "Phys. Rev. D",
    volume = "59",
    pages = "105006",
    year = "1999"
}

@article{Bellazzini:2014waa,
    author = "Bellazzini, Brando and Martucci, Luca and Torre, Riccardo",
    title = "{Symmetries, Sum Rules and Constraints on Effective Field Theories}",
    eprint = "1405.2960",
    archivePrefix = "arXiv",
    primaryClass = "hep-th",
    reportNumber = "DFPD-2014-TH-10, LPN14-072",
    doi = "10.1007/JHEP09(2014)100",
    journal = "JHEP",
    volume = "09",
    pages = "100",
    year = "2014"
}

\end{document}